\documentclass[letterpaper,twocolumn,10pt]{article}
\usepackage{usenix,epsfig,endnotes}
\usepackage{array,color,url}
\usepackage[linesnumbered,ruled,vlined]{algorithm2e}
\usepackage[thicklines]{cancel}
\usepackage{indentfirst}
\usepackage{algpseudocode}
\usepackage{tabu}
\usepackage{amsmath}
\usepackage{amsthm}
\usepackage{enumitem}
\usepackage{empheq}
\setlist[enumerate,1]{start=0}
\usepackage{amssymb}
\usepackage{bm}
\usepackage{bbm}
\usepackage{graphicx}
\usepackage{float}
\usepackage{epsfig}
\usepackage{subcaption}
\usepackage{epstopdf}
\usepackage{color}
\usepackage{comment}
\usepackage{booktabs}
\usepackage{multirow}
\usepackage{url}
\usepackage{caption}
\usepackage[capitalize]{cleveref}
\usepackage{verbatimbox}
\usepackage{comment}
\usepackage{diagbox}
\usepackage{enumitem}
\usepackage{setspace}
\usepackage[explicit]{titlesec}
\setlist[itemize]{leftmargin=*}
\usepackage{appendix}
\usepackage{tabularx}
\usepackage{color}
\usepackage{makecell}
\usepackage{bbm}
\usepackage{mathtools}
\usepackage{tikz}

\DeclareMathOperator*{\argmax}{arg\,max}

\SetCommentSty{mycommfont}

\begin{document}

\newtheorem{theorem}{Theorem}
\newtheorem{lemma}{Lemma}
\newtheorem{claim}[lemma]{Claim}
\newtheorem{proposition}{Proposition}
\newtheorem{fact}[lemma]{Fact}
\newtheorem{corollary}{Corollary}
\newtheorem{conjecture}{Conjecture}

\newtheorem{definition}{Definition}
\date{}

\title{Poisoning Attacks to Local Differential Privacy Protocols for Trajectory Data}

 \author{
 {\rm I-Jung Hsu$^*$\hspace{12pt} \rm Chih-Hsun Lin$^\dag$\hspace{12pt} \rm Chia-Mu Yu$^\dag$\hspace{12pt} \rm Sy-Yen Kuo$^\ddag$\hspace{12pt} \rm Chun-Ying Huang$^\dag$}\\\\
 $^*$National Taiwan University,\\
 $^\dag$National Yang Ming Chiao Tung University,\\
 $^\ddag$Chang Gung University
 }

\maketitle

\newcommand{\citehere}{\textcolor{red}{[citehere]}}
\newcommand{\refhere}{\textcolor{red}{[refhere]}}
\newcommand{\continuehere}{\textcolor{red}{To Be Continue}}
\newcommand{\todo}{\textcolor{red}{TODO}}

\thispagestyle{empty}

\begin{abstract}
Trajectory data, which tracks movements through geographic locations, is crucial for improving real-world applications. However, collecting such sensitive data raises considerable privacy concerns. Local differential privacy (LDP) offers a solution by allowing individuals to locally perturb their trajectory data before sharing it. Despite its privacy benefits, LDP protocols are vulnerable to data poisoning attacks, where attackers inject fake data to manipulate aggregated results. In this work, we make the first attempt to analyze vulnerabilities in several representative LDP trajectory protocols. We propose \textsc{TraP}, a heuristic algorithm for data \underline{P}oisoning attacks using a prefix-suffix method to optimize fake \underline{Tra}jectory selection, significantly reducing computational complexity. Our experimental results demonstrate that our attack can substantially increase target pattern occurrences in the perturbed trajectory dataset with few fake users. This study underscores the urgent need for robust defenses and better protocol designs to safeguard LDP trajectory data against malicious manipulation.
\end{abstract}

\section{Introduction}\label{sec: Introduction}
Today, mobile applications are a major source of extensive personal data, such as user locations and browsing histories. Analyzing this data enables service providers to refine their offerings, benefiting both the providers and their users. Trajectory data, a significant category of personal information, is especially critical as it finds diverse applications in various data mining tasks. For example, numerous studies have utilized trajectory data, primarily in taxi route planning~\cite{liao2021experience, ding2013hunts, li2024distributed} and travel itinerary optimization~\cite{puacurar2021tourist, cheng2021travel, hsieh2014traveling, zhu2019toward, bin2019personalized}. 

However, the collection and utilization of such sensitive data raise considerable privacy concerns, limiting widespread adoption. Local differential privacy (LDP)~\cite{4690986, duchi} offers a critical solution to this issue. Unlike traditional differential privacy (DP)~\cite{10.1007/11681878_14} that requires a trusted aggregator to collect raw data, LDP allows individuals to perturb their trajectory data locally before sharing it. Various LDP trajectory protocols~\cite{10.14778/3594512.3594520, 10.14778/3603581.3603597, 10.14778/3476249.3476280, 9861201} have been proposed to protect privacy while enabling valuable data collection for analysis.


The motivation to manipulate trajectory data arises when seeking monetary benefits. For example, a travel agency could skew data analysis to make their routes appear more popular than those of competitors. Similarly, a taxi company might insert fake trajectories to create a misleading impression of traffic congestion. This deceit could cause competitors' drivers to take longer routes to avoid perceived congestion, while the deceptive company uses the actual shortest route without any congestion. In this paper, we consider an attacker who attempts to promote the a set of target patterns. In particular, the attacker places fake users in the field, asking the fake users to report the maliciously crafted data to the server, while obeying the underlying LDP trajectory protocols. The fake trajectories contribute to significantly promote the frequency of the target patterns. 

Despite the extensive research on data poisoning in LDP protocols, attacking LDP trajectories is far from trivial. Unlike prior studies~\cite{Xiaoyu2021MGA, 287322, 279934}, where the victim protocols transmit basic data types such as numerical or categorical data, attacking LDP trajectory protocols presents four unique challenges. First, the output data types differ across protocols; some~\cite{10.14778/3594512.3594520, 9861201} transmit transition probabilities between points to the server, while others~\cite{10.14778/3603581.3603597, 10.14778/3476249.3476280} send complete trajectories. Second, as trajectories are sequences of points, considering reachability between points is essential. Since data is not represented by a single value, fake users reporting too many identical poisoned trajectories are easily detected. Third, the computational complexity in synthesizing trajectories poses a challenge; initializing all potential trajectories and then selecting the best attack combination is practically infeasible. Fourth, the lack of a closed-form aggregation formula causes the difficulty in the target poisoning. 

Given these challenges, through the lens of an attacker, we introduce \textsc{TraP}, a heuristic algorithm for data \underline{P}oisoning attacks to LDP \underline{Tra}jectory protocols. Since local users typically know the geographic information, \textsc{TraP} is designed to search for trajectories under the reachability constraint. Moreover, to mitigate the high computational cost of finding the optimal fake trajectory set, we propose a prefix-suffix approach, which determines the fake trajectory set without instantiating all possible trajectories. \textsc{TraP} also allows the attacker to set a hyper-parameter, $\textit{max}_\text{rep}$, controlling the maximum number of trajectory repetitions within the set of fake trajectories. This feature lets attackers balance attack effectiveness and stealthiness. Experimental results show that \textsc{TraP} significantly impacts most protocols and datasets. With just 20\% of fake users, the average scores of trajectories (calculated as the frequency of target patterns multiplied by the score of target patterns) increased more than fourfold compared to scenarios without the attack. Additionally, the percentile rank of the occurrence frequency of target patterns, compared to other patterns of the same length in perturbed trajectories, saw a substantial rise. Simple defense mechanisms proved ineffective against our attack, highlighting the severe risk posed by this form of data poisoning. This underscores the urgent need for improved protocol design or new defense methods.

Our main contributions are as follows:
\begin{enumerate}
\setlength{\topsep}{0pt}
\setlength{\partopsep}{0pt}
\setlength{\itemsep}{5pt}
\setlength{\parsep}{5pt}
\setlength{\parskip}{-2pt}
\item[\small$\bullet$] We are the first to systematically study data poisoning attacks against LDP protocols specifically for trajectory data. It identifies critical vulnerabilities in existing LDP mechanisms, demonstrating how these protocols can be compromised through targeted attacks.
\item[\small$\bullet$] We introduce \textsc{TraP}, a heuristic algorithm which utilizes a prefix-suffix approach to efficiently create attack-effective fake trajectories. \textsc{TraP} significantly reduces time complexity, making it feasible to execute sophisticated attacks on LDP protocols.
\item[\small$\bullet$] We conduct extensive experiments to evaluate the effectiveness of the proposed attacks across various protocols and datasets. It also assesses basic defense strategies, highlighting their limitations and the need for more advanced solutions to protect against data poisoning in LDP trajectory protocols.
\end{enumerate}

\section{Related Work}\label{sec:related work}
\textbf{LDP Trajectory Protocols} Given user privacy concerns, there is an urgent need for privacy-preserving solutions in the collection of trajectory data to encourage more individuals to share their trajectories. In particular, local differentially private (LDP) solutions focus on decentralized settings where noise is added to individual data points before server collection. Cunningham et al.~\cite{10.14778/3476249.3476280} propose an LDP mechanism that perturbs $n$-grams of trajectory data. PrivTC~\cite{9861201} constructs a private grid and uses a locally differentially private spectral learning approach to generate a synthetic trajectory dataset. In LDPTrace~\cite{10.14778/3594512.3594520}, an LDP trajectory synthesis framework that creates realistic trajectories without external knowledge and includes a method for selecting grid granularity. ATP~\cite{10.14778/3603581.3603597} uses direction information and introduced an anchor-based method to adaptively restrict the region of each perturbed trajectory. In contrast to non-realtime protocols~\cite{10.14778/3603581.3603597,10.14778/3594512.3594520,9861201,10.14778/3476249.3476280}, RetraSyn~\cite{hu2024retrasyn} is a realtime LDP trajectory protocol. 

\BlankLine

\noindent\textbf{Poisoning Attacks to LDP Protocols}
Although LDP protocols enable data collection by untrusted servers, they are more vulnerable to poisoning attacks due to their local implementation details, allowing attackers to craft fake values that closely resemble real ones and send them to the server.

For LDP protocols, while Cheu et al.~\cite{untargetd} propose an untargeted poisoning attack, Cao et al.~\cite{Xiaoyu2021MGA} propose the first targeted poisoning attack, MGA, aiming to increase the estimated frequencies of attacker-chosen target items. More targeted poisoning attacks are proposed subsequently. For example, M2GA~\cite{279934} formulate a poisoning attack to LDP key-value estimation protocols as a two-objective optimization problem, raising both the estimated frequencies and mean values of target keys. Li et al.~\cite{287322} present a fine-grained poisoning attack on LDP mean/variance estimation protocols, showing an accurate manipulation of statistical estimates. Very recently, Tong et al.~\cite{itemset} poison LDP frequent itemset mining. 

\section{Background}\label{sec:background}
\subsection{Local Differential Privacy}
In the local differential privacy (LDP) model~\cite{4690986, duchi}, we consider a scenario with $n$ local users and a remote, untrusted data collector. Each user holds private data $x$, which is of interest to the collector. To safeguard their privacy, users employ an algorithm $\mathcal{M}$ to randomly perturb $x$ before reporting the altered data $\mathcal{M}(x)$ to the collector. The protection conferred by $\mathcal{M}$ qualifies as LDP if it adheres to the established definition.

\begin{definition} [$\varepsilon$-Local Differential Privacy]
A randomized mechanism $\mathcal{M}:\mathcal{X} \rightarrow \mathcal{Y}$ is $\varepsilon$-\textit{LDP} if and only if, for any two inputs $x$, $x^\prime \in \mathcal{X}$ and any possible output $y \in \mathcal{Y}$, $\text{Pr}[\mathcal{M}(x)=y] \le \exp(\varepsilon) \cdot \text{Pr}[\mathcal{M}(x^\prime)=y]$.
\end{definition}

A smaller \textit{privacy budget}, $\varepsilon$, tightens the level of privacy but diminishes the utility of the data.

\subsection{Victim LDP Trajectory Protocols}\label{sec: Victim LDP Trajectory Protocols}
Four non-realtime LDP trajectory protocols are chosen as victims: $n$-gram~\cite{10.14778/3476249.3476280}, PrivTC~\cite{9861201}, LDPTrace~\cite{10.14778/3594512.3594520}, and ATP~\cite{10.14778/3603581.3603597}. All of these protocols can be divided into the following steps: (1) reducing the perturbation domain size, (2) performing local perturbation processes, and (3) synthesizing the trajectory dataset. We also apply \textsc{TraP} to attack real-time LDP trajectory protocol, RetraSyn~\cite{hu2024retrasyn}; the result is presented in Section~\ref{sec: Evaluation}.

\BlankLine

\noindent\textbf{Perturbation domain size reduction} 
The term "point" refers to a location (any latitude and longitude), a POI or a region along the trajectory, depending on the protocol. The perturbation domain is formed by the collection of all points that can be perturbed into. Despite richer information in location or POI-level trajectories, a larger  domain results in higher computational costs and more severe privacy budget splitting (i.e., significant noise). Hence, each protocol has its own method to reduce the size of the perturbation domain. 

\BlankLine

\noindent\textbf{Local perturbation process} 
After reducing the domain size, each protocol perturbs the information from users' trajectories. Users may report either a full trajectory, a single point, an adjacent pair (2 points), or an adjacent triplet (3 points) within the trajectory to the server. 

\BlankLine

\noindent\textbf{Synthesizing the trajectory dataset (optional)} 
In some protocols, users do not send full trajectories. Instead, the server performs synthesizing steps to generate the trajectory dataset. 

\subsubsection{$n$-gram}

\begin{enumerate}
\setlength{\topsep}{0pt}
\setlength{\partopsep}{0pt}
\setlength{\itemsep}{5pt}
\setlength{\parsep}{5pt}
\setlength{\parskip}{-2pt}
\item[\small$\bullet$] \textbf{Perturbation domain size reduction:}
    \begin{itemize}
    \setlength{\itemsep}{0.5pt}
    \setlength{\parsep}{0.5pt}
    \setlength{\parskip}{0.5pt}
        \item[$\circ$] Partition the geographic space into regions based on the space, time, and category distribution of public POIs.
        \item[$\circ$] Transform POI-level trajectories into region-level trajectories.
    \end{itemize}
\item[\small$\bullet$] \textbf{Local perturbation process:}
    \begin{itemize}
    \setlength{\itemsep}{0.5pt}
    \setlength{\parsep}{0.5pt}
    \setlength{\parskip}{0.5pt}
        \item[$\circ$] Extract $n$-grams from the region-level trajectory and use the exponential mechanism~\cite{4389483} to perturb each $n$-gram $n$ times.
        \item[$\circ$] Reconstruct a region-level trajectory optimally from the perturbed $n$-grams.
        \item [$\circ$] Sample a POI-level trajectory from the region-level trajectory, ensuring reachability.
    \end{itemize}
\item[\small$\bullet$] \textbf{Synthesizing the trajectory dataset:}
    \begin{itemize}
    \setlength{\itemsep}{0.5pt}
    \setlength{\parsep}{0.5pt}
    \setlength{\parskip}{0.5pt}
        \item[$\circ$] No synthesizing steps exist in $n$-gram since users send complete trajectories after perturbation. The server only needs to gather all trajectories to create a dataset.
    \end{itemize}    
\end{enumerate}

\subsubsection{PrivTC}
\begin{enumerate}
\setlength{\topsep}{0pt}
\setlength{\partopsep}{0pt}
\setlength{\itemsep}{5pt}
\setlength{\parsep}{5pt}
\setlength{\parskip}{-2pt}
\item[\small$\bullet$] \textbf{Perturbation domain size reduction:}
    \begin{itemize}
    \setlength{\itemsep}{0.5pt}
    \setlength{\parsep}{0.5pt}
    \setlength{\parskip}{0.5pt}
        \item[$\circ$] Divide users into two groups, $U_1$ and $U_2$. Partition the geographic space into regions based on the location distribution in $U_1$ and broadcast it to $U_2$.
        \item[$\circ$] Transform location-level trajectories into region-level trajectories.
    \end{itemize}
\item[\small$\bullet$] \textbf{Local perturbation process:}
    \begin{itemize}
    \setlength{\itemsep}{0.5pt}
    \setlength{\parsep}{0.5pt}
    \setlength{\parskip}{0.5pt}
        \item[$\circ$] Divide users in group $U_2$ into three subgroups to report probabilities for $S_1$, $S_2$, and $S_3$: single points, adjacent pairs, and adjacent triplets, respectively.
        \item[$\circ$] Each user in subgroups 1, 2, and 3 randomly samples a point, adjacent pair, and adjacent triplet, respectively. Encode and perturb the sampled data using OLH~\cite{203872}.
    \end{itemize}
\item[\small$\bullet$] \textbf{Synthesizing the trajectory dataset:}
    \begin{itemize}
    \setlength{\itemsep}{0.5pt}
    \setlength{\parsep}{0.5pt}
    \setlength{\parskip}{0.5pt}
        \item[$\circ$] After collecting probabilities in $S_1 \cup S_2 \cup S_3$ via OLH, the server removes negativity and inconsistency, then computes the spectral learning parameters.
        \item[$\circ$] To synthesize a trajectory, the server first creates a region-level trajectory $\tau_r$ from the learned HMM~\cite{baum1967HMM} parameters. For each $i$-th location in $\tau_l$, $\tau_l[i]$ is set to the coordinates of a point randomly sampled from region $\tau_r[i]$.
    \end{itemize}
\end{enumerate}

\subsubsection{LDPTrace}
\begin{enumerate}
\setlength{\topsep}{0pt}
\setlength{\partopsep}{0pt}
\setlength{\itemsep}{5pt}
\setlength{\parsep}{5pt}
\setlength{\parskip}{-2pt}
\item[\small$\bullet$] \textbf{Perturbation domain size reduction:}
    \begin{itemize}
    \setlength{\itemsep}{0.5pt}
    \setlength{\parsep}{0.5pt}
    \setlength{\parskip}{0.5pt}
        \item[$\circ$] Partition the geographic space into regions based on grid granularity $N$, selected to minimize the estimated error of range queries.
        \item[$\circ$] Transform location-level trajectories into region-level trajectories.
    \end{itemize}
\item[\small$\bullet$] \textbf{Local perturbation process:}
    \begin{itemize}
    \setlength{\itemsep}{0.5pt}
    \setlength{\parsep}{0.5pt}
    \setlength{\parskip}{0.5pt}
        \item[$\circ$] Encode and perturb the trajectory length and send it to the server, which broadcasts the $k$-th quantile of the estimated length distribution $L_k$ to users. Users upload at most $L_k$ transition states, omitting excess transitions. $k$ balances noise and bias error.
        \item[$\circ$] Encode and perturb intra-trajectory transitions with OUE~\cite{203872}. These transitions involve moving from the $i$-th to $(i+1)$-th point where $i \in [1, \min(L_k, \text{trajectory length}-1)]$.
        \item [$\circ$] Add virtual start point $R_s$ and end point $R_e$ to the trajectory. Perturb beginning and terminated transitions with OUE. The beginning transition is $R_s \rightarrow$ the first region, and the terminated transition is the last region $\rightarrow R_e$.
    \end{itemize}
\item[\small$\bullet$] \textbf{Synthesizing the trajectory dataset:}
    \begin{itemize}
    \setlength{\itemsep}{0.5pt}
    \setlength{\parsep}{0.5pt}
    \setlength{\parskip}{0.5pt}
        \item[$\circ$] For each trajectory, the server determines its length $L$ by sampling from the collected length distribution.
        \item[$\circ$] Samples the first point with probabilities from the virtual start point to others.
        \item [$\circ$] Extends the trajectory by adding new points with probabilities from the current point. This repeats until reaching the virtual end point or $L$.
    \end{itemize}
\end{enumerate}

\subsubsection{ATP}
\begin{enumerate}
\setlength{\topsep}{0pt}
\setlength{\partopsep}{0pt}
\setlength{\itemsep}{5pt}
\setlength{\parsep}{5pt}
\setlength{\parskip}{-2pt}
\item[\small$\bullet$] \textbf{Perturbation domain size reduction:}
    \begin{itemize}
    \setlength{\itemsep}{0.5pt}
    \setlength{\parsep}{0.5pt}
    \setlength{\parskip}{0.5pt}
        \item[$\circ$] Restrict the perturbation domain using an anchor $\alpha$ and a radius $r$.
        \item[$\circ$] Determine the anchor and radius adaptively for each user's trajectory. Only include POIs within the circle centered at $\alpha$ with radius $r$ in the perturbation domain.
    \end{itemize}
\item[\small$\bullet$] \textbf{Local perturbation process:}
    \begin{itemize}
    \setlength{\itemsep}{0.5pt}
    \setlength{\parsep}{0.5pt}
    \setlength{\parskip}{0.5pt}
        \item[$\circ$] Duplicate the trajectory: one using odd points as pivots, the other using even points.
        \item[$\circ$] Perturb pivot points with EM, limited to the circle centered at $\alpha$ with radius $r$.
        \item [$\circ$] Discretize directions with chosen granularity. Encode and perturb directions from perturbed pivot points to unperturbed neighbor points using $k$-RR~\cite{NIPS2014_86df7dcf}.
        \item [$\circ$] Perturb neighbor points with EM, with the domain set by nearby perturbed pivot points and directions.
        \item [$\circ$] Reconstruct a trajectory optimally from the two perturbed trajectories.
    \end{itemize}
\item[\small$\bullet$] \textbf{Synthesizing the trajectory dataset:}
    \begin{itemize}
    \setlength{\itemsep}{0.5pt}
    \setlength{\parsep}{0.5pt}
    \setlength{\parskip}{0.5pt}
        \item[$\circ$] No synthesizing steps exist in ATP since users send complete trajectories after perturbation. The server only needs to gather all trajectories to create a dataset.
    \end{itemize} 
\end{enumerate}

Users in various LDP trajectory protocols send different types of data to the server. For example, full POI-level trajectories are sent in $n$-gram and ATP. To attack these protocols, we create a set of poisonous trajectories. These trajectories act as the direct output of the LDP protocol and bypass the LDP processes. In cases where users provide partial perturbed trajectories, like in PrivTC and LDPTrace, crafting the optimal fake trajectories for the fake users to attack PrivTC and LDPTrace can be challenging. To address this, we can generate POI-level or region-level trajectories and use them as input for these LDP protocols to obtain the required data. In particular, the fake trajectories will be fed to the fake users in PrivTC and LDPTrace, which then go through the local perturbation and forward the perturbed results to the server. This approach allows \textsc{TraP} to generate diverse data types for attacking the specified protocols.

Here, we simplify the victim protocols and put them in a three-step unified framework. This is more convenient for the design of a single poisoning attack. However, when we evaluate the effectiveness of the poisoning attack in Section~\ref{sec: Evaluation}, all the victim protocols will be implemented with all their details.

\section{Attacking LDP Trajectory Protocols}
\subsection{System Model}
We describe the problem setting, the attacker's goal, and the attacker's capability in this section. The notation table is shown in Table~\ref{table1} in Appendix~\ref{sec: Notation Table}.

\BlankLine

\noindent\textbf{Problem Setting} 
Consider an LDP trajectory protocol, where $n$ real users individually interact with a server, who aims to synthesize the user trajectories. A point $p$ refers to a location, a POI, or a region along the trajectory, depending on the victim protocol. The point domain $\mathcal{P}$ consists of all the points. The size of $\mathcal{P}$ is denoted as $|\mathcal{P}|$. A \textit{trajectory}~\cite{wang2023privtrace, gursoy2018utility, he2015dpt, jin2022frequency, he2016demonstration, zhao2022group} is defined as an ordered sequence of points that describe the path of a moving object over time. 

\BlankLine

A \textit{pattern} is formed by concatenating several points. In other words, a pattern represents an arbitrary segment of a trajectory. Through the lens of an attacker, we aim to promote a pre-defined set $\textit{TP}$ of target patterns, where each pattern $tp\in TP$ is associated with an importance score $\text{score}(\textit{tp})$. Here we assume that a longer pattern is associated with a higher score. The attacker's goal is to generate a set $\mathcal{T}^*$ of fake trajectories that maximizes the total importance score.

\noindent\textbf{Attacker's Capability} 
An attacker can inject up to $m < n$ fake users, controlling their data submissions and tailoring them to the chosen LDP protocol. This enables an output poisoning attack (OPA), as discussed in prior work~\cite{Xiaoyu2021MGA,279934,287322,itemset}. We particularly note that due to the complex nature of LDP trajectory protocols (compared to the other LDP protocols such frequency estimation), we consider $m/(m+n)$ up to $0.2$ in Section~\ref{sec: Evaluation}. Such a setting can be justified because fake accounts are cheap—about \$0.3 for Facebook or X (Twitter) and \$0.03 for phone-verified Google—allowing a large $m$.  However, an attacker typically purchases only as many as needed to balance cost and attack effectiveness.

To simplify the explanation of \textsc{TraP}, we initially assume the attacker has prior knowledge of $L_{\text{min}}$ and $L_{\text{min}}$, where $L_{\text{min}}$ and $L_{\text{min}}$ are the minimum and maximum lengths of real trajectories, respectively. In this scenario, the attacker can easily determine the number $m_{L=i}$ of length-$i$ fake trajectories (i.e., the fake trajectories' length distribution), where $\sum_{i=L_{\text{min}}}^{L_{\text{max}}} m_{L=i} = m$, by performing $m$ samplings (rounded to integers) from Gaussian distribution with the mean $(L_{\text{min}}+L_{\text{max}})/2$ and standard deviation $(L_{\text{max}}-L_{\text{min}})/5$ to enhance the stealthiness of the fake trajectories. However, since it is unrealistic to assume the attacker knows $L_{\text{min}}$ and $L_{\text{min}}$, we will relax this assumption in Section~\ref{sec: Comparison with different length distributions}.

\noindent\textbf{Attacker's Goal} The attacker's goal is to generate the set $\mathcal{T}^*$ of attack-effective fake trajectories that maximize the total importance score. In this sense, a trajectory is realistic if it obeys the reachability constraint, ensuring neighboring points are not too distant. Each LDP trajectory protocol typically defines its own reachability constraint by setting a time granularity, which determines the recording intervals from the maximum movement speed of a person or vehicle. If the physical distance between two points divided by the movement speed is less than or equal to the time interval, the points meet the reachability constraint. On the other hand, to make the fake trajectories stealthy, we use a hyperparameter $\textit{max}_{\text{rep}}$ to limit the maximum permissible trajectory repetitions in the fake trajectory set, because it is unlikely to have many users sending identical trajectories. 

Let $\mathcal{T}=\{\tau_1,\dots,\tau_n\}\cup \mathcal{T}^*$ be the poisoned trajectory set, where $\{\tau_1,\dots,\tau_n\}$ denotes the set of real trajectories and $\mathcal{T}^*=\{\tau^*_1,\dots,\tau^*_m\}$ denotes the fake trajectory set. Define the total gain $G(\mathcal{T},TP)=\frac{1}{n+m}\sum_{i=1}^{n+m}\text{TrajScore}(\tau_i)-\frac{1}{n}\sum_{i=1}^{n}\text{TrajScore}(\tau_i)$, where $\text{TrajScore}(\cdot)$ is calculated by

\begin{equation}
\label{TrajScore}
\text{TrajScore}(\tau) = \sum_{\textit{tp}\in \textit{TP}} \text{score}(\textit{tp}) \cdot \text{CountPattern}(\textit{tp},\tau),
\end{equation}where $\text{CountPattern}(\textit{tp},\tau)$ is the number of subarrays in trajectory $\tau$ that match the target pattern $\textit{tp}$ and can be implemented with Algorithm~\ref{alg:CountPattern} in Section~\ref{sec: Supplementary Algorithms}. In other words, $\text{TrajScore}(\tau)$ calculates the scores that the trajectory $\tau$ contributes to the target pattern set $TP$. The higher $\text{TrajScore}(\tau)$, the greater the number of target patterns and the higher importance scores assigned to those target patterns on that trajectory.

Given the real trajectories $\{\tau_1, \cdots, \tau_n\}$ and target pattern set $TP$, the problem of poisoning the LDP trajectory protocols can be formulated as a constrained optimization below,

\begin{equation}
\label{maximization_problem}
\mathcal{T}^*=\argmax\limits_{\tau_{n+1}, \cdots, \tau_{n+m}} G(\mathcal{T}, TP)
\end{equation}

\begin{footnotesize}
\begin{empheq}[left=\text{s.t.}\empheqlbrace]{align}
|\mathcal{T}^*_{L=i}| = m_{L=i} \quad \hspace{6em} \forall L_{\text{min}} \le i \le L_{\text{max}}  \label{eq: 3}\\
\text{count}(\tau^*_i, \mathcal{T}^*) \le \textit{max}_{\text{rep}} \quad \hspace{4.7em} \forall 1\le i \le m  \label{eq: 4} \\
\sum_{j=1}^{|\tau_i|-1}\text{\small{reachable}}(\tau^*_{i, j}, \tau^*_{i, j+1})=|\tau_i|-1 \quad \forall 1\le i \le m \label{eq: 5}
\end{empheq}\end{footnotesize}where $\mathcal{T}^* = \{ \tau^*_1, \dots, \tau^*_m \}$ is the set of fake trajectories, $\mathcal{T}^*_{L=i}\subseteq \mathcal{T}^*$ is a subset of the fake trajectory set, which contains the fake trajectories with length $i$, $\text{count}(\tau_i, \mathcal{T}^*)$ counts how many times $\tau_i$ appears in $\mathcal{T}^*$, $\text{reachable}(a, b)$ returns $1$ ($0$) if point $b$ is (not) reachable from point $a$, $\tau_{i,j}$ ($\tau^*_{i,j}$) denotes the the $j$-th point of $\tau_i$ ($\tau^*_i$). Equation~(\ref{eq: 4}) is defined as $\text{count}(\tau_i, \mathcal{T}^*)$, rather than $\text{count}(\tau_i, \mathcal{T})$, because the attacker cannot eavesdrops on the real trajectories $\tau_1, \dots, \tau_n$. The constrained optimization from Equations~(\ref{maximization_problem})$\sim$(\ref{eq: 5}) is defined to find an optimal set $\mathcal{T}^*$ of fake trajectories such that, the set $\mathcal{T}$ of poisoned trajectories will lead to the maximum gain of score.

\subsection{Baselines}\label{sec: Baselines}

\BlankLine
\noindent\textbf{Brute-force Approach} To solve the optimization problem in Equation~(\ref{maximization_problem}), a brute-force approach is outlined in Algorithm~\ref{alg:brute-force} in Section~\ref{sec: Supplementary Algorithms}. More specifically, this method instantiates all length-$i$ trajectories that satisfy the reachability constraint for $i \in [L_{\min}, L_{\max}]$. Within each equal-length trajectory set, trajectories are sorted in descending order based on their scores. Recall that at most $\textit{max}_{\text{rep}}$ identical trajectories are allowed in the fake set. To maximize the total score of fake trajectories with length $i$, the top $\lfloor m_{L=i} / \textit{max}_{\text{rep}} \rfloor$ trajectories should be included $\textit{max}_{\text{rep}}$ times in the fake set, and the $\left( \lfloor m_{L=i} / \textit{max}_{\text{rep}} \rfloor + 1 \right)$-th trajectory should be included $\left( m_{L=i} - \lfloor m_{L=i} / \textit{max}_{\text{rep}} \rfloor \cdot \textit{max}_{\text{rep}} \right)$ times.

Although this solution is straightforward, it may be infeasible due to its high time complexity. To evaluate this, consider that on average, only $1/r$ of the points are reachable. The instantiation of all trajectories incurs a cost of $O\left(|\mathcal{T}_{L=i}|*|\tau|\right)=O\left(\left(\frac{|\mathcal{P}|}{r}\right)^{L_{\max}}*L_{\max}\right)$. Executing $\text{CountPattern}(\textit{tp}, \tau)$ for a trajectory and target pattern (Algorithm~\ref{alg:CountPattern}) costs $O(|\tau|*|\textit{tp}|)=O(L_{\max}*k_{\max})$, and a total of $O\left(\left(\frac{|\mathcal{P}|}{r}\right)^{L_{\max}}*|\textit{TP}|*L_{\max}*k_{\max}\right)$ for all scores. Sorting trajectories by score costs $O\left(\left(\frac{|\mathcal{P}|}{r}\right)^{L_{\max}}\log{\left(\frac{|\mathcal{P}|}{r}\right)^{L_{\max}}}\right)$. Finally, putting $m$ trajectories into the fake set $\mathcal{T}^*$, each with a maximum length of $L_{\max}$, costs $O\left(m*L_{\max}\right)$.

The execution time of this brute-force approach grows exponentially with $L_{\max}$. In their experiments, both $n$-gram~\cite{10.14778/3476249.3476280} and ATP~\cite{10.14778/3603581.3603597} set the maximum trajectory length to $8$. However, LDPTrace~\cite{10.14778/3594512.3594520} considers longer trajectories. According to Table 2 in LDPTrace, the Hangzhou dataset has an average length of $125.02$, causing the time complexity to skyrocket. The experiments in Appendix~\ref{sec:execution_time_comparison} demonstrate that in practical settings, the brute-force method is limited to generating trajectories with a maximum length of 4.

\BlankLine

\noindent\textbf{Frequency-based Approach} When considering attacks to the LDP trajectory protocol, the initial instinct is to elevate the frequency of a single point. However, the victim protocols do not merely concentrate on the frequency of individual points; they also factor in the interrelationships between points. Simply disrupting the frequency of single points does not equate to a full compromise of the protocol. Hence, this method is not deemed valid. Hence, we do not compare \textsc{TraP} to the frequency-based method. For a more detailed explanation and experiments, please refer to Appendix~\ref{sec:Baseline_single_point}.

\subsection{\textsc{TraP}: A Prefix-Suffix Approach}\label{sec: TraP: A Prefix-Suffix Approach} 

Given the inefficiency of brute-force methods, a more scalable solution is required. The challenge lies in instantiating all trajectories, which introduces significant overhead for subsequent steps. This raises a critical question: Can we compute the optimal fake trajectory set $\mathcal{T}^*$ without generating every trajectory? We address this by proposing \textsc{TraP}, a heuristic algorithm.

\paragraph{Main Idea} We observed that the brute-force method generates trajectories of varying lengths, selecting the highest-scoring ones only after listing all trajectories of a given length. \textsc{TraP} is designed to avoid the inefficiency of regenerating trajectories for each length by explicitly extending them, reducing computational overhead. Simultaneously, \textsc{TraP} aims to retain high-scoring trajectories across multiple lengths. Specifically, before extending to the next length, \textsc{TraP} filters out low-scoring trajectories, keeping the candidate pool within manageable bounds and minimizing unnecessary computations. A toy example showing how \textsc{TraP} works is shown in Section~\ref{sec: An Example of TraP}.

\begin{algorithm}[h]
    \caption{\textsc{TraP}: Prefix-suffix Approach}
    \label{alg:prefix-suffix}
    \SetKwInOut{Input}{Input}
    \SetKwInOut{Output}{Output}
    \Input{Point domain $\mathcal{P}$, target pattern set $\textit{TP}$, length distribution $m_{L=i} \hspace{0.5em} \forall L_{\min} \le i \le L_{\max}$, maximum permissible trajectory repetitions $\textit{max}_{\text{rep}}$}
    \Output{A trajectory set $\mathcal{T}^*$ containing $m$ trajectories}  
    \tcc{Preprocess to get the reachable point set for each point, denoted as $\textit{rps}[\cdot]$}

    \tcc{Obtain the prefix set $\textit{PREF}$ from $\textit{TP}$, and sort $\textit{PREF}$ by criteria=length}
    
    Initialize $\mathcal{T}^{*}=\emptyset$ and $\Omega_{1}$ = $\emptyset$\\
    
    \For{$i=1$ \rm{to} $L_{max}$}{
        Initialize $\mathcal{M}_i=\{u_{\tau}:[\enspace] | u_{\tau}\in PREF\}$ and $SC=\{\}$\\
        \tcc{Instantiate length-(i+1) trajectories}
        $\Omega_{i+1}$ = extend one point $\forall$ $\tau \in \Omega_{i}$ using $\textit{rps}[\tau]$\\
        
        \For{$\tau \in \Omega_{i+1}$}{
            \tcc{Prefix category refers to the longest suffix of a given trajectory that matches the category within the set PREF}
            Append $\tau$ to $\mathcal{M}_i[$the prefix category of $\tau]$\\
            $SC[\tau] =$ TrajScore$(\tau)$
        }
        \tcc{Determine the maximal trajectory score set}
        $\mathcal{T}^{*} = \mathcal{T}^{*}\cup$ Pick-High($\Omega_{i+1}$, $m_{L=i}$, $\textit{max}_{\textnormal{rep}}$, $SC$)\\        
        \tcc{Delete hopeless trajectories}
        $m_{\text{max}} = max(m_{L=i+1}, \cdots, m_{L=L_\text{max}})$\\
        Delete($\Omega_{i+1}$, $i$, $m_{\text{max}}$, $\textit{max}_\text{rep}$, $\textit{rps}$, $\textit{SC}$, $\textit{PREF}$, $\mathcal{M}_i$)\\

    }
    Return $\mathcal{T}^{*}$
\end{algorithm}

\paragraph{Main Algorithm of \textsc{TraP}} The pseudocode for \textsc{TraP} is outlined in Algorithm~\ref{alg:prefix-suffix}. Before its execution, preprocessing is performed, including creating a dictionary $rps$ to store reachable points for each position and determining prefix categories by taking the longest proper prefix of each target pattern. Line 1 initializes $\mathcal{T}^{*}$ to store the generated trajectories for various length requirements, and $\Omega_{1} = \emptyset$ to handle the first round where length 0 trajectories are stored. Lines 2-10 manage trajectory generation and collection for different lengths. Although Line 2 begins with length 1, trajectories are collected only when a length requires a specific number $m_{L=i}$ for $L_{\min} \le i \le L_{\max}$. Line 4 extends candidate trajectories by adding one position based on the next step from the $rps$ list. Lines 5-7 categorize extended trajectories by their prefix and compute scores, assigning the longest suffix matching the prefix category as its new prefix category. Line 8 invokes \verb"Pick-High" (Algorithm~\ref{alg:pick-high} in Section~\ref{sec: Missing Pseudocode}) to select the highest-scoring trajectories to meet the $m_{L=i}$ requirement. Line 9 determines the maximum number $m_{\text{max}}$ for the next round, and Line 10 invokes \verb"Delete" (Algorithm~\ref{alg:Delete} in Section~\ref{sec: Missing Pseudocode}) to filter candidates for the subsequent round.

\paragraph{Generating High-Scoring Trajectories} \verb"Pick-High" (Algorithm~\ref{alg:pick-high}), a subroutine of \textsc{TraP}, selects trajectories from $\Omega$ based on descending scores until the required number of trajectories, $m_{L=i}$, is obtained. Line 2 sorts $\Omega$ in descending order by score, while Lines 3-9 manage the selection of high-scoring trajectories. The parameter $\textit{max}_{\text{rep}}$ in Line 5 limits how many times the same trajectory can be repeated. This design enhances the stealthiness of fake trajectories, preventing the server from detecting and defending against a large volume of identical trajectories.

\paragraph{Removing Hopeless Trajectories} To limit the number of candidate trajectories in each round, \textsc{TraP} introduces a parameter $m_{\text{max}}$, which sets the requirement for the number of trajectories needed in the next round. $m_{\text{max}}$ corresponds to the largest required number of trajectories across all future rounds. Once at least $m_{\text{max}}$ trajectories are selected, the current trajectory is filtered. This is because only $m_{\text{max}}$ trajectories are needed in the next round, and any trajectory with a score no higher than the lowest of the selected ones will not affect the correctness of high-scoring trajectories. However, filtering purely by score risks trapping the solution in local optima. Therefore, the filtering also considers target pattern length, as longer patterns generally yield higher rewards. \textsc{TraP} triggers filtering under the following conditions:

\textbf{Condition 1 (Lines 6-18 in Algorithm~\ref{alg:Delete}):} Consider trajectories $\tau_1$ and $\tau_2$ ending at "a", where $score(\tau_1) \leq score(\tau_2)$. If $\tau_1$ can satisfy "a→b" and $\tau_2$ can satisfy "c→a→b" in the next round, $\tau_2$ will inevitably outscore $\tau_1$ due to its longer, higher-reward pattern.

\textbf{Condition 2 (Lines 20-25 in Algorithm~\ref{alg:Delete}):} If $\tau_1$ and $\tau_2$ have equal scores, end at "a", and both satisfy "a→b" in the next round, $\tau_1$'s score cannot exceed $\tau_2$'s. Thus, $\tau_1$ can be removed if either condition is met.

In this sense, the trajectory is hopeless and can be removed if one of the conditions is met. \textsc{TraP} employs a prefix category to classify trajectories based on their potential next-round target patterns. For a trajectory satisfying "$p_1\rightarrow p_2\rightarrow \dots \rightarrow p_k$", its prefix category is $(p_1, p_2, \dots, p_k)$. Situation-1 applies when (1) the checked trajectory's prefix category is a suffix of other categories $ancestor$, and (2) at least $m_{\text{max}}$ trajectories in $ancestor$ have scores $\geq$ the checked trajectory. Situation-2 applies when (1) the prefix category $u_\tau$ is not a suffix of any other category, and (2) at least $m_{\text{max}}$ trajectories in $u_\tau$ have scores $\geq$ the checked trajectory.

Algorithm~\ref{alg:Delete} initializes $\mathcal{M}_{i}^{selected}$ to record selected trajectories per prefix category. It processes categories from longest to shortest, applying Situation-1 (Lines 6-18) or Situation-2 (Lines 20-25) based on condition (1). Lines 7-8 retain trajectories violating Situation-1's condition (1), while Lines 10-18 filter trajectories based on Situation-1's condition (2).

\subsection{Time Complexity Analysis}
In the following analysis, we use a hash-based data structure for the elements: $\textit{TP}$ (target pattern set), $\textit{PREF}$ (prefix set), $\mathcal{R}$ (recordings of trajectories with suffixes matching prefixes in $\textit{PREF}$), and $\textit{SC}$ (trajectory scores).

In Algorithm~\ref{alg:prefix-suffix}, before generating length-$(i+1)$ trajectories, \verb"Delete" removes length-$i$ trajectories that cannot extend to have top-$\lceil m_{L=i+1} / \textit{max}_\text{rep} \rceil$ scores, disqualifying them as candidates. Thus, the number of trajectories retained is limited to $\lceil m / \textit{max}_\text{rep} \rceil$ for each prefix category $\textit{pref} \in \textit{PREF}$. The size of $\textit{PREF}$ is constrained by $|\textit{TP}| * k_\text{max}$. The total number of trajectories processed per iteration is bounded by $\lceil m / \textit{max}_\text{rep} \rceil \cdot |\textit{TP}| \cdot k_\text{max}$, denoted as $N$.

In Algorithm~\ref{alg:pick-high}, sorting $\Omega_{i+1}$ depends on the number of length-$i$ trajectories, with a maximum of $N$, yielding a complexity of $O(N \log N)$. The outer loop runs at most $\lfloor m_{L=i} / \textit{max}_\text{rep} \rfloor + 1$ times, and the inner loop runs up to $\textit{max}_\text{rep}$ times, totaling $O(m_{L=i})$. In the worst case, $O(m_{L=i}) = O(m)$, so the overall time complexity of Algorithm~\ref{alg:pick-high} is dominated by the sorting step, $O(N \log N)$.

In Algorithm~\ref{alg:Delete}, searching for $u_\tau$'s ancestor has a complexity of $O(|\textit{TP}|)$. With $|ancestor|=0$, the remaining instructions have a complexity of $O(N)$. The complexity of the if statement (Lines 7-8) is $O(|\textit{TP}|)$, while the else statement (Lines 10-18) is $O(N)$. Thus, the overall complexity of Algorithm~\ref{alg:Delete} is $O(|\textit{TP}|\cdot N^2)$.

Lastly, the core operations in Algorithm~\ref{alg:prefix-suffix} include:
\begin{enumerate}
\setlength{\topsep}{0pt}
\setlength{\partopsep}{0pt}
\setlength{\itemsep}{5pt}
\setlength{\parsep}{5pt}
\setlength{\parskip}{-2pt}
\item[\small$\bullet$] Appending points to trajectories (Lines 4-7 in Algorithm~\ref{alg:prefix-suffix}): Extending trajectories is $O(N\cdot k_{\max})$, while appending trajectories into $\mathcal{M}_i$ and calculating the score is $O(|\textit{TP}|)$ per trajectory. Repeated for $1 \le i \le L_{\max}$, the total complexity is $O(L_{\max} \cdot N \cdot (k_{\max} + |\textit{TP}|))$.

\item[\small$\bullet$] Selecting maximal trajectory score set (Algorithm~\ref{alg:pick-high}): The sorting step is $O(N \log N)$ per trajectory length. Repeated for each $L_{\min} \le i \le L_{\max}$, the total complexity is $O(L_{\max} \cdot N \log N)$.

\item[\small$\bullet$] Deleting trajectories (Algorithm~\ref{alg:Delete}): Applied across all trajectory lengths, the total complexity is $O(L_{\max} \cdot |\textit{TP}| \cdot N^2)$.
\end{enumerate}

Combining these, the total complexity for Algorithm~\ref{alg:prefix-suffix} is $O(L_{\max} \cdot N \cdot (\log N + k_{\max} + |\textit{TP}| \cdot N))$, where $N = \lceil m / \textit{max}_\text{rep} \rceil \cdot |\textit{TP}| \cdot k_\text{max}$ in the worst case.

\section{Evaluation}\label{sec: Evaluation}
We conduct experiments on the four non-realtime LDP trajectory protocols in Section~\ref{sec: Victim LDP Trajectory Protocols}. In addition, we will present experimental results for RetraSyn~\cite{hu2024retrasyn}, a realtime LDP trajectory protocol. Since RetraSyn requires datasets with timestamps, it cannot be evaluated using the same datasets as the other four protocols. Consequently, we have included the RetraSyn evaluation in the Appendix~\ref{sec:RetraSyn}. 

\subsection{Experimental Setup}\label{sec:exp_setup}
We describe the datasets, the protocol implementation details, the parameters, and the attack types used in our experiments.

\subsubsection{Datasets}\label{sec:datasets} 
We consider the datasets used in the four victim protocols, including CHI, CLE, CPS, Gowalla, NYC, Oldenburg, and Porto. Below is an overview of these datasets.
\begin{enumerate}
\setlength{\topsep}{0pt}
\setlength{\partopsep}{0pt}
\setlength{\itemsep}{5pt}
\setlength{\parsep}{5pt}
\setlength{\parskip}{-2pt}
\item[\small$\bullet$] \textbf{CHI, CLE:} These datasets comprise check-in trajectories in Chicago and Cleveland from the Gowalla dataset~\cite{Cho2011Gowalla} and are used in ATP~\cite{10.14778/3603581.3603597}.
\item[\small$\bullet$] \textbf{CPS (Campus):} These trajectories are based on 262 buildings at the University of British Columbia~\cite{ubcgeodata} and used in $n$-gram~\cite{10.14778/3476249.3476280}, LDPTrace~\cite{10.14778/3594512.3594520}, and ATP~\cite{10.14778/3603581.3603597}.
\item[\small$\bullet$] \textbf{Gowalla:} This dataset spans global latitude and longitude, beyond just Chicago and Cleveland, and used in PrivTC~\cite{9861201}.
\item[\small$\bullet$] \textbf{NYC:} This dataset is formed by combining Foursquare check-in~\cite{Yang2015FourSquareCheckin} and historic taxi trip~\cite{historicTaxiTrip} data from New York City and is used in $n$-gram~\cite{10.14778/3476249.3476280} and ATP~\cite{10.14778/3603581.3603597}.
\item[\small$\bullet$] \textbf{Oldenburg:} This dataset is simulated by Brinkhoff’s network-based moving objects generator~\cite{brinkOffGenerator} based on the map of Oldenburg city and is used in LDPTrace~\cite{10.14778/3594512.3594520}.
\item[\small$\bullet$] \textbf{Porto (Taxi)}~\cite{Starace2020Porto}\textbf{:} This dataset contains taxi traces in Porto, Portugal and is used in PrivTC~\cite{9861201} and LDPTrace~\cite{10.14778/3594512.3594520}.
\end{enumerate}

Since these protocols are designed for various types of trajectory data, we apply different preprocessing steps for each protocol. In particular, $n$-gram and ATP work with discrete trajectories, where points represent Points of Interest (POIs) rather than arbitrary (latitude, longitude) pairs. Thus, continuous trajectory datasets like Oldenburg and Porto are converted into discrete trajectories. To manage time complexity, we cluster the points based on their latitudes and longitudes into 1000 clusters and replace each point with its nearest cluster point.

PrivTC and LDPTrace handle continuous trajectories by segmenting the map into coarse grids to reduce time complexity, making them capable of managing discrete trajectories as well. In LDPTrace, the reachability constraint for a grid only includes its 8 neighboring grids, excluding the grid itself. When multiple consecutive points fall within the same grid, we retain only one point. If two adjacent points in a trajectory are not in neighboring grids, we use interpolation to connect them.

For $n$-gram, trajectory lengths range from 2 to 6. PrivTC requires all trajectories to be of the same length for unbiased results, so we set this length to 6. LDPTrace handles trajectories from 2 to 15 in length, while ATP ranges from 3 to 8. For protocols other than LDPTrace, the datasets' speed is set to 50 km/hr. If two adjacent points in a trajectory do not meet this reachability constraint, the trajectory is excluded. For datasets with more than 5,000 trajectories, we randomly sample approximately 5,000; otherwise, all trajectories are included.

\subsubsection{Implementations of Protocols}\label{sec:impl_protocols} 
We generally use the official code in our experiments. For ATP, we make no significant modifications. For LDPTrace, we set the grid granularity to $16\times16$.

For PrivTC, the official code includes two versions: one using OUE~\cite{203872} as the LDP protocol and the other using OLH~\cite{203872}. We use the OUE version but also demonstrate the effectiveness of \textsc{TraP} on the OLH version in Appendix~\ref{sec:OUE_OLH_PrivTC}. To enhance PrivTC's pattern preservation capability, we adjust its parameters for determining grid granularity to $\alpha_1=0.5$ and $\alpha_2=0.01$. This adjustment will result in finer grids and further improve pattern preservation. We also restrict the grid granularity to a maximum of $16\times16$ to avoid excessive memory usage.

On the contrary, as no official code is available for $n$-gram, we implement $n$-gram by our own. Since many datasets lack details on opening times and categories of locations, we simplified the STC region merging step. Our implementation merges based on the spatial dimension only, excluding time and category dimensions.

\subsubsection{Parameters}\label{sec:exp_parameters} 
In our experiments, we set the minimum target pattern length to $k_{\text{min}} = 1$ and the maximum to $k_{\text{max}} = 6$. Five patterns of each target length are randomly selected from the dataset. Sampling from the dataset ensures that target patterns already exist, albeit at lower frequencies, allowing us to compare results before and after the attack. Randomly generating target patterns might result in patterns with zero frequency without an attack.

The scoring policy for target patterns is length-based, with each pattern's score equal to its length. The maximum number of repetitions $\max_\text{rep}$ for the same trajectory is set to $1$, ensuring no identical trajectories appear in the poisoned dataset. The proportion of fake users to total users (fake + real) is $0.2$.

We assume that the attacker only knows the shortest length $L_\text{min}$ and the longest length $L_\text{max}$ of the real trajectory dataset. The attacker generates $m$ random numbers from a Gaussian distribution with a mean of $(L_\text{min} + L_\text{max}) / 2$ and a standard deviation of $(L_\text{max} - L_\text{min}) / 5$. These random numbers are then rounded to the nearest integer and kept within the range $[L_\text{min}, L_\text{max}]$. Each random number represents the length of a trajectory. By determining how many random numbers are equal to a specific length $i$, we can calculate $m_{L=i}$. We remove this assumption in Section~\ref{sec: Dataset Distribution Mismatch}.

We set the granularity by dividing the space into $16\times 16$ grids. Target patterns are evaluated at the grid(region) level, while trajectory input requires (latitude, longitude) points. Locations/POIs within each grid are chosen as close to the center as possible to represent the grid.

\subsubsection{Attack Types}\label{sec:attack types} 
Depending on whether the poisoned trajectory passes through LDP protocols, \textsc{TraP} can be classified into two types: input poisoning attack (IPA)~\cite{287322} and output poisoning attack (OPA)~\cite{287322}.
\begin{enumerate}
\setlength{\topsep}{0pt}
\setlength{\partopsep}{0pt}
\setlength{\itemsep}{5pt}
\setlength{\parsep}{5pt}
\setlength{\parskip}{-2pt}
\item[\small$\bullet$] \textbf{Input Poisoning Attack (IPA):} Fake users input the poisoned trajectories generated by Algorithm~\ref{alg:prefix-suffix} into the LDP protocols, like real users. These trajectories undergo perturbation before being sent to the server. IPA typically has lower effectiveness but is harder to defend against.
\item[\small$\bullet$] \textbf{Output Poisoning Attack (OPA):} Fake users directly send crafted perturbed data to the server by bypassing the LDP protocols. For $n$-gram and ATP, they output POI-level trajectories generated by Algorithm~\ref{alg:prefix-suffix}. For PrivTC and LDPTrace, which output transition probabilities of points, pairs, or triplets, we make some adaptations. In particular, using OUE as the LDP protocol, all bits for poisoned trajectories' points, pairs, or triplets are set to 1. We also randomly sample non-target bits to match the expected number of 1's in a genuine user's perturbed binary vector. This technique, similar to MGA~\cite{Xiaoyu2021MGA}, makes detection more difficult.
\end{enumerate}

\subsection{Evaluation Metrics}
Two metrics are used to evaluate the attack performance.

\subsubsection{Average Score of Trajectories} 
For a perturbed trajectory dataset $\hat{\mathcal{T}}$, the average score of trajectories, $\text{AvgScore}(\hat{\mathcal{T}})$, is calculated as

\begin{small}
\begin{equation}
\label{AvgScore}
\text{AvgScore}(\hat{\mathcal{T}}) = \frac{\sum_{\textit{tp}\in \textit{TP}} \sum_{\tau \in \hat{\mathcal{T}}} \text{score}(\textit{tp}) \cdot \text{CountPattern}(\textit{tp},\tau)}{|\hat{\mathcal{T}}|}.
\end{equation}
\end{small}

\noindent A higher $\text{AvgScore}(\hat{\mathcal{T}})$ is achieved when the product of the average number of target patterns per trajectory and their corresponding scores in the perturbed trajectory dataset $\hat{\mathcal{T}}$ is larger. The \textit{gain} in average score is defined as the difference between the average score after the attack and the average score before the attack. A larger gain indicates better attack performance.

\subsubsection{Average Percentile Rank of Target Patterns} 
This metric compares how frequently each target pattern appears relative to patterns of the same length by calculating the percentile rank (PR) for each target pattern and averaging these PR values. For a perturbed trajectory dataset $\hat{\mathcal{T}}$, the average PR is calculated as

\begin{equation}
\label{AvgPR}
\text{AvgPR}(\hat{\mathcal{T}}) = \frac{\sum_{\textit{tp}\in \textit{TP}} \text{GetPR}(\textit{tp}, \hat{\mathcal{T}})}{|\textit{TP}|},
\end{equation}

\noindent where $\text{GetPR}(\textit{tp}, \hat{\mathcal{T}})$ calculates the percentile rank of \linebreak $\sum_{\tau \in \hat{\mathcal{T}}}\text{CountPattern}(\textit{tp}, \tau)$ within the set $\sum_{\tau \in \hat{\mathcal{T}}}\text{CountPattern}(\textit{tp}, \tau)$ within the set $\left\{\sum_{\tau \in \hat{\mathcal{T}}}\text{CountPattern}(\textit{pattern}, \tau) \, \middle| \, |\textit{pattern}| = |\textit{tp}|\right\}$.\linebreak Overall, a higher $\text{AvgPR}(\hat{\mathcal{T}})$ indicates that target patterns rank higher in frequency among patterns of the same length. The \textit{gain} in average PR is calculated as the difference between the average PR after the attack and the average PR before the attack. A larger gain signifies improved attack performance.

\subsection{Results}\label{sec: Results}
We report the experimental results here. Each value is derived by averaging the results from five independent experiments. 

\subsubsection{Results for IPA and OPA}\label{sec: Results for IPA and OPA}
In this section, we examine the effectiveness of our input poisoning attack (IPA), and output poisoning attack (OPA) across all datasets and protocols with a privacy budget of $\varepsilon = 1$.

\begin{figure}[htbp]
\centering
\captionsetup{font=small}
\includegraphics[trim=10 2 10 10, clip, width=8cm, height=6cm]{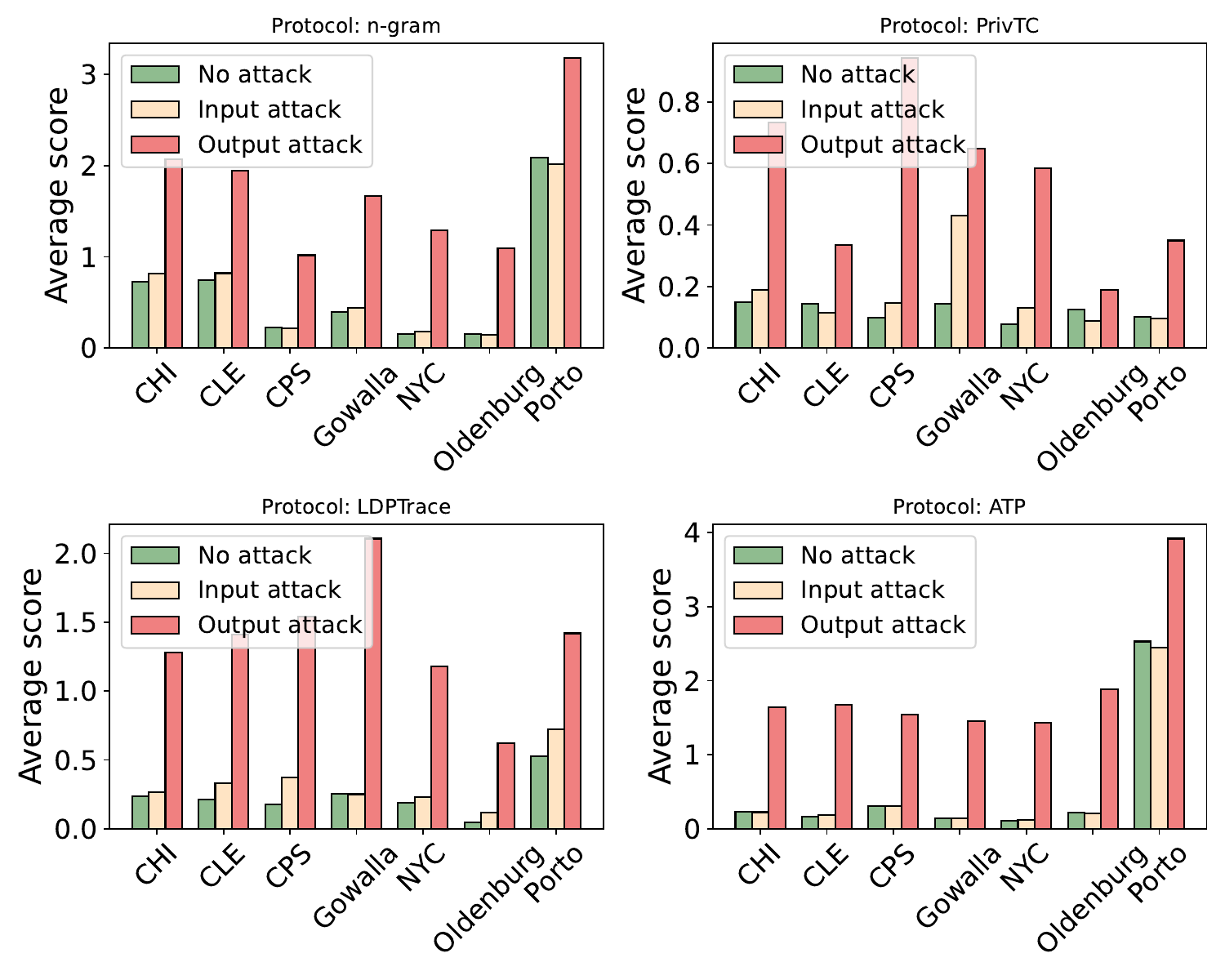}
\vspace{-10pt}
\caption{Average score of IPA and OPA with $\varepsilon = 1$ (higher score indicates greater attack effectiveness).}
\label{Fig:avgscore_all_protocol_all_dataset_eps1}
\end{figure}

Figure~\ref{Fig:avgscore_all_protocol_all_dataset_eps1} compares the average scores for each protocol and dataset under three conditions: no attack, IPA, and OPA. For most datasets using $n$-gram, PrivTC, LDPTrace, and ATP, the average scores for IPA are slightly higher than those with no attacks. Moreover, the average scores for OPA are significantly higher than both IPA and no attacks.

\begin{figure}[htbp]
\captionsetup{font=small}
\centering
\includegraphics[trim=10 2 10 10, clip, width=8cm, height=6cm]{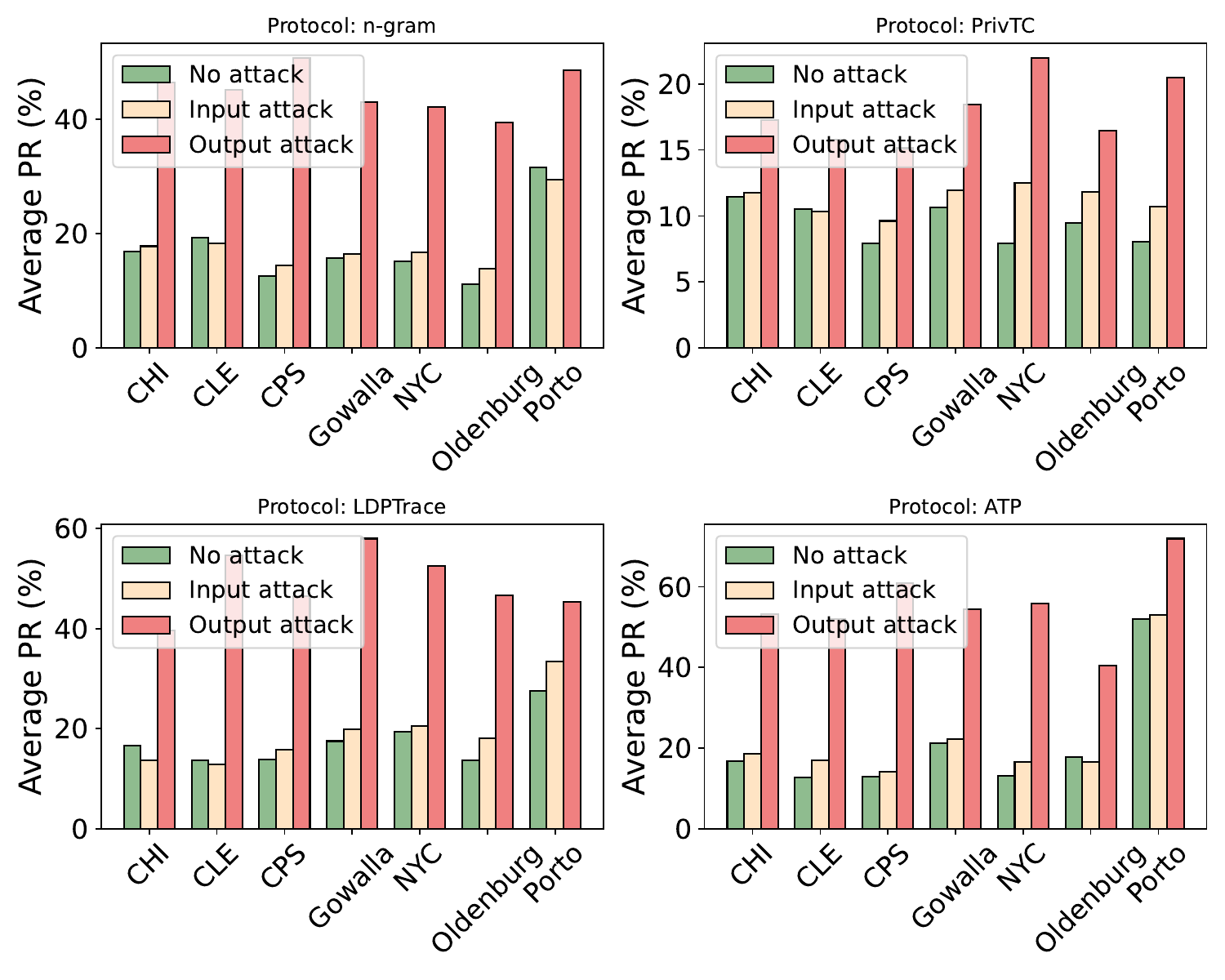}
\vspace{-10pt}
\caption{Average percentile rank (PR) of IPA and OPA with $\varepsilon = 1$ (higher score indicates greater attack effectiveness).}
\label{Fig:avgpr_all_protocol_all_dataset_eps1}
\end{figure}

Figure~\ref{Fig:avgpr_all_protocol_all_dataset_eps1} compares the average percentile ranks among different protocols and datasets in scenarios without attacks, with IPA, and with OPA. A higher average percentile rank indicates that target patterns occur more frequently, increasing their recommendation likelihood. Similar to the average score results, IPA shows a slight increase, and OPA shows a significant increase in PR value for all protocols.

Note that the average scores and average PRs for PrivTC's no attack, IPA and OPA are lower than other three protocols. This could be due to PrivTC's limited capability in preserving patterns with $\varepsilon=1$. The grid granularity of PrivTC is also influenced by $\varepsilon$, where a larger $\varepsilon$ leads to a finer granularity. Upon investigating the granularity with $\varepsilon$ set to 1, we observed that it typically varies from $8\times 8$ to $10\times 10$ across most datasets. These granularities are too coarse compared to the $16\times 16$ granularity of the target pattern, leading to reduced pattern preservation capability.

\begin{figure}[htbp]
\captionsetup{font=small}
\centering
\begin{minipage}[t]{0.21\textwidth}
\centering
\includegraphics[trim=5 5 2 5, clip, width=3.8cm, height=2.8cm]{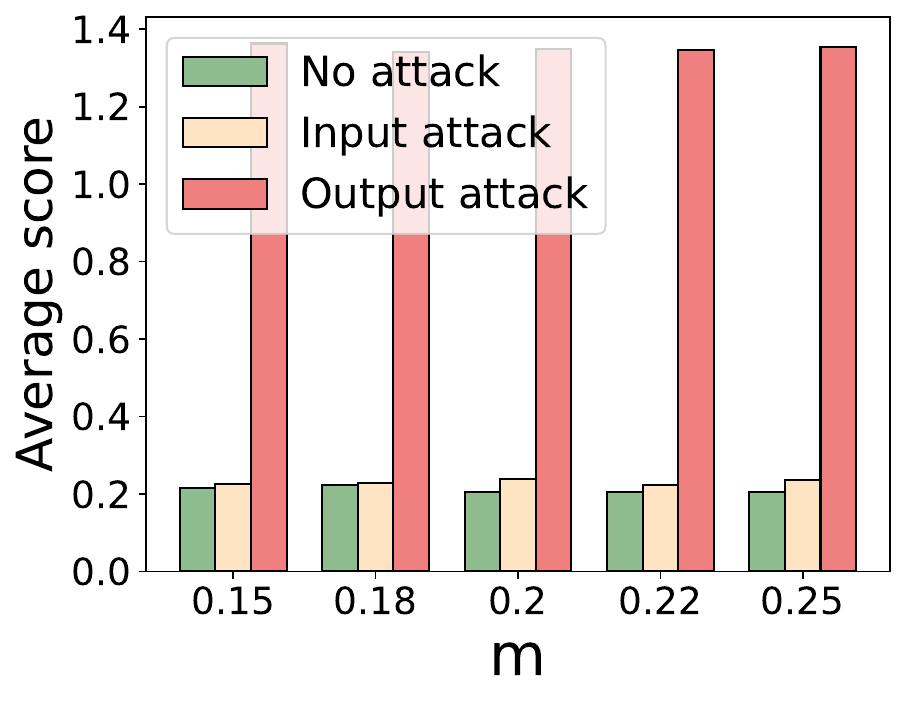}
\subcaption{ATP on CHI}
\end{minipage}
\begin{minipage}[t]{0.21\textwidth}
\centering
\includegraphics[trim=5 5 2 5, clip, width=3.8cm, height=2.8cm]{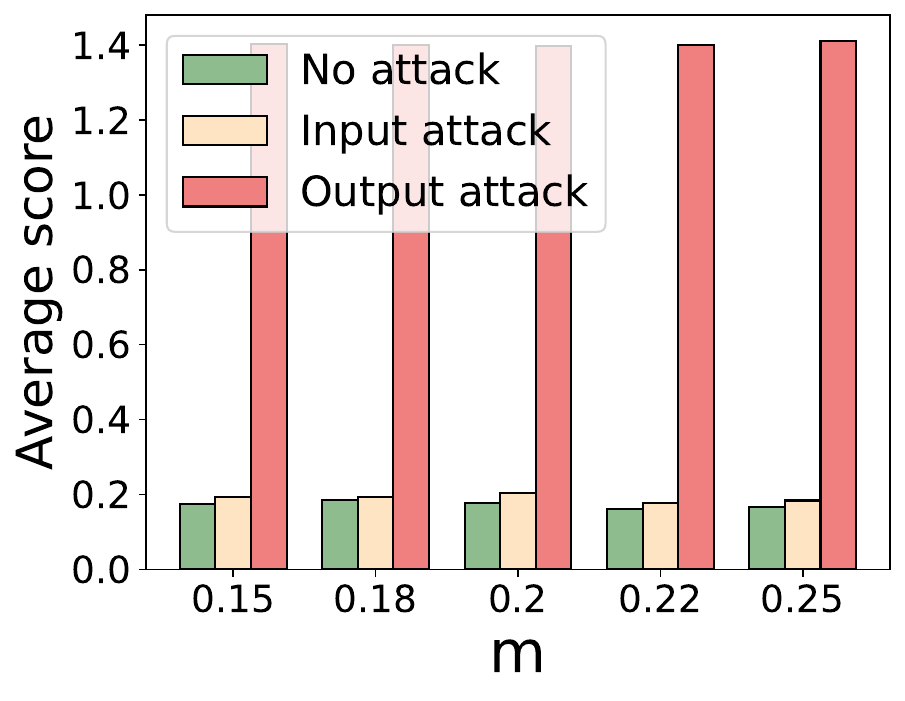}
\subcaption{ATP on CLE}
\end{minipage}
\begin{minipage}[t]{0.21\textwidth}
\centering
\includegraphics[trim=5 5 2 5, clip, width=3.8cm, height=2.8cm]{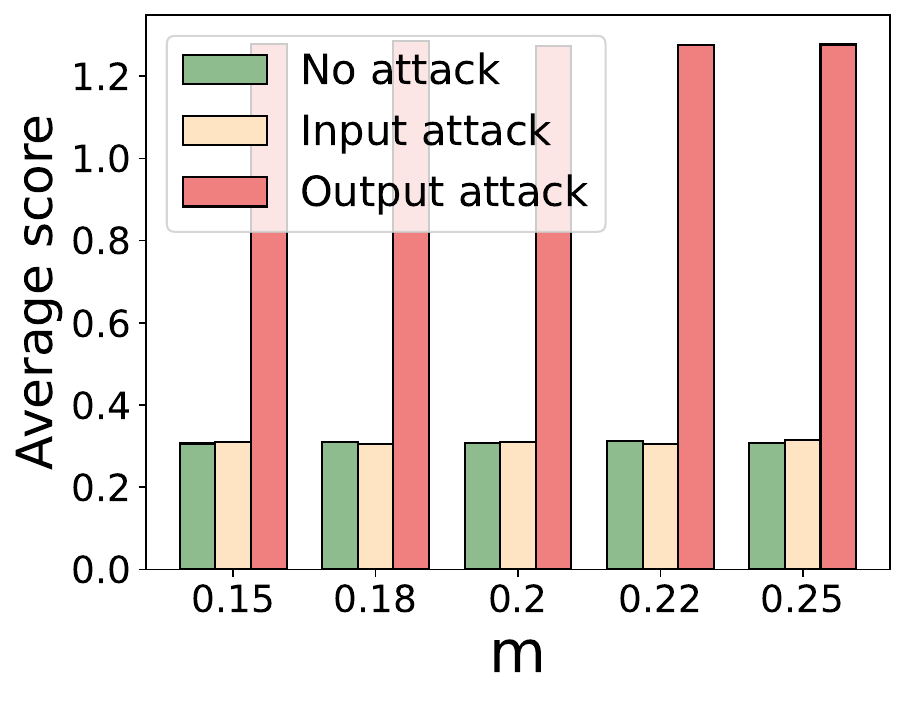}
\subcaption{ATP on CPS}
\end{minipage}
\begin{minipage}[t]{0.21\textwidth}
\centering
\includegraphics[trim=5 5 2 5, clip, width=3.8cm, height=2.8cm]{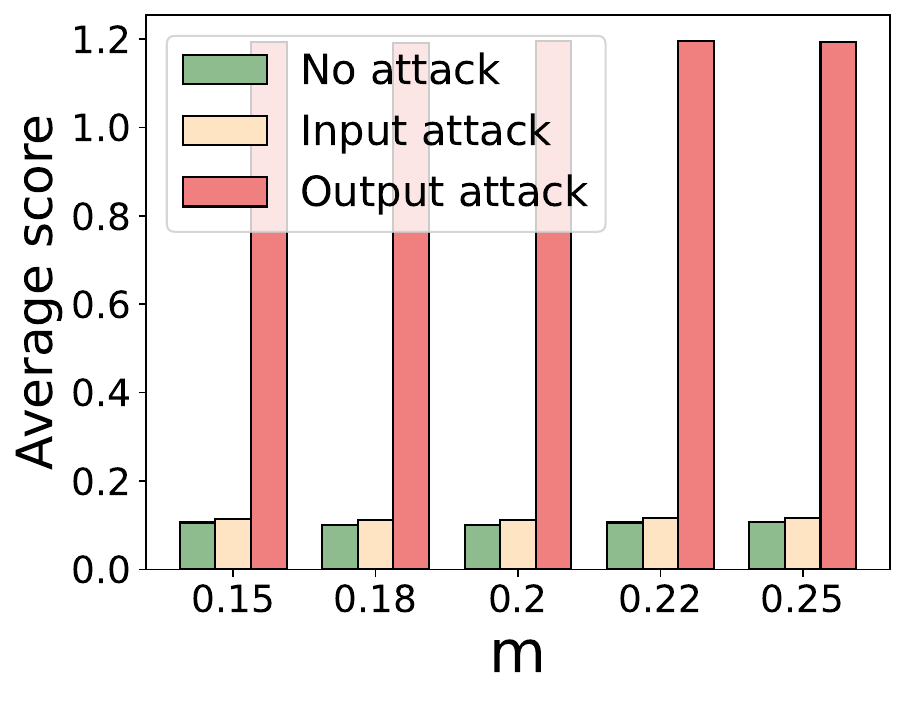}
\subcaption{ATP on NYC}
\end{minipage}
\vspace{-8pt}
\caption{Average scores of ATP on CHI, CLE, CPS, and NYC with different fake user ratios ($\frac{m}{m+n}$).}
\label{ATP_score_with_ms}
\end{figure}

\begin{figure}[htbp]
\captionsetup{font=small}
\centering
\begin{minipage}[t]{0.15\textwidth}
\centering
\includegraphics[trim=0 13 0 2, clip, width=2.8cm]{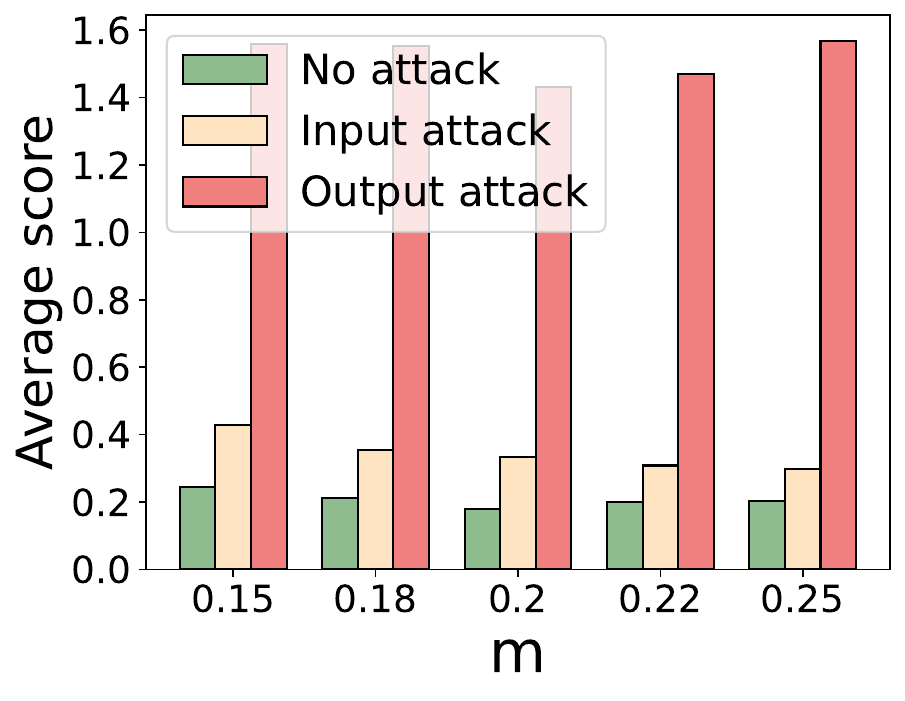}
\subcaption{\scriptsize LDPTrace on CPS}
\end{minipage}
\begin{minipage}[t]{0.15\textwidth}
\centering
\includegraphics[trim=0 13 0 2, clip, width=2.8cm]{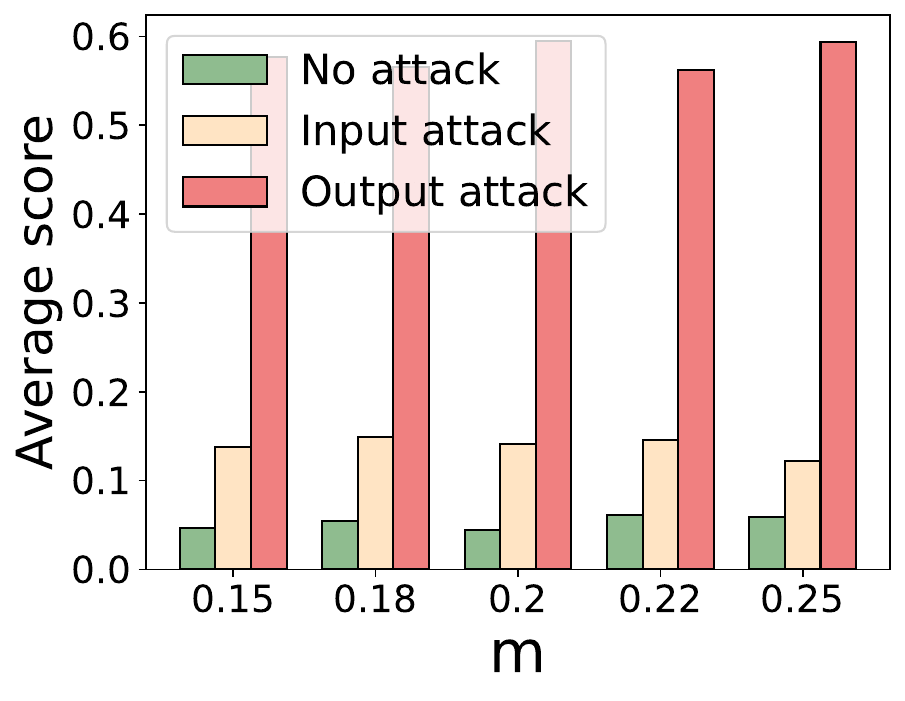}
\subcaption{\scriptsize LDPTrace on Oldenburg}
\end{minipage}
\begin{minipage}[t]{0.15\textwidth}
\centering
\includegraphics[trim=0 13 0 2, clip, width=2.8cm]{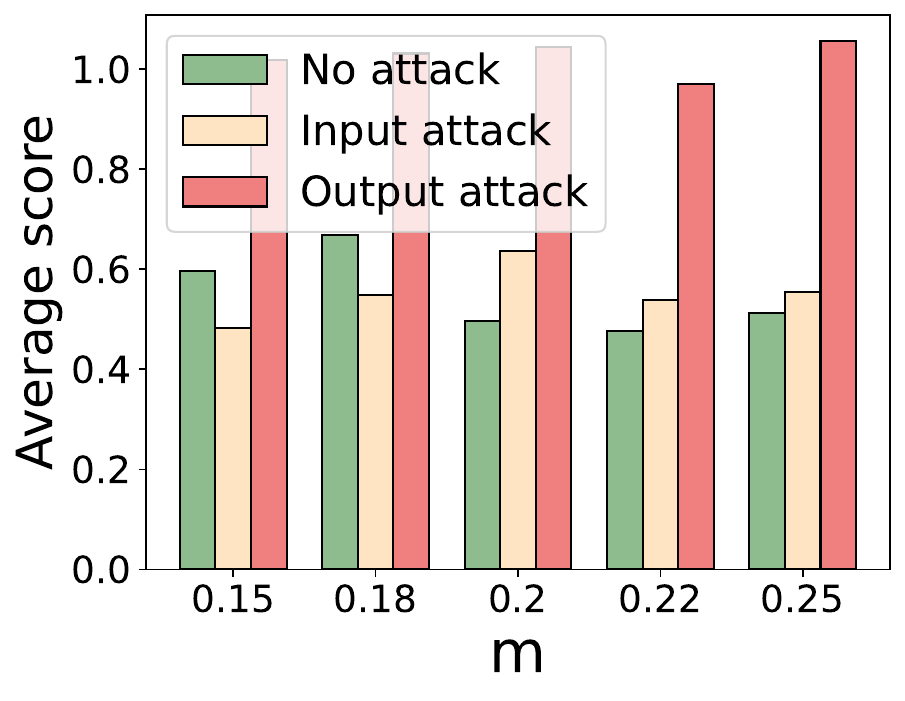}
\subcaption{\scriptsize LDPTrace on Porto}
\end{minipage}
\vspace{-13pt}
\caption{\footnotesize Average scores of LDPTrace with different fake user ratios ($\frac{m}{m+n}$).}
\label{LDPTrace_score_with_ms}
\end{figure}

\subsubsection{Impact of Different Privacy Budgets}\label{5.3.2}
In this section, we evaluate the performance of our IPA and OPA under various privacy budgets. Due to space constraints, we focus on the datasets analyzed by the original protocol developers. Specifically, for $n$-gram, the datasets include CPS and NYC; for PrivTC, the datasets are Gowalla and Porto; for LDPTrace, the datasets include CPS, Oldenburg, and Porto; for ATP, the datasets cover CHI, CLE, CPS, and NYC.

As illustrated in Figure~\ref{n_gram_All_eps} in the Appendix~\ref{sec: More Experimental Results} and Figure~\ref{ATP_All_eps} in the Appendix~\ref{sec: More Experimental Results}, for the OPA on $n$-gram and ATP, since they transmit fake trajectories directly to the server, the attack's effectiveness is relatively independent of $\varepsilon$. Regardless of $\varepsilon$, the increases in scores and PRs remain consistent. When $\varepsilon$ increases, there is a slight improvement in the performance of the IPA. This is because with a higher privacy budget, target patterns in fake trajectories can be better preserved. In some cases, such as ATP on CHI, the IPA shows a negative average score gain, meaning that the average score for IPA is slightly lower than that of no attack. This may be due to the $max_\text{rep}$ parameter, which restricts the same poisoning trajectory from appearing more than once. Therefore, the poisoning trajectories generated will encompass both shorter and longer target patterns more evenly. The frequency of shorter target patterns in the poisoned dataset might be lower than that of the real dataset. However, longer target patterns are less likely to persist after LDP, so the impact on the score increase is minimal. The proportion of shorter target patterns in poisoned trajectories is lower, resulting in a slight decrease in the average score. However, when assessing the average PRs, IPA generally outperforms no attack. This is because even if only a small fraction of longer target patterns are preserved, they can substantially enhance its ranking among patterns of the same length.

The results of LDPTrace are illustrated in Figure~\ref{LDPTrace_All_eps} in the Appendix~\ref{sec: More Experimental Results}. On the CPS, Oldenburg and Porto datasets, privacy budgets seem to have little impact on the average scores and percentile ranks of OPA and IPA. However, this might be due to the insufficient difference between the $\varepsilon$ values. For further experiments and analysis, please refer to Section~\ref{Security-Privacy Trade-off or Consistency}.

Shown in Figure~\ref{PrivTC_All_eps} in the Appendix~\ref{sec: More Experimental Results}, for PrivTC tested on the Gowalla and Porto datasets, the results for IPA and OPA tend to perform better with larger $\varepsilon$. This is because a larger $\varepsilon$ results in finer grid granularity, which helps in preserving patterns. However, when $\varepsilon=5$, the attack performance seems to be worse compared to $\varepsilon=4$. This may be due to having already reached the maximum granularity of $16\times 16$. Additionally, at $\varepsilon=5$, non-target patterns in real data are better preserved, making them less susceptible to the influence of target patterns in fake trajectories during the trajectory synthesis process, leading to a decrease in attack performance.

\subsubsection{Security-Privacy Trade-off or Consistency?}\label{Security-Privacy Trade-off or Consistency} 
Previous studies on OPA reveal conflicting findings. Some research~\cite{Xiaoyu2021MGA, 279934} observed that a smaller $\varepsilon$ (higher privacy) improves data poisoning attacks (lower security), termed "privacy-security trade-off". Conversely, another study~\cite{287322} found that a smaller $\varepsilon$ reduces attack performance, known as "privacy-security consistency".

The main reason for this difference lies in the former's goal is maximizing the frequency or value of specific items, while the latter aims to make the result as close to the target value as possible. For the former scenario, a smaller $\varepsilon$ introduces more noise and dispersion in real user data, while the poisoned data is unaffected and contributing significantly to estimated item frequencies, thereby yielding a higher attack effectiveness. Conversely, in the latter case, a small $\varepsilon$ introduces more noise, complicating the estimation of real user data and causing the final attack outcome to deviate from the target value.

In Section~\ref{5.3.2}, the results for $\varepsilon$ values ranging from 0.1 to 5 do not demonstrate a clear security-privacy trade-off or consistency. We attribute this to insufficient differentiation between these $\varepsilon$ values. Consequently, here we conduct extra experiments using a broader range of $\varepsilon$ values: 1, 10, 100, and 1000, for all protocols except PrivTC. For PrivTC, the minimum $\varepsilon$ is set to 5 to achieve a grid granularity of $16\times 16$, while the maximum $\varepsilon$ is limited to 500, as a value of 1000 would exceed the computable numerical range during the LDP calculation process. To optimize space and improve clarity, we calculate the average results from all seven datasets. The results are shown in Figure~\ref{More_eps} in the Appendix~\ref{sec: More Experimental Results}.

According to our experimental results of OPA, different phenomena emerged across various trajectory protocols, defying simple categorization under security-privacy trade-off or consistency. In protocols like $n$-gram and ATP (Figures~\ref{n_gram_more_eps} and \ref{ATP_more_eps} in the Appendix~\ref{sec: More Experimental Results}), where trajectories are directly sent to the server, the impact of OPA remains unaffected by $\varepsilon$; we term this as \textit{security-privacy independence}. This is a unique characteristic in LDP trajectory protocols. Poisoned trajectories are filled with target patterns and are transmitted directly to the server in both protocols, which is independent of the privacy budget, so the overall attack performance is unrelated to $\varepsilon$. On the other hand, the OPA results of PrivTC and LDPTrace (Figures~\ref{PrivTC_more_eps} and \ref{LDPTrace_more_eps} in the Appendix~\ref{sec: More Experimental Results}) exhibit a security-privacy trade-off, which is in line with previous studies~\cite{Xiaoyu2021MGA, 279934}. This is because these two protocols use OUE or OLH to collect transition pairs or triplets, and perform aggregation and trajectory synthesis on the server side, thus being affected by $\varepsilon$. In contrast, $n$-gram and ATP do not have aggregation and trajectory synthesis steps on the server side, so they are not affected by $\varepsilon$.

IPA demonstrates security-privacy consistency (Figures~\ref{n_gram_more_eps}$\sim$\ref{ATP_more_eps} in the Appendix~\ref{sec: More Experimental Results}). This implies that as $\varepsilon$ increases (indicating lower privacy), the attack performance improves (indicating lower security). This is because the target patterns in the poisonous trajectories can be better preserved after LDP, outnumbering most of the non-target patterns.

\begin{figure}[t]
\captionsetup{font=small}
\centering
\begin{minipage}[t]{0.23\textwidth}
\centering
\includegraphics[trim=5 14 2 5, clip, width=4cm]{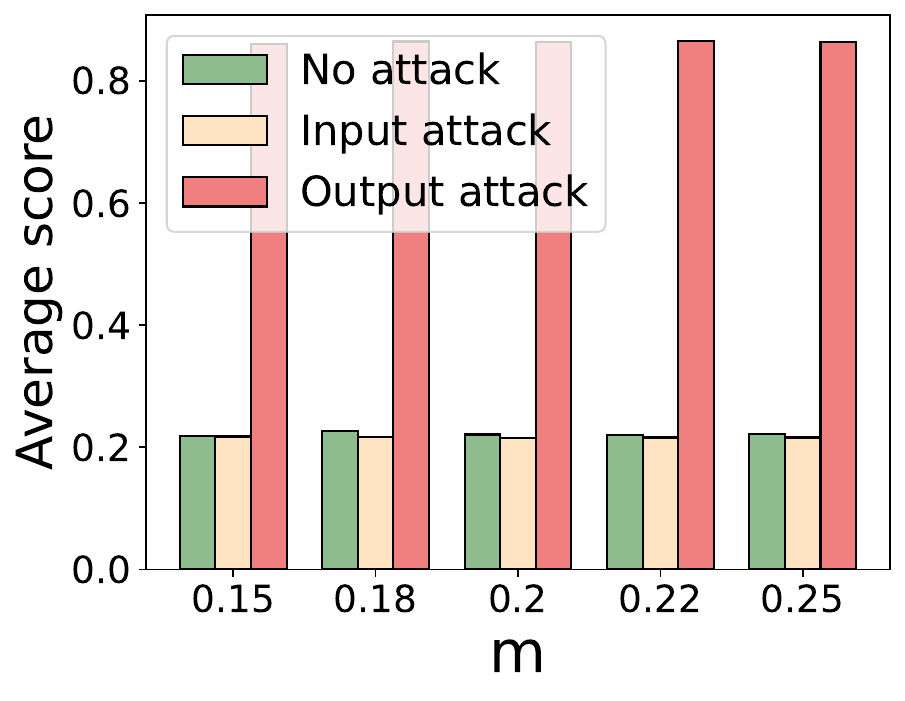}
\subcaption{$n$-gram on CPS}
\end{minipage}
\begin{minipage}[t]{0.23\textwidth}
\centering
\includegraphics[trim=5 14 2 5, clip, width=4cm]{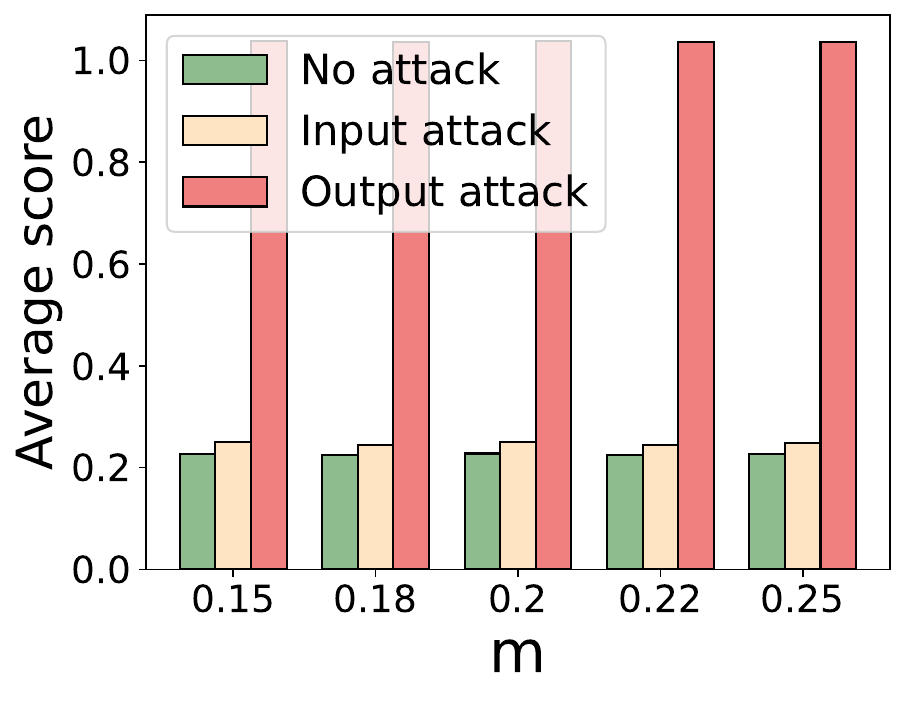}
\subcaption{$n$-gram on NYC}
\end{minipage}
\vspace{-13pt}
\caption{\footnotesize Average scores of $n$-gram with different fake user ratios ($\frac{m}{m+n}$).}
\label{n_gram_score_with_ms}
\end{figure}

\begin{figure}[t]
\captionsetup{font=small}
\centering
\begin{minipage}[t]{0.23\textwidth}
\centering
\includegraphics[trim=5 14 2 5, clip, width=4cm]{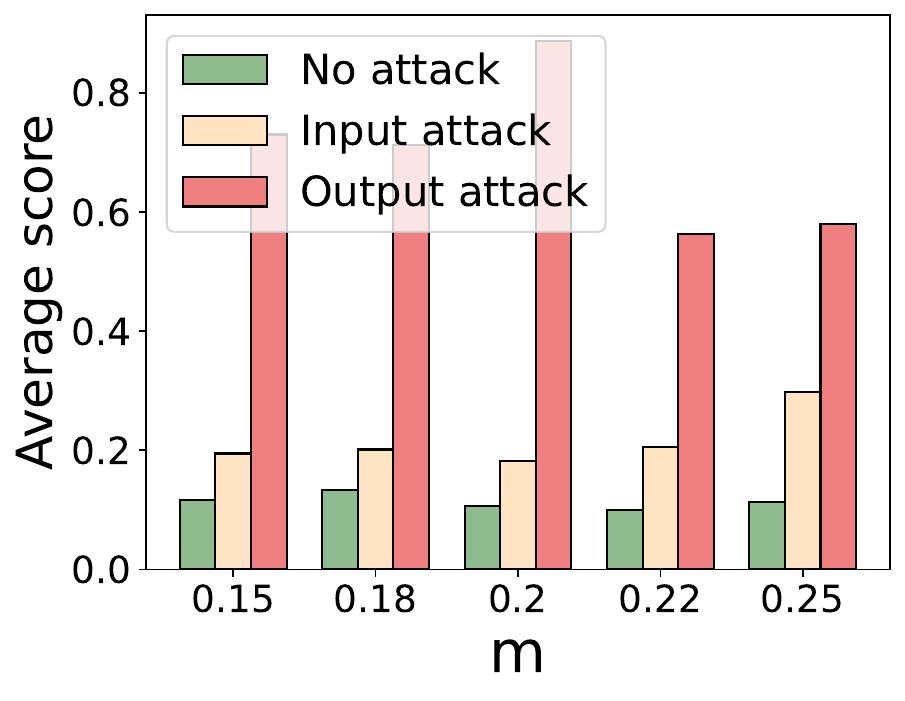}
\subcaption{PrivTC on Gowalla}
\end{minipage}
\begin{minipage}[t]{0.23\textwidth}
\centering
\includegraphics[trim=5 14 2 5, clip, width=4cm]{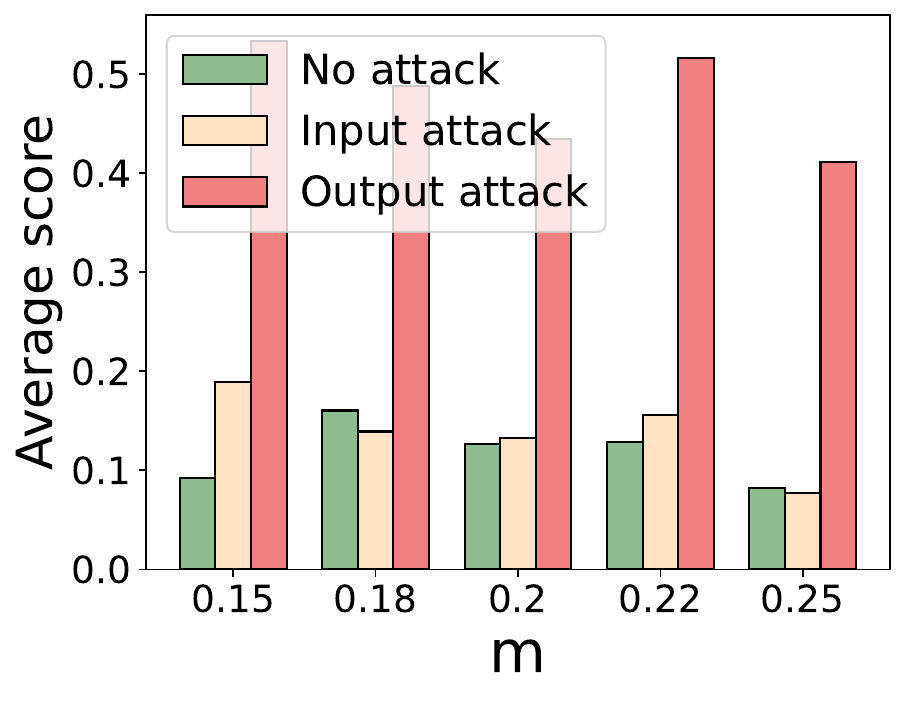}
\subcaption{PrivTC on Porto}
\end{minipage}
\vspace{-13pt}
\caption{\footnotesize Average scores of PrivTC with different fake user ratios ($\frac{m}{m+n}$).}
\label{PrivTC_score_with_ms}
\end{figure}

\subsection{Dataset Distribution Mismatch}\label{sec: Dataset Distribution Mismatch}

We evaluate our attack's robustness under various degrees of mismatch between the attacker's assumptions and the actual dataset distribution. By adjusting parameters in Section~\ref{sec:exp_parameters}, we simulate real-world discrepancies to assess the attack's effectiveness across different mismatch conditions.

\subsubsection{Comparison with different fake user ratios}

In Section~\ref{sec:exp_parameters}, we initially set the fake user ratio $\frac{m}{m+n}$ to $0.2$. Nevertheless, in the real-world case, the attacker might not be aware of the accurate number $n$. To further investigate the impact of $n$, we conduct comparative experiments by adjusting the value of $m$ to achieve fake user ratios of $0.15$, $0.18$, $0.2$, $0.22$, and $0.25$. The results of these experiments are presented in Figures~\ref{n_gram_score_with_ms}, ~\ref{PrivTC_score_with_ms}, ~\ref{LDPTrace_score_with_ms}, ~\ref{ATP_score_with_ms}. The results show the attack remains highly effective across various fake user ratios, demonstrating the method's robustness.

\subsubsection{Comparison with different length distributions}\label{sec: Comparison with different length distributions}

\begin{figure}[htbp]
\captionsetup{font=small}
\centering
\begin{minipage}[t]{0.23\textwidth}
\centering
\includegraphics[trim=5 13 2 5, clip, width=4cm]{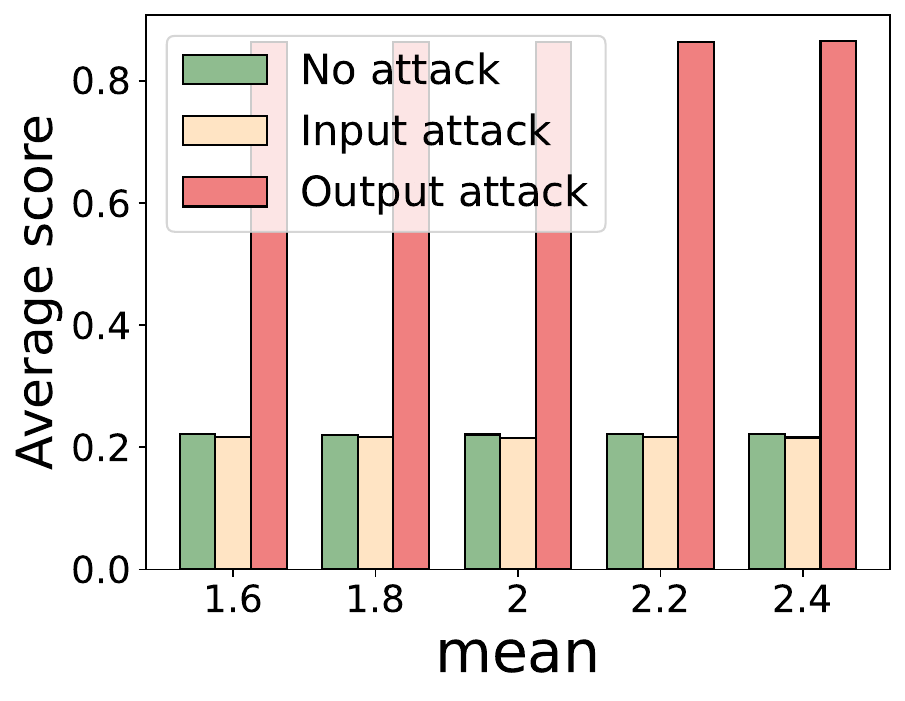}
\subcaption{$n$-gram on CPS}
\end{minipage}
\begin{minipage}[t]{0.23\textwidth}
\centering
\includegraphics[trim=5 13 2 5, clip, width=4cm]{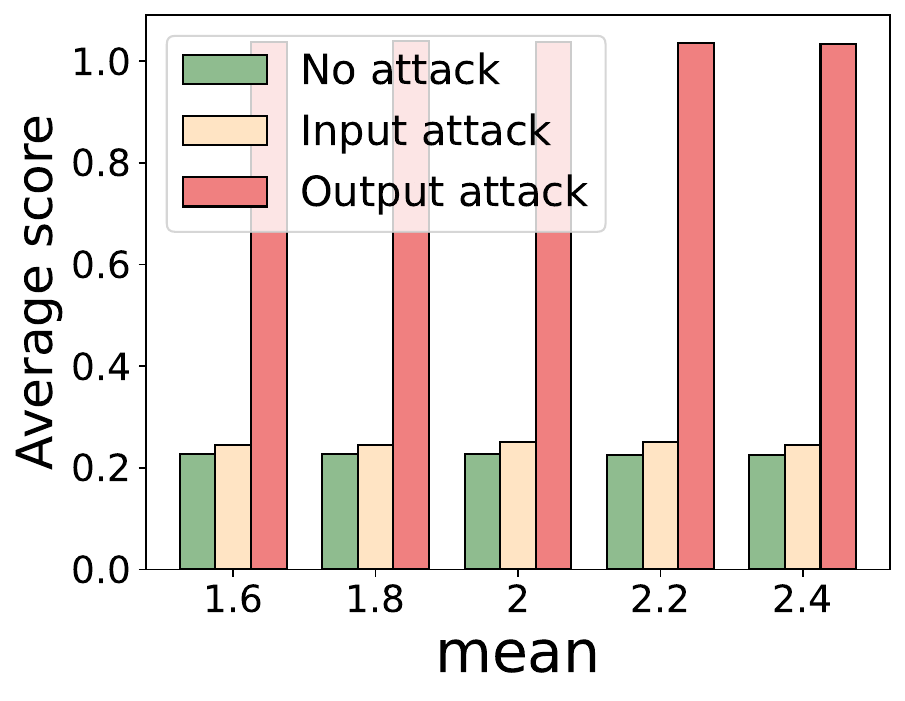}
\subcaption{$n$-gram on NYC}
\end{minipage}
\vspace{-13pt}
\caption{\footnotesize Average scores of $n$-gram with different means.}
\label{n_gram_score_with_means}
\end{figure}

\begin{figure}[htbp]
\captionsetup{font=small}
\centering
\begin{minipage}[t]{0.23\textwidth}
\centering
\includegraphics[trim=5 13 2 5, clip, width=4cm]{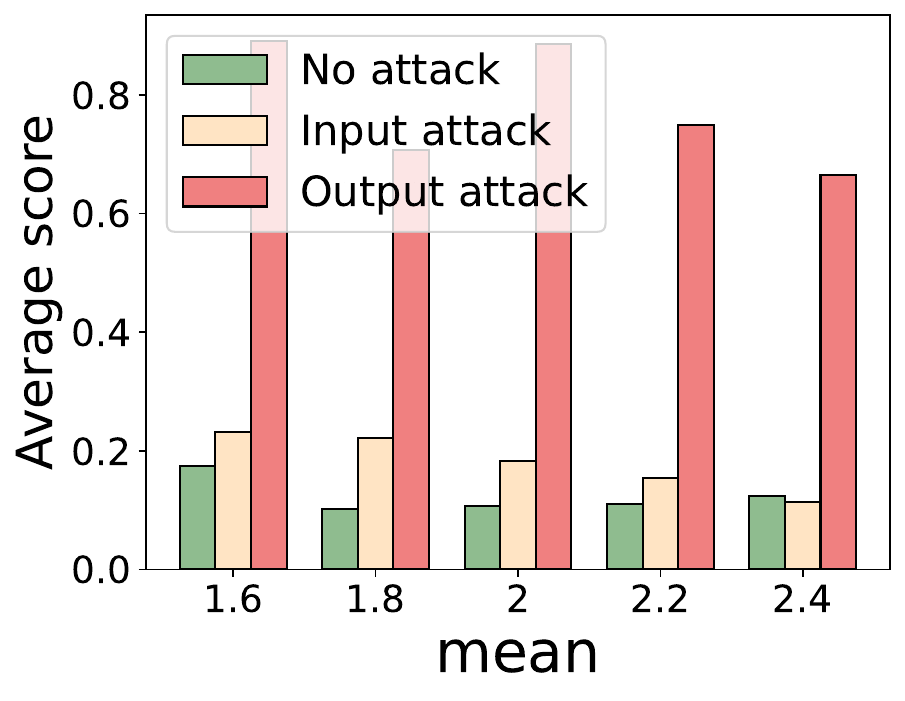}
\subcaption{PrivTC on Gowalla}
\end{minipage}
\begin{minipage}[t]{0.23\textwidth}
\centering
\includegraphics[trim=5 13 2 5, clip, width=4cm]{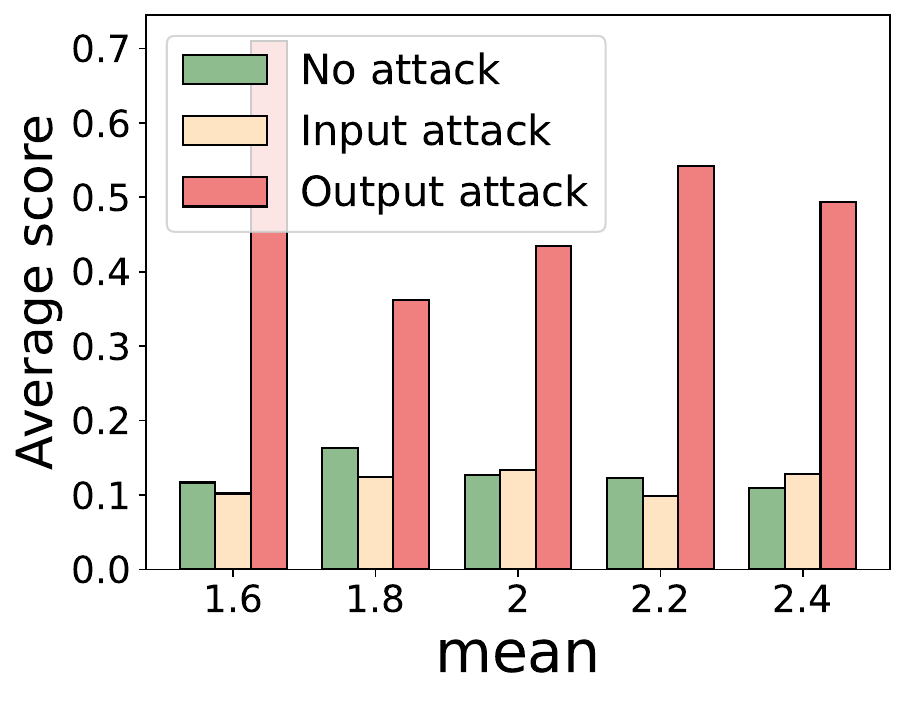}
\subcaption{PrivTC on Porto}
\end{minipage}
\vspace{-13pt}
\caption{\footnotesize Average scores of PrivTC with different means.}
\label{PrivTC_score_with_means}
\end{figure}

\begin{figure}[htbp]
\captionsetup{font=small}
\centering
\begin{minipage}[t]{0.15\textwidth}
\centering
\includegraphics[trim=0 13 0 2, clip, width=2.8cm]{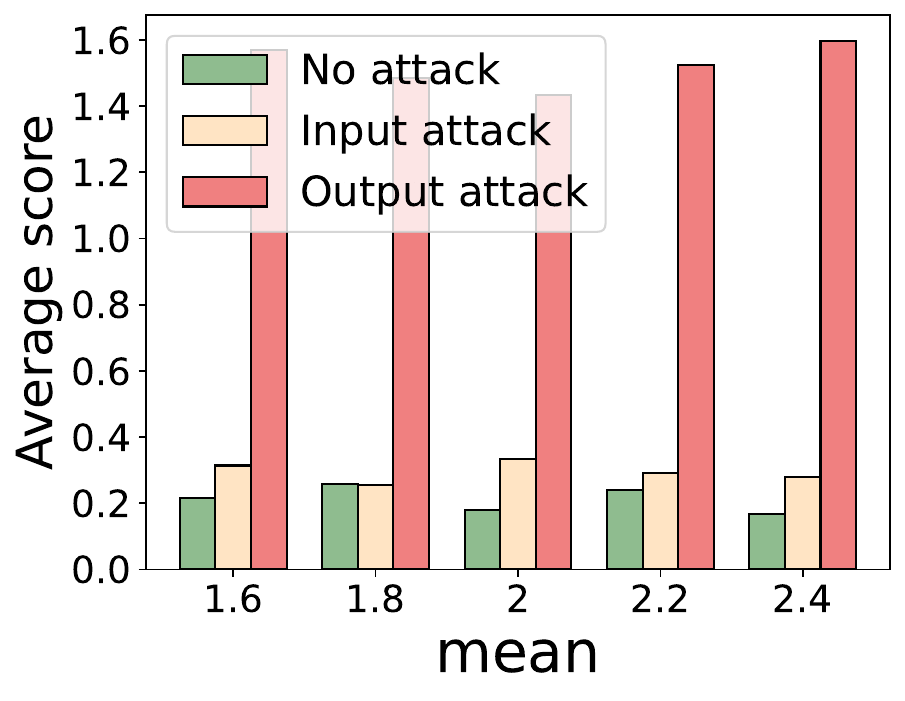}
\subcaption{\scriptsize LDPTrace on CPS}
\end{minipage}
\begin{minipage}[t]{0.15\textwidth}
\centering
\includegraphics[trim=0 13 0 2, clip, width=2.8cm]{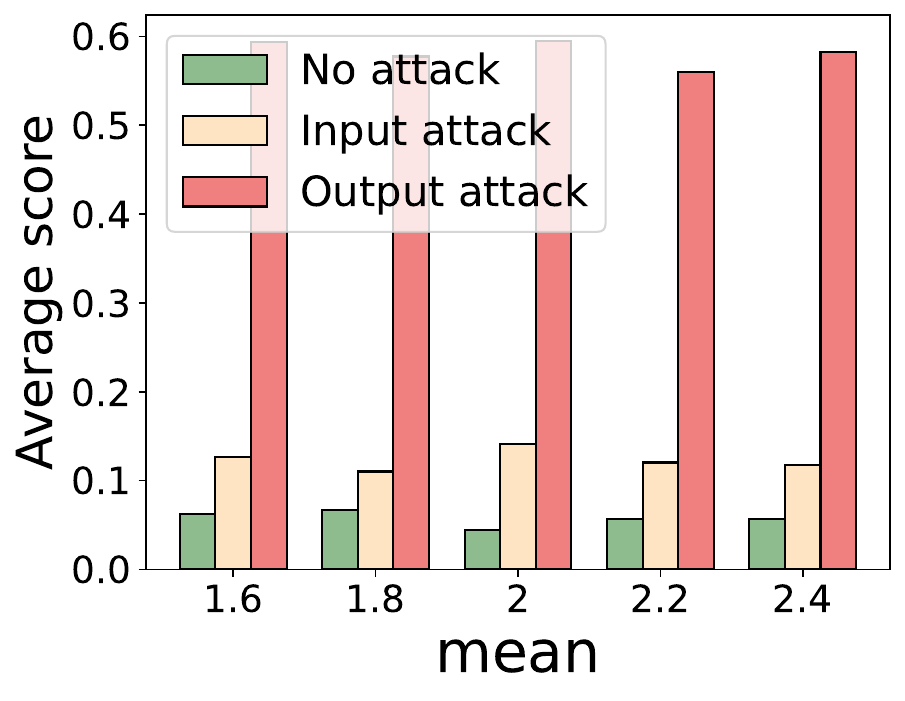}
\subcaption{\scriptsize LDPTrace on Oldenburg}
\end{minipage}
\begin{minipage}[t]{0.15\textwidth}
\centering
\includegraphics[trim=0 13 0 2, clip, width=2.8cm]{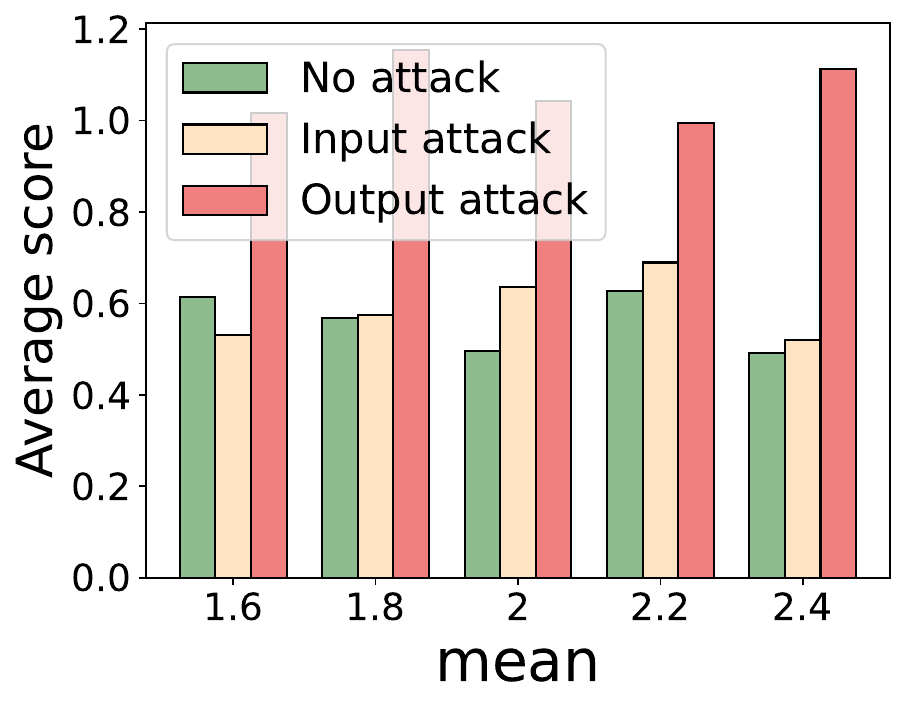}
\subcaption{\scriptsize LDPTrace on Porto}
\end{minipage}
\vspace{-13pt}
\caption{\footnotesize Average scores of LDPTrace with different means.}
\label{LDPTrace_score_with_means}
\end{figure}

\begin{figure}[htbp]
\captionsetup{font=small}
\centering
\begin{minipage}[t]{0.22\textwidth}
\centering
\includegraphics[trim=5 5 2 5, clip, width=3.8cm, height=2.8cm]{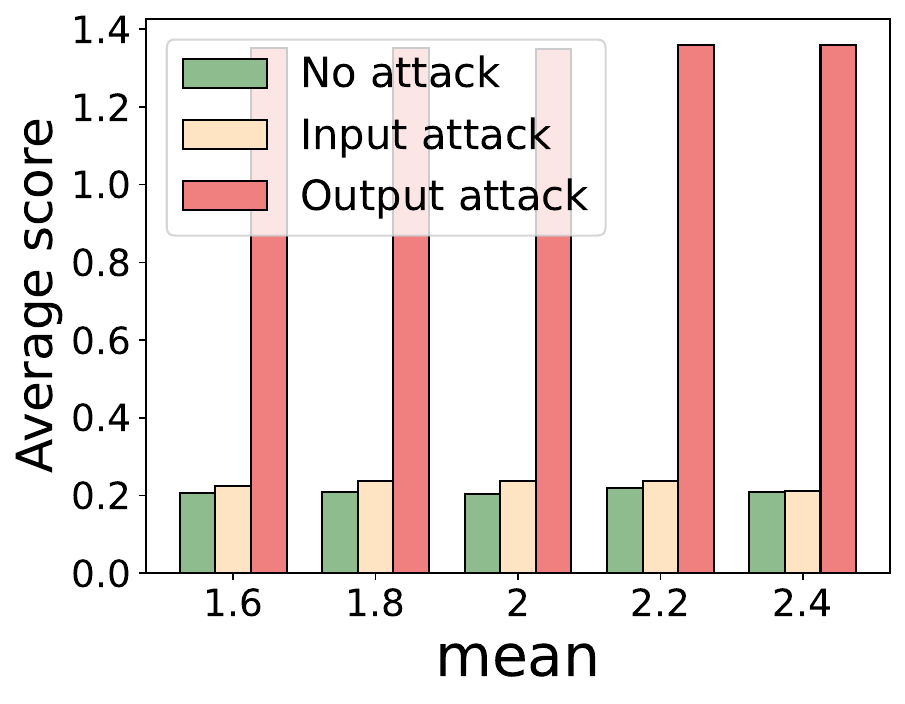}
\subcaption{ATP on CHI}
\end{minipage}
\begin{minipage}[t]{0.22\textwidth}
\centering
\includegraphics[trim=5 5 2 5, clip, width=3.8cm, height=2.8cm]{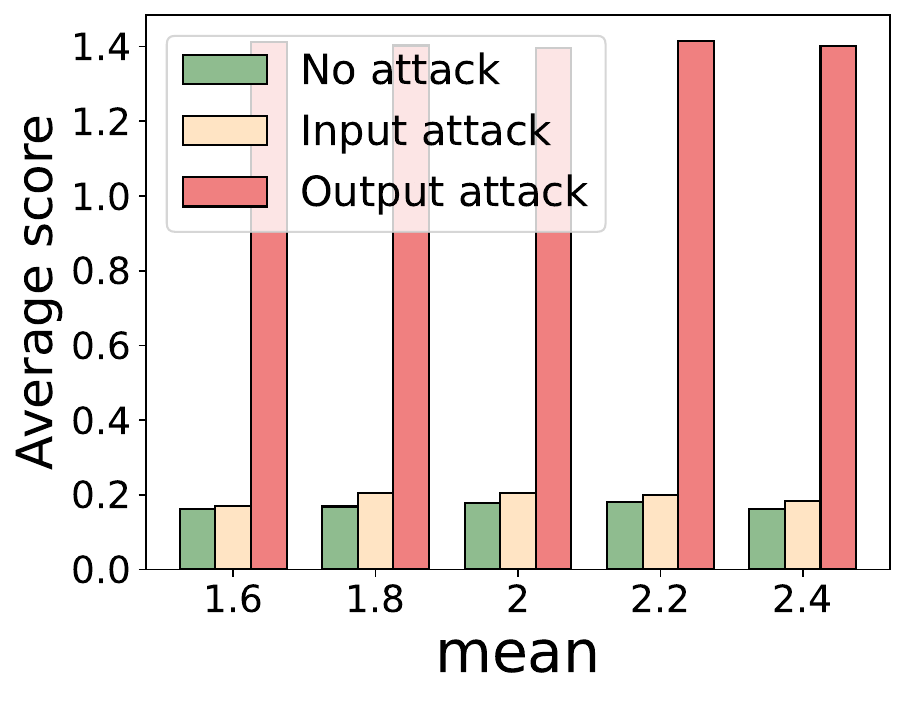}
\subcaption{ATP on CLE}
\end{minipage}
\begin{minipage}[t]{0.22\textwidth}
\centering
\includegraphics[trim=5 5 2 5, clip, width=3.8cm, height=2.8cm]{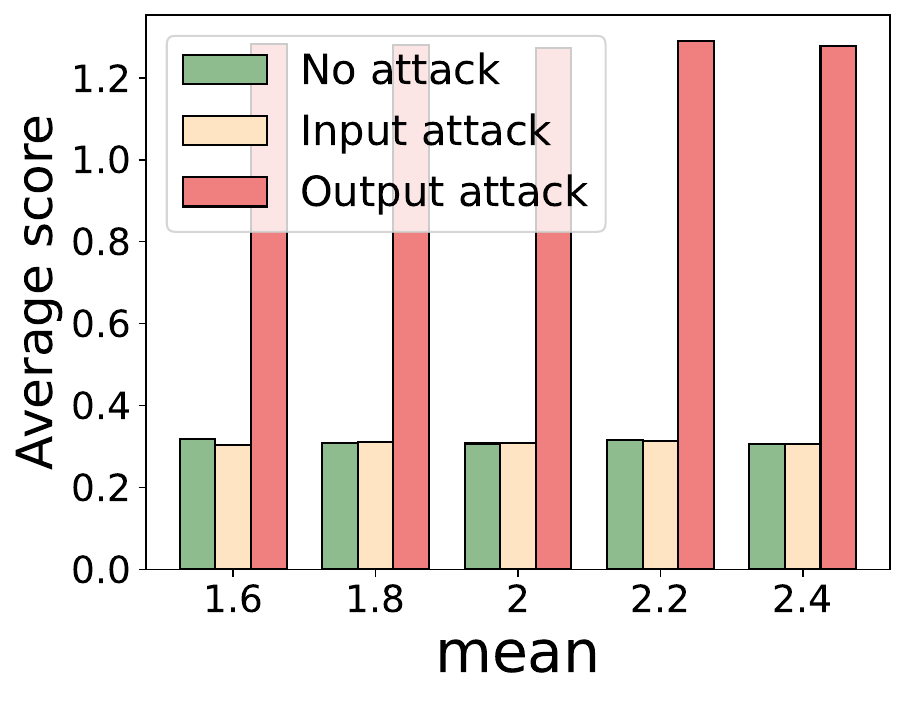}
\subcaption{ATP on CPS}
\end{minipage}
\begin{minipage}[t]{0.22\textwidth}
\centering
\includegraphics[trim=5 5 2 5, clip, width=3.8cm, height=2.8cm]{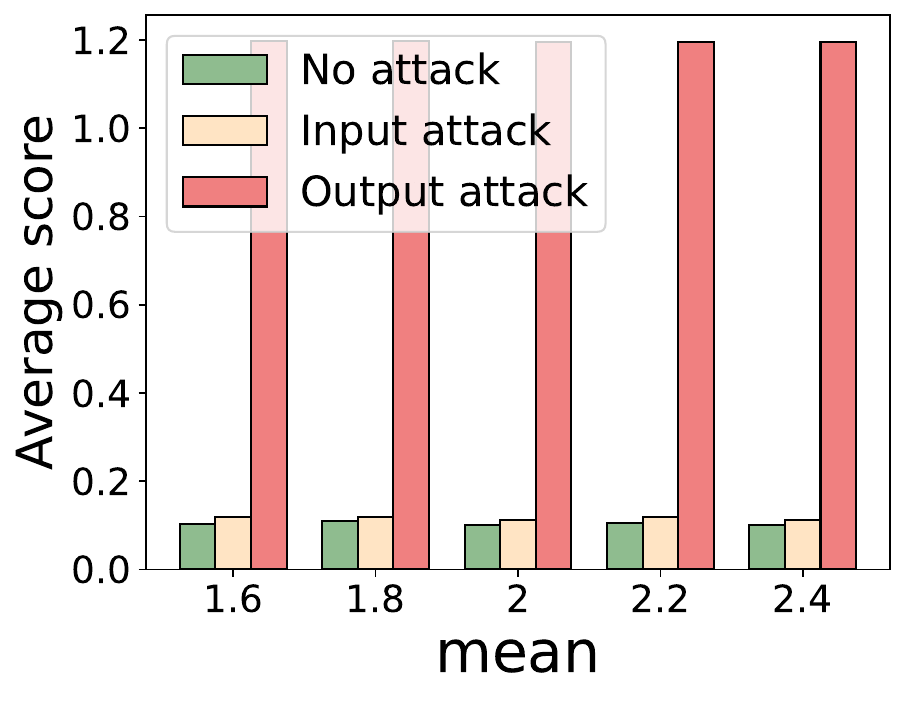}
\subcaption{ATP on NYC}
\end{minipage}
\vspace{-13pt}
\caption{Average scores of ATP with different means.}
\label{ATP_score_with_means}
\end{figure}

\begin{figure}[htbp]
\captionsetup{font=small}
\centering
\begin{minipage}[t]{0.23\textwidth}
\centering
\includegraphics[trim=5 13 2 5, clip, width=4cm]{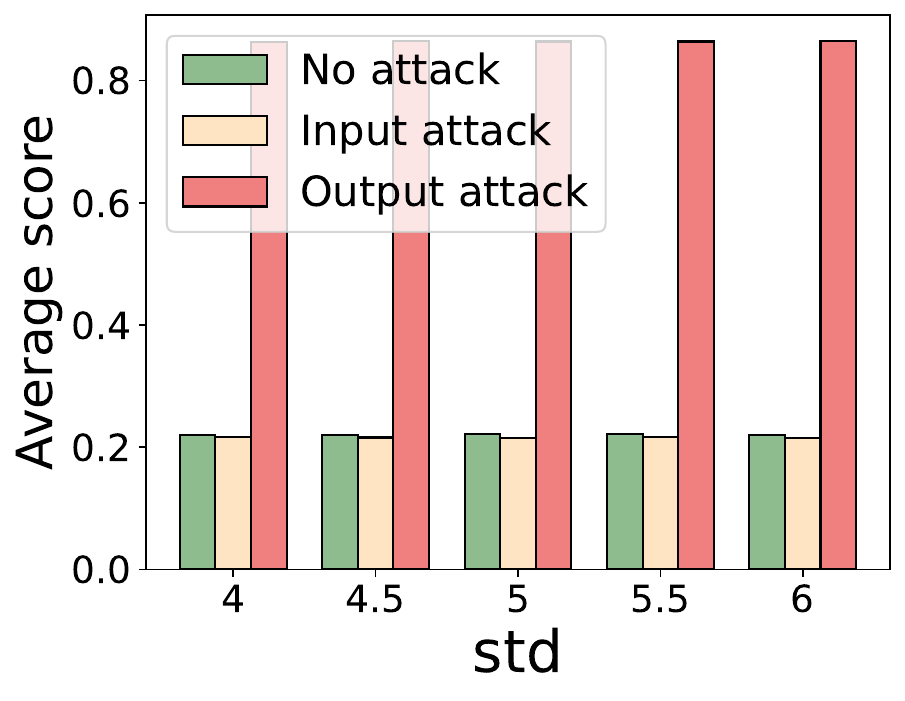}
\subcaption{$n$-gram on CPS}
\end{minipage}
\begin{minipage}[t]{0.23\textwidth}
\centering
\includegraphics[trim=5 13 2 5, clip, width=4cm]{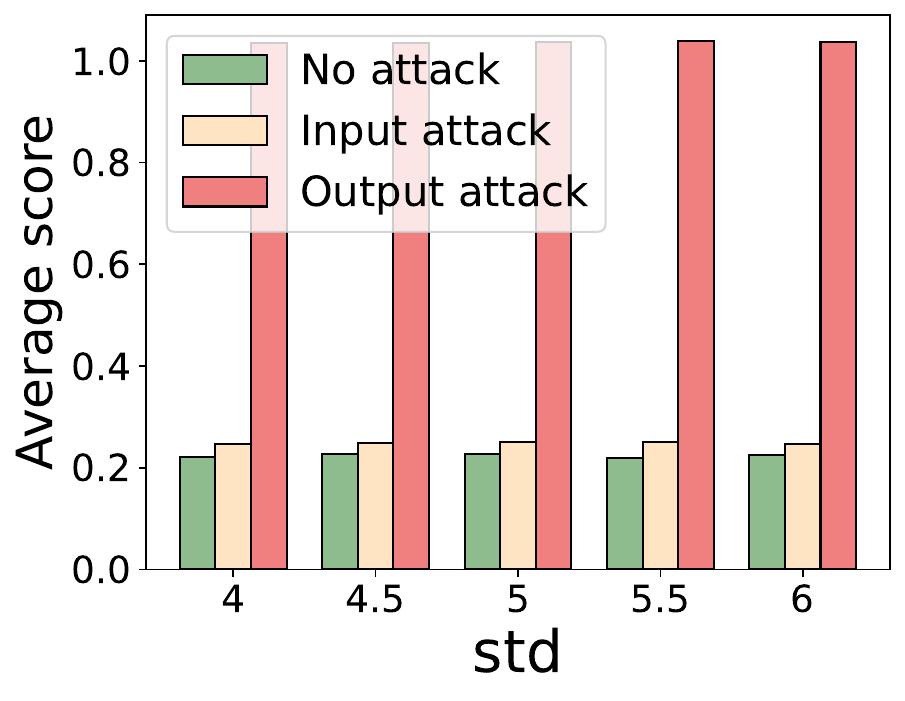}
\subcaption{$n$-gram on NYC}
\end{minipage}
\vspace{-13pt}
\caption{\footnotesize Average scores of $n$-gram with different stds.}
\label{n_gram_score_with_stds}
\end{figure}

\begin{figure}[htbp]
\captionsetup{font=small}
\centering
\begin{minipage}[t]{0.23\textwidth}
\centering
\includegraphics[trim=5 13 2 5, clip, width=4cm]{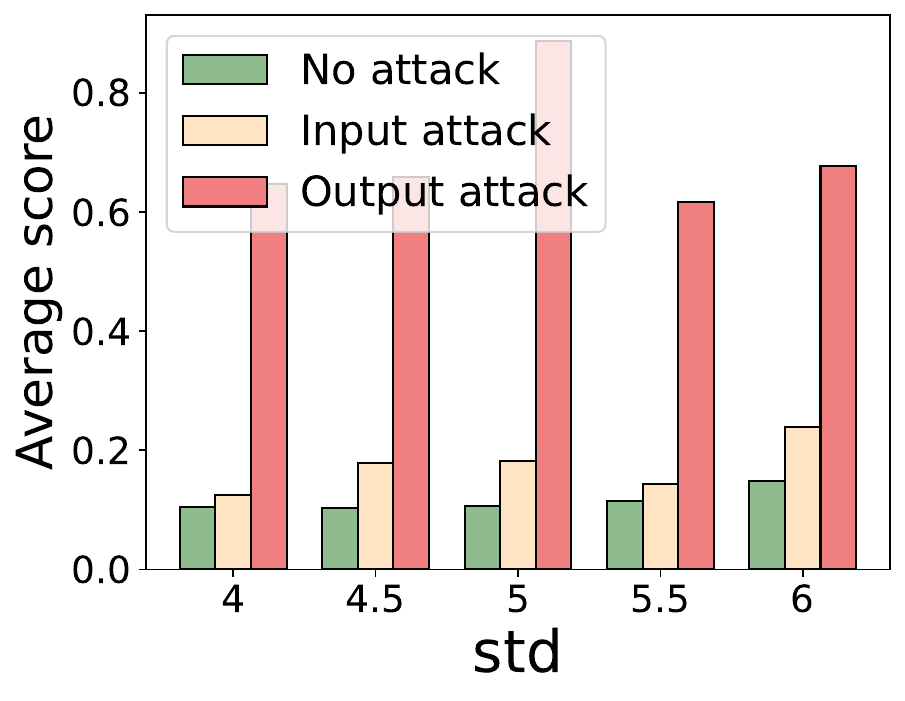}
\subcaption{PrivTC on Gowalla}
\end{minipage}
\begin{minipage}[t]{0.23\textwidth}
\centering
\includegraphics[trim=5 13 2 5, clip, width=4cm]{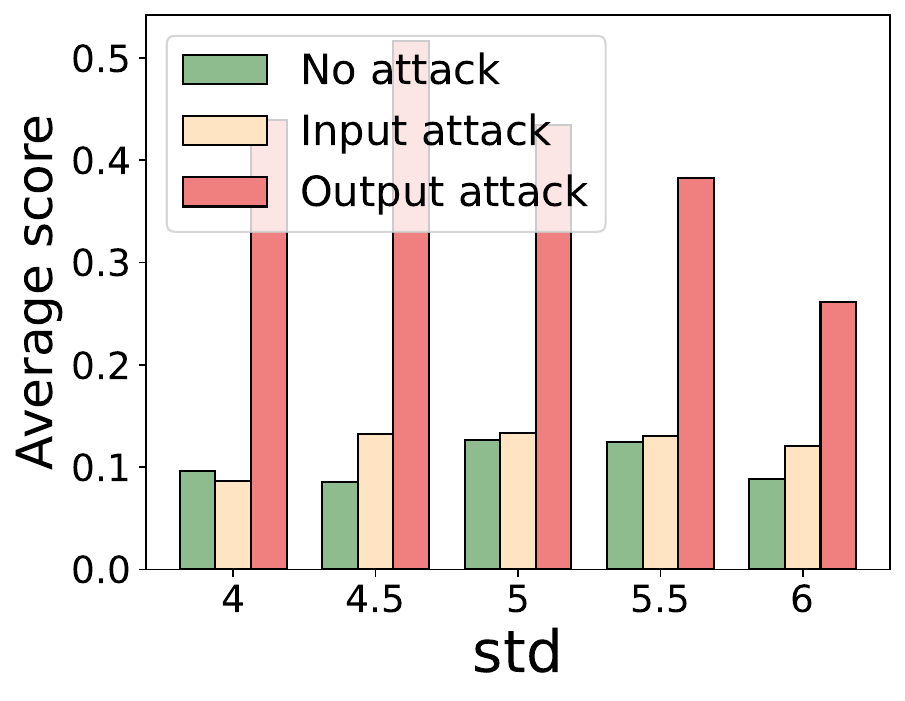}
\subcaption{PrivTC on Porto}
\end{minipage}
\vspace{-13pt}
\caption{\footnotesize Average scores of PrivTC with different stds.}
\label{PrivTC_score_with_stds}
\end{figure}

\begin{figure}[t!]
\captionsetup{font=small}
\centering
\begin{minipage}[t]{0.15\textwidth}
\centering
\includegraphics[trim=0 13 0 2, clip, width=2.8cm]{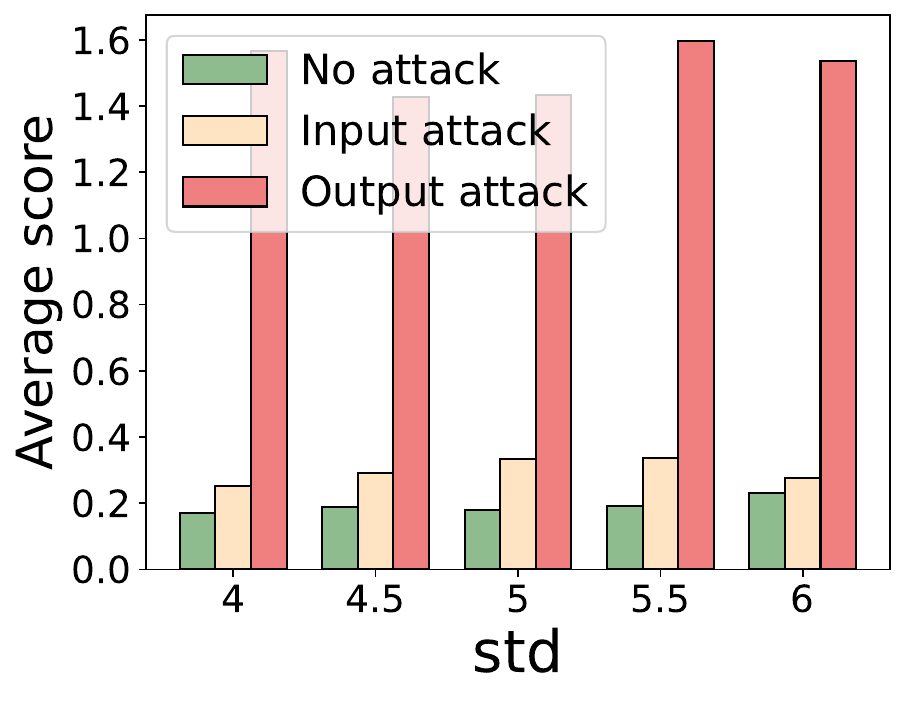}
\subcaption{\scriptsize LDPTrace on CPS}
\end{minipage}
\begin{minipage}[t]{0.15\textwidth}
\centering
\includegraphics[trim=0 13 0 2, clip, width=2.8cm]{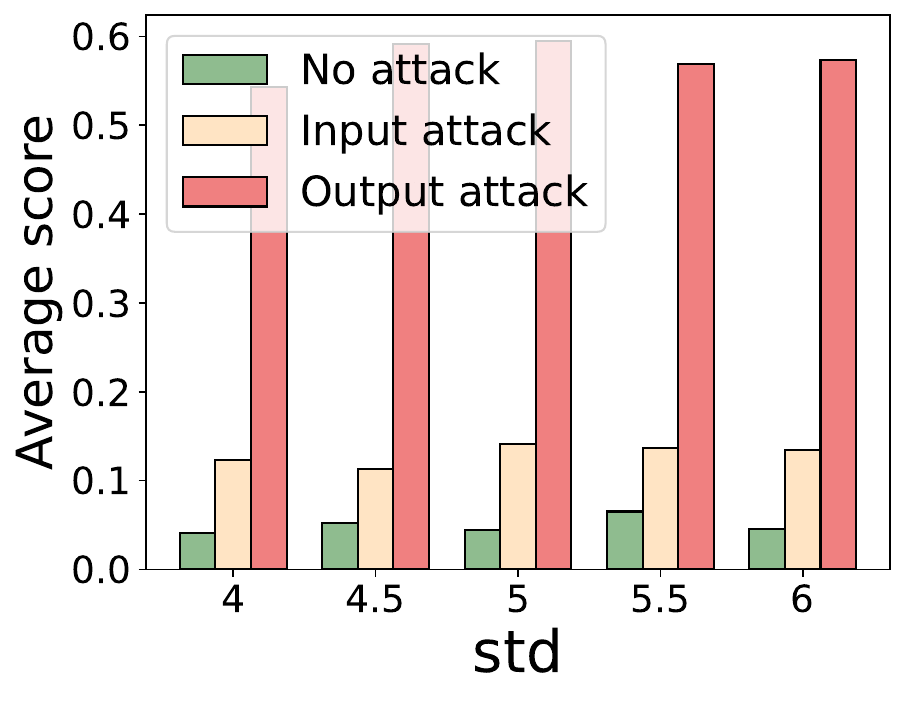}
\subcaption{\scriptsize LDPTrace on Oldenburg}
\end{minipage}
\begin{minipage}[t]{0.15\textwidth}
\centering
\includegraphics[trim=0 14 0 2, clip, width=2.8cm]{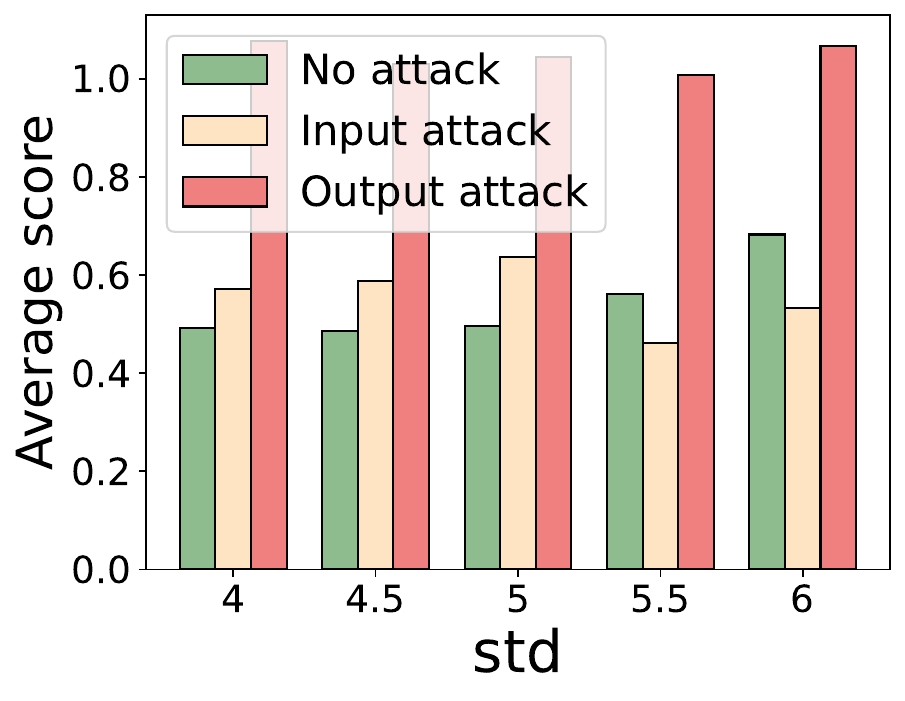}
\subcaption{\scriptsize LDPTrace on Porto}
\end{minipage}
\vspace{-13pt}
\caption{\footnotesize Average scores of LDPTrace with different stds.}
\label{LDPTrace_score_with_stds}
\end{figure}

\begin{figure}[htbp]
\captionsetup{font=small}
\centering
\begin{minipage}[t]{0.22\textwidth}
\centering
\includegraphics[trim=5 5 2 5, clip, width=3.8cm, height=2.8cm]{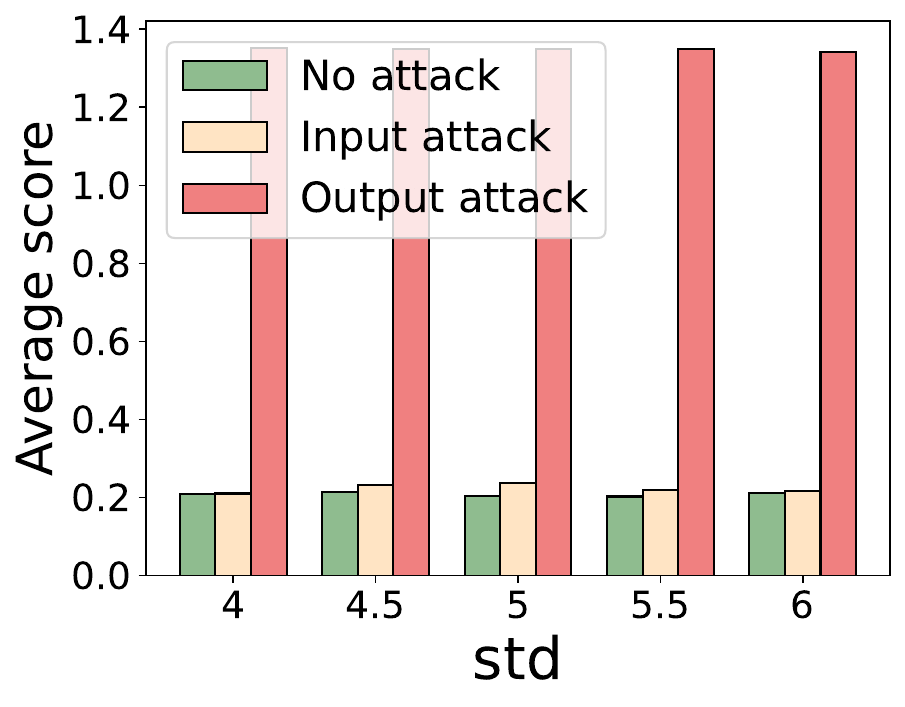}
\subcaption{ATP on CHI}
\end{minipage}
\begin{minipage}[t]{0.22\textwidth}
\centering
\includegraphics[trim=5 5 2 5, clip, width=3.8cm, height=2.8cm]{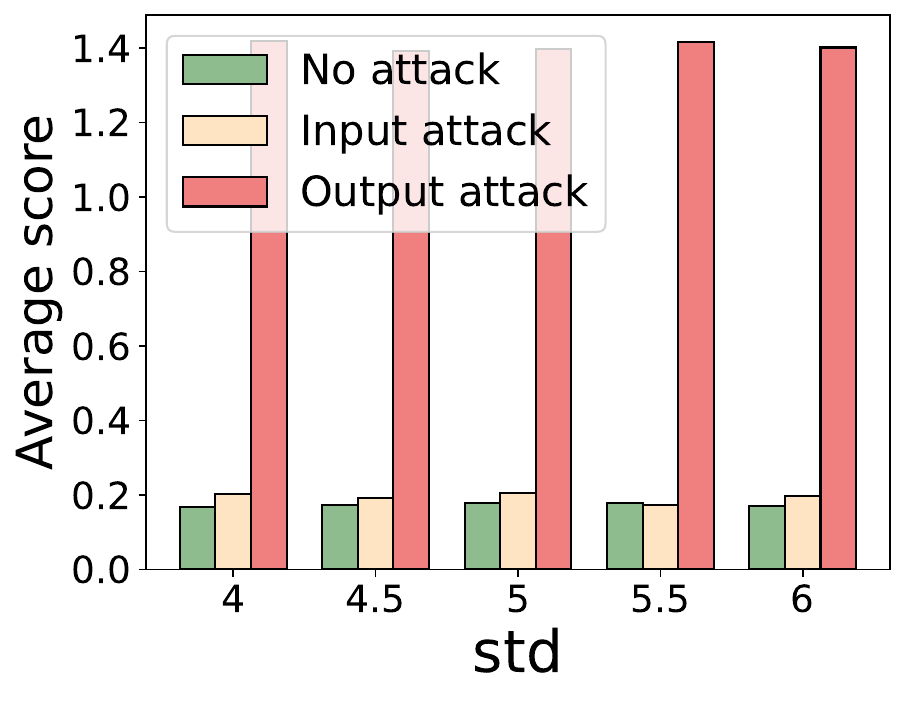}
\subcaption{ATP on CLE}
\end{minipage}
\begin{minipage}[t]{0.22\textwidth}
\centering
\includegraphics[trim=5 5 2 5, clip, width=3.8cm, height=2.8cm]{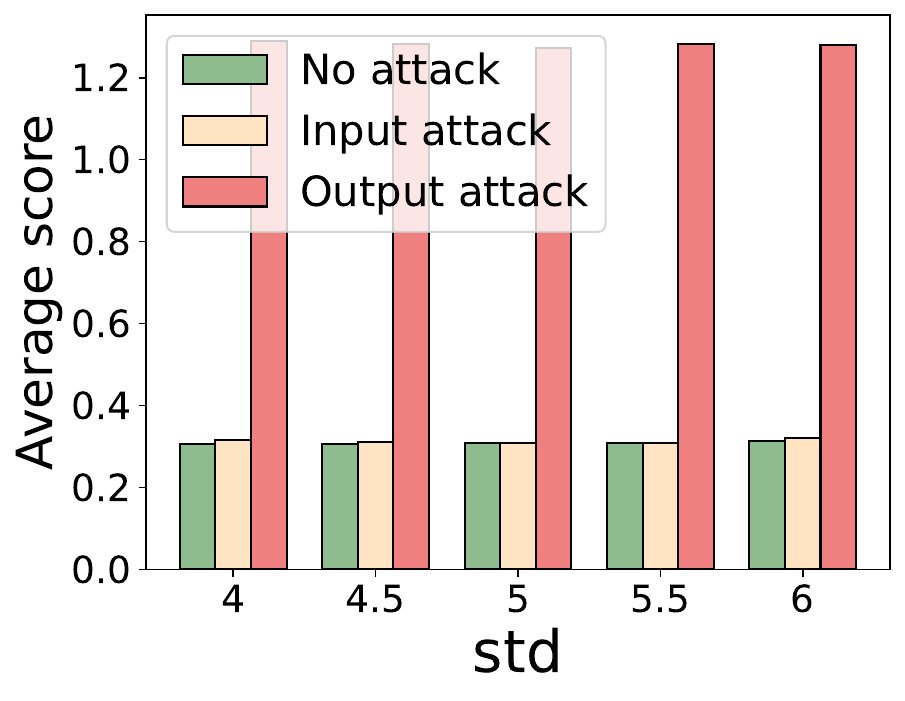}
\subcaption{ATP on CPS}
\end{minipage}
\begin{minipage}[t]{0.22\textwidth}
\centering
\includegraphics[trim=5 5 2 5, clip, width=3.8cm, height=2.8cm]{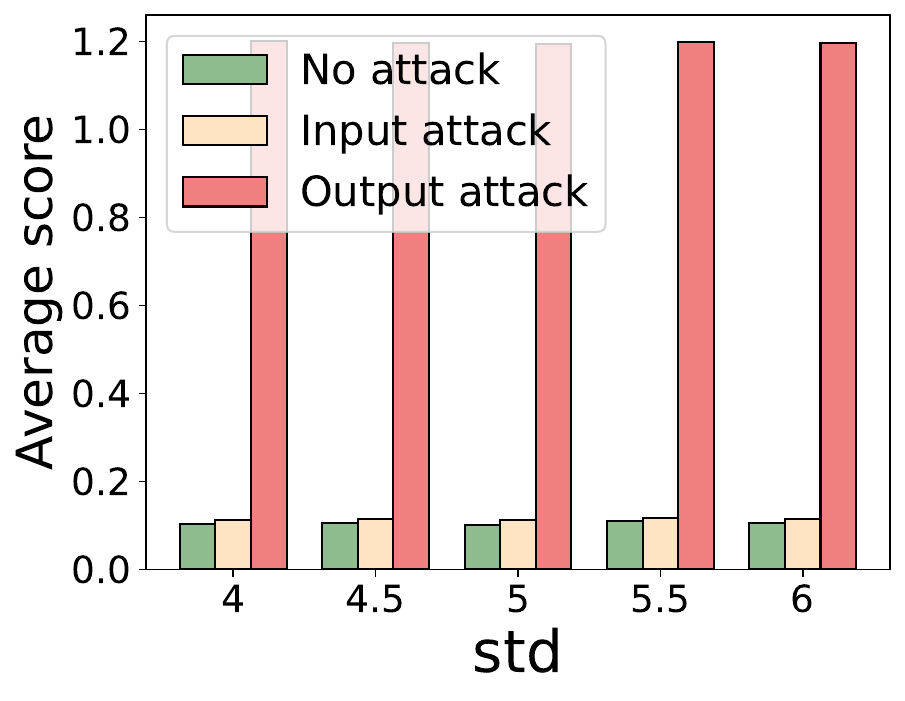}
\subcaption{ATP on NYC}
\end{minipage}
\vspace{-13pt}
\caption{Average scores of ATP with different stds.}
\label{ATP_score_with_stds}
\end{figure}

In Section~\ref{sec:exp_parameters}, we employ a Gaussian distribution to determine the length of each trajectory, using a mean of $(L_\text{min} + L_\text{max}) / 2$ and a standard deviation of $(L_\text{max} - L_\text{min}) / 5$. However, in real-world scenarios, the attacker may not have access to the precise values of $L_\text{min}$ and $L_\text{max}$ (e.g., the attacker might rely on public information or $k_{\text{min}}/k_{\text{max}}$ to estimate $L_\text{min}/L_\text{max}$), leading to inaccuracies in both the mean and standard deviation estimates. To simulate this uncertainty, we conduct two sets of experiments:

\begin{enumerate}
\setlength{\topsep}{0pt}
\setlength{\partopsep}{0pt}
\setlength{\itemsep}{5pt}
\setlength{\parsep}{5pt}
\setlength{\parskip}{-2pt}
\item[\small$\bullet$] We vary the mean by using the formula $mean = (L_\text{min} + L_\text{max}) / a$, where $a$ take values of 1.6, 1.8, 2, 2.2, and 2.4. The results are presented in Figures~\ref{n_gram_score_with_means}, ~\ref{PrivTC_score_with_means},~\ref{LDPTrace_score_with_means},~\ref{ATP_score_with_means}. 
\item[\small$\bullet$] We adjust the standard deviation using the formula $std = (L_\text{max} - L_\text{min}) / b$, where b ranged from 4 to 6 in increments of 0.5 (i.e., 4, 4.5, 5, 5.5, and 6). The results of these experiments are presented in Figures~\ref{n_gram_score_with_stds}, ~\ref{PrivTC_score_with_stds}, ~\ref{LDPTrace_score_with_stds},~\ref{ATP_score_with_stds}.
\end{enumerate}

Our findings demonstrate that even when there is a discrepancy between the attacker's assumed length distribution and the actual distribution, the attack remains effective. This resilience suggests that the attack method is robust to imperfect knowledge of the underlying trajectory length parameters.

\section{Countermeasures}
On the defense side, three efforts~\cite{ldpguard, tifs, mdai, ontherobustness} have been made to secure LDP frequency estimation protocols. 

In this section, we discuss two basic countermeasures. The first is frequent itemset mining, applicable to all four protocols. The second is normalization, applicable only to PrivTC and LDPTrace. Due to space constraints, we only display the results for average scores.

\subsection{Frequent Itemset Mining}
Frequent itemset mining (FIM)~\cite{a20b7577c86d4387add5d5e161d7c665} identifies sets of items that often co-occur in a dataset. If an itemset appears too frequently in a trajectory dataset, it is likely a target itemset and should be removed for more accurate estimates.

In our experiments, for $n$-gram and ATP, a point is ``frequent'' if its occurrence in the perturbed dataset exceeds 90 percent of other points. If a trajectory consists of more than 90\% frequent points, the server considers it fake and removes it from the dataset. For PrivTC and LDPTrace, a binary vector index of OUE (used for collecting the probability of points, pairs, or triplets) is ``frequent'' if its occurrence exceeds 90 percent of all indices. If a binary vector is composed of more than $90\%$ frequent indices, the server will consider it poisoned and remove it before aggregation.

\begin{figure}[htbp]
\captionsetup{font=small}
\centering
\begin{minipage}[t]{0.23\textwidth}
\centering
\includegraphics[trim=5 13 2 5, clip, width=4cm]{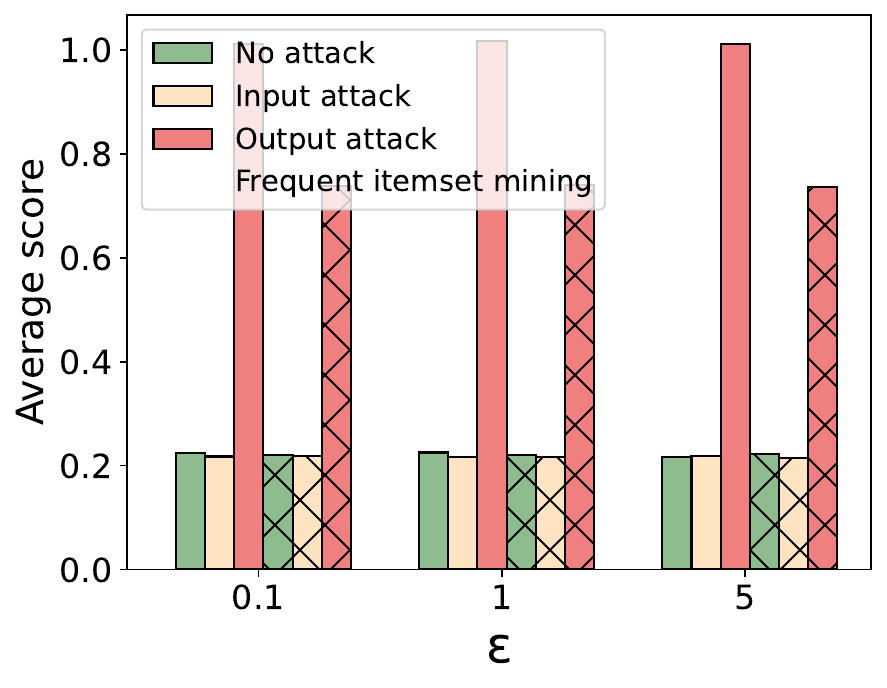}
\subcaption{$n$-gram on CPS}
\end{minipage}
\begin{minipage}[t]{0.23\textwidth}
\centering
\includegraphics[trim=5 13 2 5, clip, width=4cm]{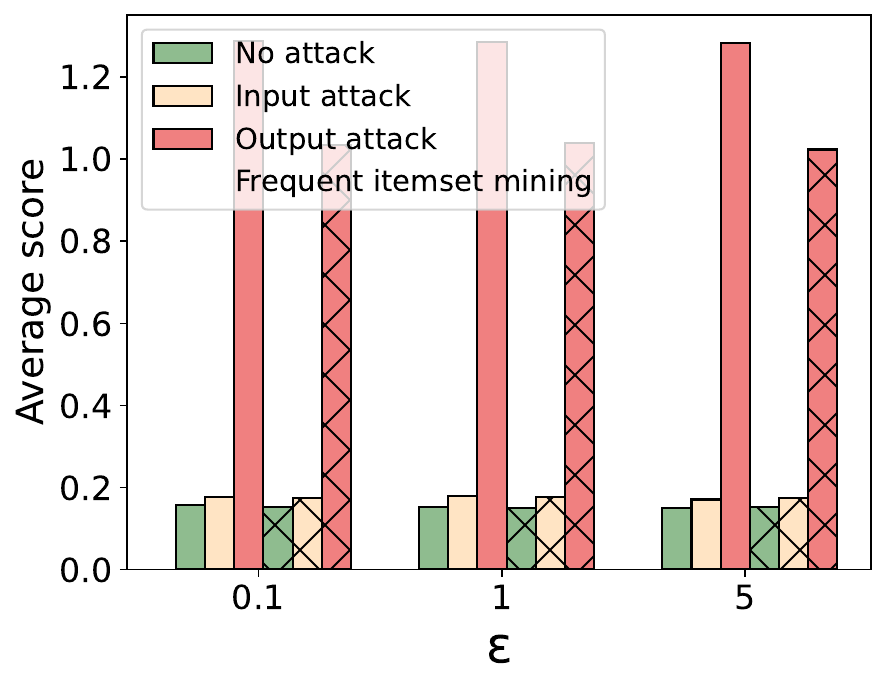}
\subcaption{$n$-gram on NYC}
\end{minipage}
\vspace{-13pt}
\caption{Average scores of $n$-gram with defenses.}
\label{n_gram_with_defense}
\end{figure}

\begin{figure}[htbp]
\captionsetup{font=small}
\centering
\begin{minipage}[t]{0.23\textwidth}
\centering
\includegraphics[trim=5 13 2 5, clip, width=4cm]{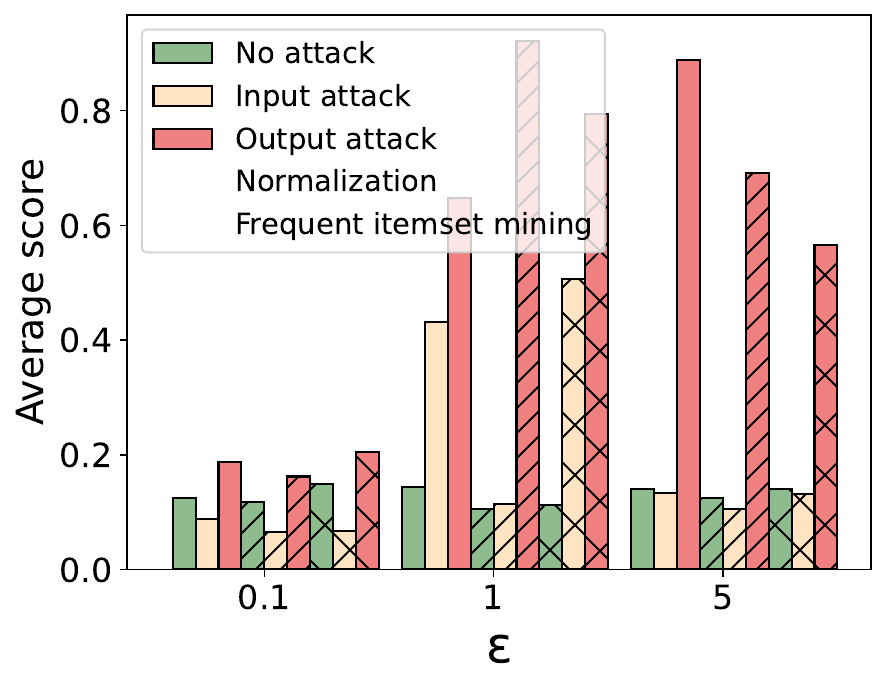}
\subcaption{PrivTC on Gowalla}
\label{PrivTC_with_defense_Gowalla}
\end{minipage}
\begin{minipage}[t]{0.23\textwidth}
\centering
\includegraphics[trim=5 13 2 5, clip, width=4cm]{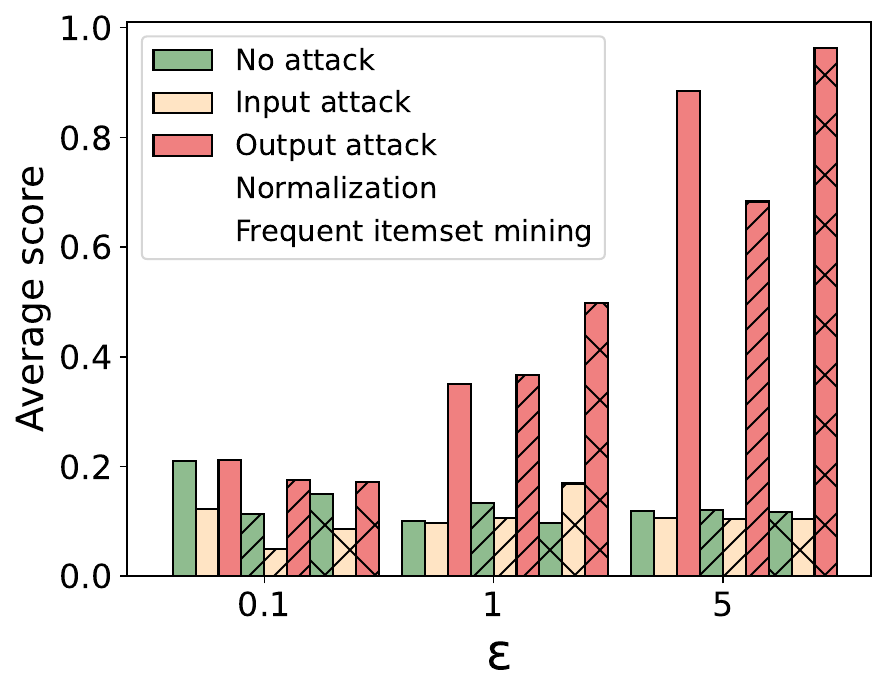}
\subcaption{PrivTC on Porto}
\label{PrivTC_with_defense_Porto}
\end{minipage}
\vspace{-13pt}
\caption{Average scores of PrivTC with defenses.}
\label{PrivTC_with_defense}
\end{figure}

\begin{figure}[htbp]
\captionsetup{font=small}
\centering
\begin{minipage}[t]{0.22\textwidth}
\centering
\includegraphics[trim=5 13 2 5, clip, width=3.8cm, height=2.8cm]{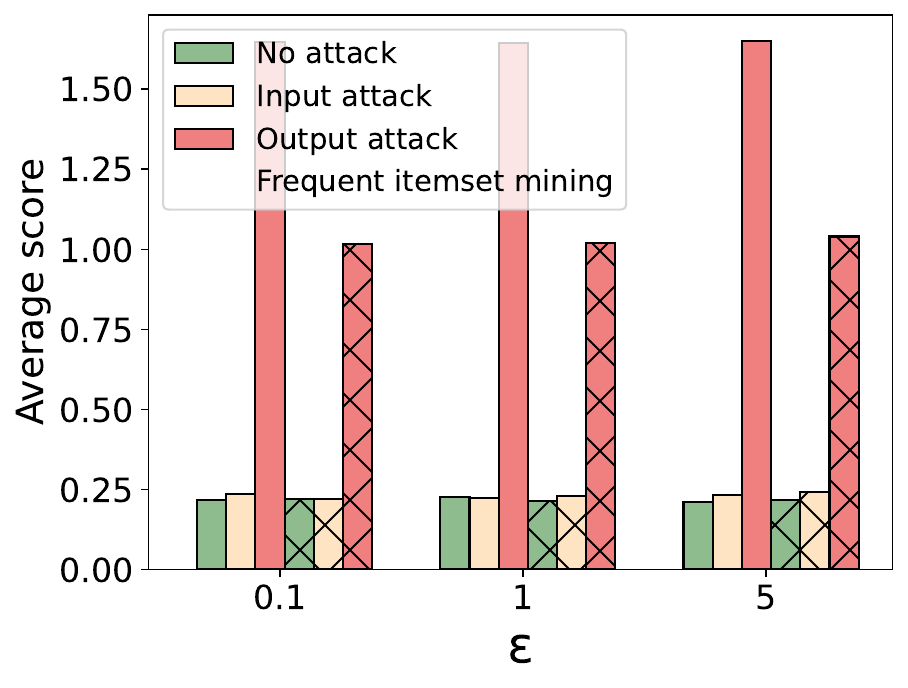}
\subcaption{ATP on CHI}
\end{minipage}
\begin{minipage}[t]{0.22\textwidth}
\centering
\includegraphics[trim=5 13 2 5, clip, width=3.8cm, height=2.8cm]{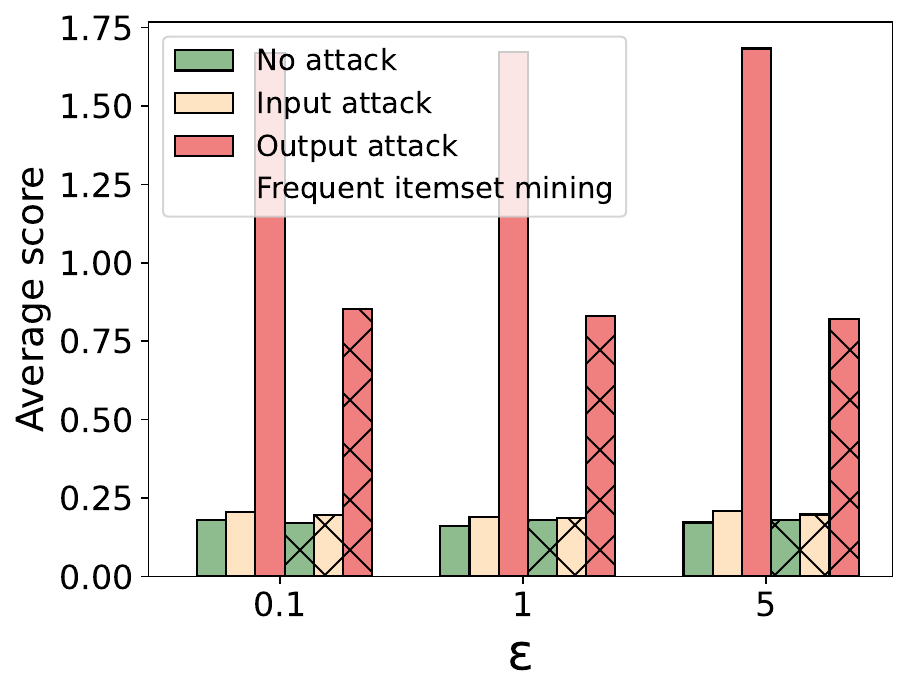}
\subcaption{ATP on CLE}
\end{minipage}
\begin{minipage}[t]{0.22\textwidth}
\centering
\includegraphics[trim=5 13 2 5, clip, width=3.8cm, height=2.8cm]{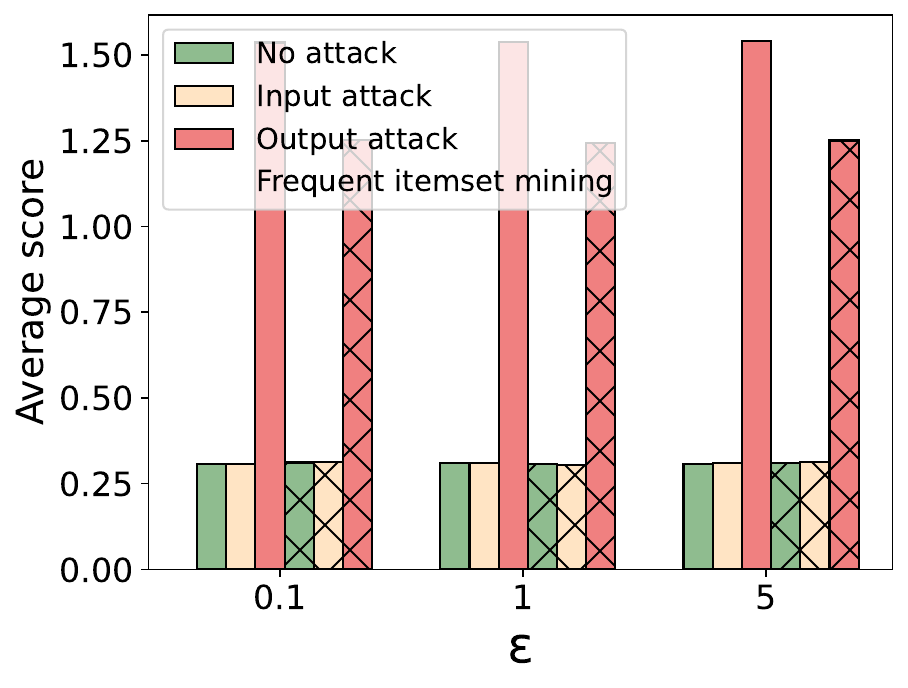}
\subcaption{ATP on CPS}
\end{minipage}
\begin{minipage}[t]{0.22\textwidth}
\centering
\includegraphics[trim=5 13 2 5, clip, width=3.8cm, height=2.8cm]{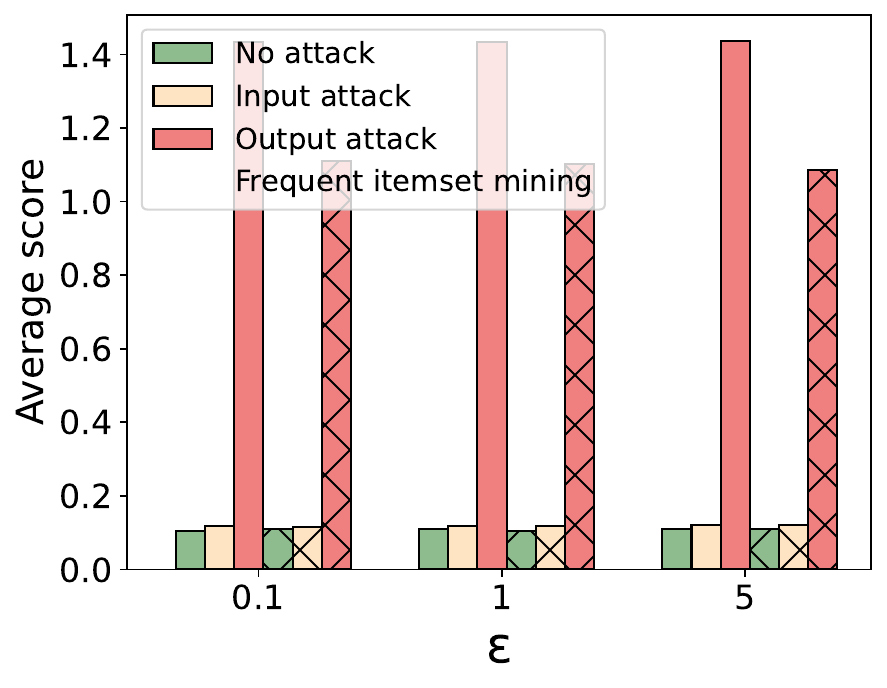}
\subcaption{ATP on NYC}
\end{minipage}
\vspace{-13pt}
\caption{Average scores of ATP with defenses.}
\label{ATP_with_defense}
\end{figure}

\begin{figure}[htbp]
\captionsetup{font=small}
\centering
\begin{minipage}[t]{0.15\textwidth}
\centering
\includegraphics[trim=0 13 0 2, clip, width=2.8cm]{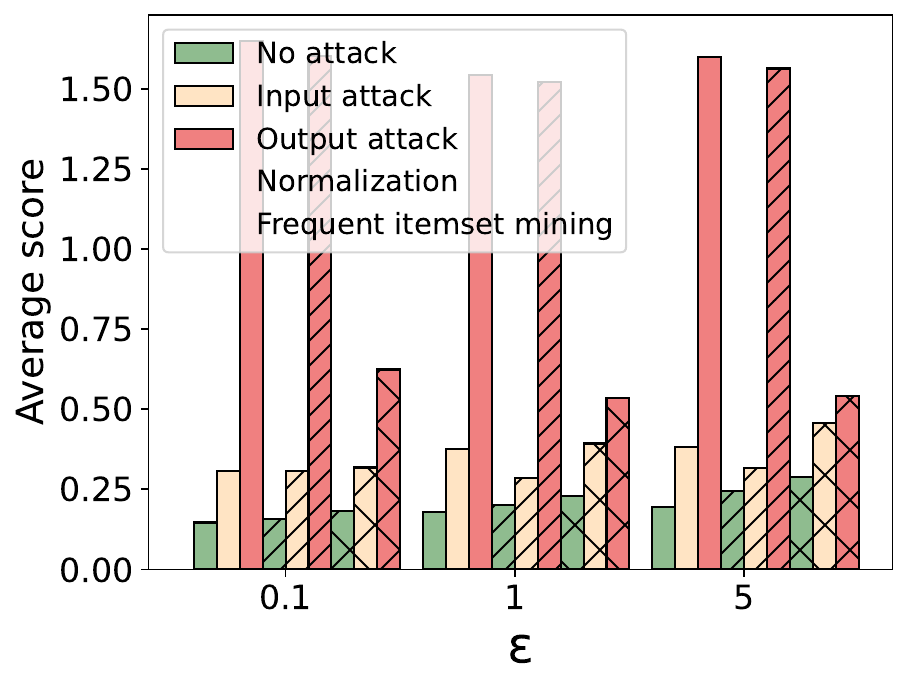}
\subcaption{\scriptsize LDPTrace on CPS}
\label{LDPTrace_with_defense_CPS}
\end{minipage}
\begin{minipage}[t]{0.15\textwidth}
\centering
\includegraphics[trim=0 13 0 2, clip, width=2.8cm]{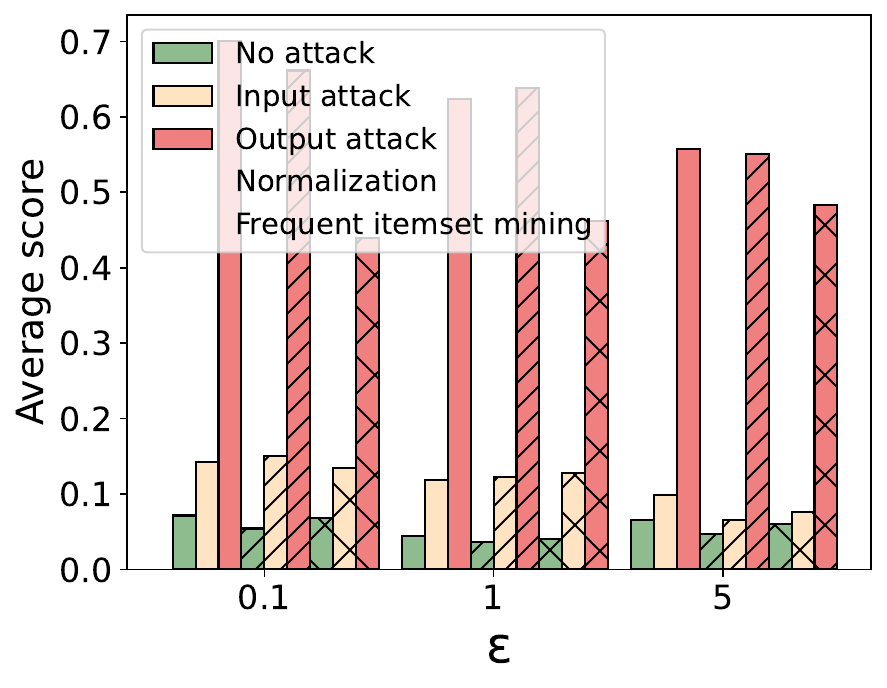}
\subcaption{\scriptsize LDPTrace on Oldenburg}
\end{minipage}
\begin{minipage}[t]{0.15\textwidth}
\centering
\includegraphics[trim=0 13 0 2, clip, width=2.8cm]{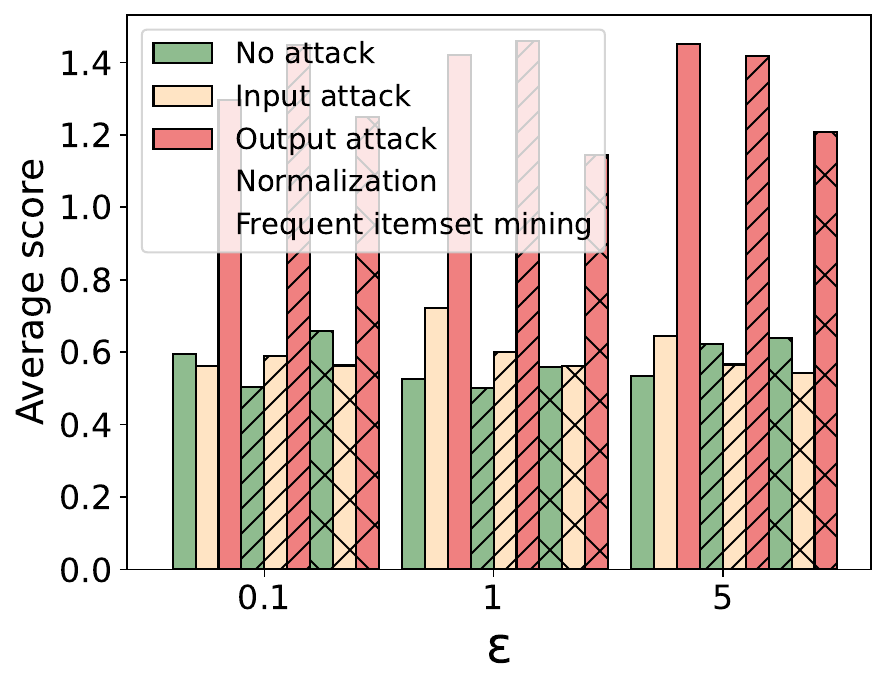}
\subcaption{\scriptsize LDPTrace on Porto}
\label{LDPTrace_with_defense_Porto}
\end{minipage}
\vspace{-13pt}
\caption{Average scores of LDPTrace with defenses.}
\label{LDPTrace_with_defense}
\end{figure}

\begin{figure}[htbp]
\captionsetup{font=small}
\centering
\begin{minipage}[t]{0.15\textwidth}
\centering
\includegraphics[trim=18 8 10 15, clip, width=3cm]{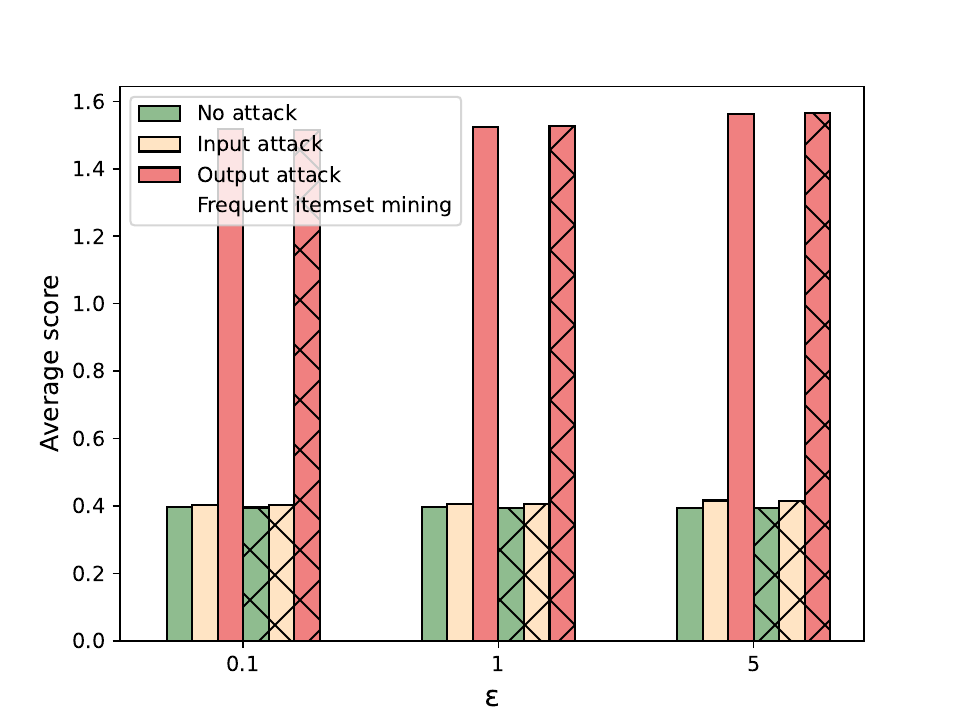}
\subcaption{$\text{RetraSyn}^\text{b}$ with defense on T-Drive}
\end{minipage}
\begin{minipage}[t]{0.15\textwidth}
\centering
\includegraphics[trim=18 8 10 15, clip, width=3cm]{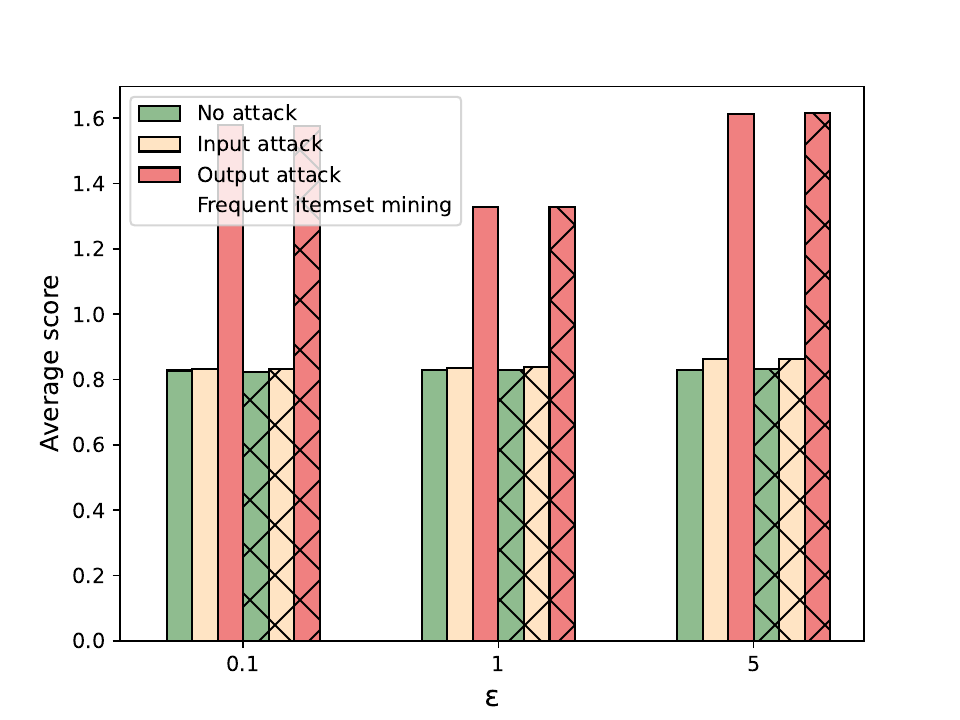}
\subcaption{$\text{RetraSyn}^\text{b}$ with defense on Oldenburg}
\end{minipage}
\begin{minipage}[t]{0.15\textwidth}
\centering
\includegraphics[trim=18 8 10 15, clip, width=3cm]{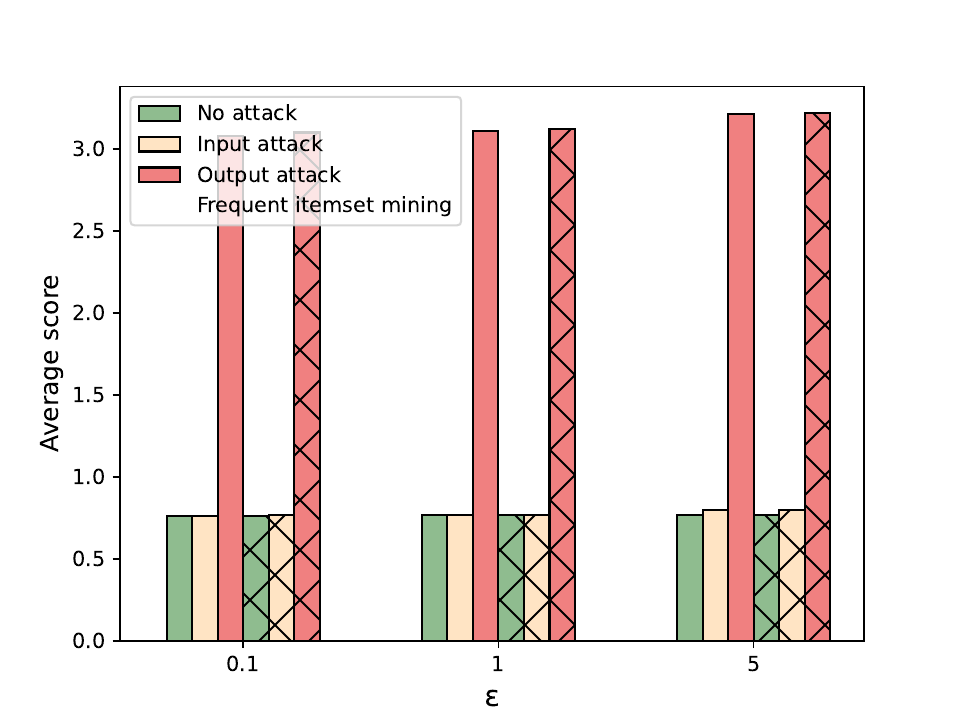}
\subcaption{$\text{RetraSyn}^\text{b}$ with defense on SanJoaquin}
\end{minipage}
\vspace{-13pt}
\caption{Average scores of $\text{RetraSyn}^\text{b}$ on T-Drive, Oldenburg and SanJoaquin datasets with defenses.}
\label{RetraSyn_b_with_defense}
\end{figure}

As illustrated in Figures~\ref{n_gram_with_defense},~\ref{PrivTC_with_defense}, ~\ref{ATP_with_defense}, ~\ref{LDPTrace_with_defense}, the grid-patterned bars represent the results of incorporating FIM on the server side. Generally, for all of the four protocols, this defense can slightly diminish both attacks' effectiveness. However, all the OPAs and some IPAs still show a significant impact.

Surprisingly, in some cases, the use of FIM  can, in turn, improve the attack performance. For example, the IPA and OPA attacks with $\varepsilon=1$ in Figure~\ref{PrivTC_with_defense_Gowalla} and Figure~\ref{PrivTC_with_defense_Porto}, as well as the OPA attack with $\varepsilon=5$ in Figure~\ref{PrivTC_with_defense_Porto}, all show higher scores when the FIM defense is applied. Additionally, even without any attack, applying frequent itemset mining can slightly increase the average scores, as demonstrated in Figure~\ref{LDPTrace_with_defense_CPS} and Figure~\ref{LDPTrace_with_defense_Porto} for the case of $\varepsilon=5$. This side effect likely occurs when the real dataset is heavily concentrated on certain non-target pairs, points, or triplets. The majority of points, pairs or triplets in the real data (non-targets) are mistakenly identified as fake and removed, while the minority of points, pairs or triplets in the real data or fake data (targets) remain, increasing the likelihood of the target pattern appearing.

These results demonstrate that FIM is not a perfect defense strategy and that more robust measures are required to withstand \textsc{TraP}.

\subsection{Normalization}
Since PrivTC and LDPTrace utilize OUE to estimate frequencies of points, pairs, and triplets, the estimated results may not always form a proper probability distribution. Some item probabilities might be negative, others might exceed 1, and the total sum might not equal 1. Normalization can address these anomalies, making it a potential method for mitigating attacks in PrivTC and LDPTrace.

In particular, normalization identifies the smallest item probability, $\text{prob}_{\min}$, subtracting $\text{prob}_{\min}$ from all item probabilities, and then dividing all probabilities by their adjusted total sum. The results of counteracting \textsc{TraP} by using normalization in PrivTC and LDPTrace are shown in the striped bars of Figure~\ref{PrivTC_with_defense} and Figure~\ref{LDPTrace_with_defense}, respectively. Essentially, the normalization defense does not significantly mitigate the attacks and, in some cases, even enhances the effectiveness of both IPA and OPA for reasons previously discussed.

Overall, both frequent itemset mining and normalization fail to effectively defend against our attacks, highlighting the need for more robust defense mechanisms. 

\section{Conclusion}
This paper presents the first study on data poisoning attacks against LDP protocols for trajectory data. Our \textsc{TraP} algorithm, which adopts a prefix-suffix approach, significantly reduces the time complexity, making it feasible to efficiently generate attack-effective fake trajectories to promote the attacker-specified target patterns. Additionally, we evaluate basic defense strategies, demonstrating their suboptimal effectiveness and highlighting the urgent need for more advanced defenses. This work not only exposes potential risks in LDP trajectory protocols but also lays a foundation for future research to develop more secure and resilient LDP trajectory protocols.

\clearpage
\newpage
\section{Ethics Considerations}
This research adheres to the ethical guidelines outlined by USENIX Security '25 and carefully considers the potential risks and impacts of our work. The study focuses on analyzing vulnerabilities in locally differentially private (LDP) trajectory protocols and proposing a heuristic algorithm (TRAP) for data poisoning attacks. While our work exposes weaknesses in existing protocols, it is conducted with the intent of improving the security and robustness of LDP systems. We ensured that no real-world systems or user data were harmed during the research. All experiments were conducted in a controlled environment using synthetic datasets, and no unauthorized access or violation of terms of service occurred. Additionally, we disclose the potential risks of our findings, emphasizing the need for stronger defenses and better protocol designs to mitigate malicious exploitation. By highlighting these vulnerabilities, we aim to contribute to the development of more secure and privacy-preserving systems, aligning with the principles of beneficence, respect for persons, justice, and respect for law and public interest.

\section{Open Science}
In alignment with the principles of open science and the guidelines set forth by USENIX Security 2025, we have made efforts to ensure the reproducibility and transparency of our research. All datasets, code, and experimental configurations used in this study are publicly available [provide repository link or specify availability]. Where data sharing is restricted due to privacy concerns, we have provided detailed descriptions of the methodologies and synthetic data to facilitate reproducibility. This commitment reflects our dedication to advancing collaborative research and fostering trust within the scientific community.
\clearpage
\newpage
\bibliographystyle{plain}
\bibliography{sample-base}

\clearpage
\newpage
\appendix
\SetKw{Continue}{Continue}
\SetKw{Break}{Break}

\section{Appendix}
\label{sec: Appendix}

\subsection{Notation Table}\label{sec: Notation Table}
The notation table is shown in Table~\ref{table1}.

\begin{table}[htbp]
    \setlength{\belowcaptionskip}{3px}
    \centering
    \caption{Notation}
    \label{table1}
    \begin{tabularx}
            {\columnwidth}{|p{2cm}|X|}
            \hline
			\textbf{Notation} & \textbf{Meaning} \\
			\hline
	        $n$, $m$ & Number of real users and fake users \\
	        \hline
	        $p$, $\mathcal{P}$ & Point, and point domain \\
	        \hline
	        $|\cdot|$ & Size of $\cdot \kern0.5pt$, e.g., $|\mathcal{P}|$ is the size of point domain; $|\tau|$ is the length of trajectory $\tau$\\
	        \hline
	        $\mathcal{T}^*$ & The set of fake trajectories $\{\tau^*_1, \dots, \tau^*_m\}$, and  $\mathcal{T}_{L=i}^{*}$ indicates those with length $i$\\
			\hline
	        $\mathcal{T}$ & The set of poisoned trajectories $\{\tau_1, \dots, \tau_n\}\cup\{\tau^*_1, \dots, \tau^*_m\}$\\
			\hline
	        $\textit{tp}$, $\textit{TP}$ & Target pattern, and set of target patterns \\
	        \hline
	        $k_{\min}$, $k_{\max}$ & Minimum and maximum target pattern lengths \\
	        \hline
	        $\text{score}(\textit{tp})$ & Score of target pattern \textit{tp} \\
	        \hline
	        $L_{\min}$, $L_{\max}$ & Minimum and maximum real trajectory lengths \\
	        \hline
	        $m_{L=i}$ & Number of fake trajectories with length $L$ equal to $i$; $\sum_{i=L_{\text{min}}}^{L_{\text{max}}} m_{L=i} = m$ \\
	        \hline
	        $\textit{max}_{\text{rep}}$ & Maximum permissible trajectory repetitions \\
            \hline
            $SC$ & The dictionary used to record trajectories with their corresponding scores\\
            \hline
            $\Omega, \Omega_{i},\Omega_{i}^{sorted}$ & The set includes trajectories awaiting check/pick, with the subscript $i$ indicating the $i$-th round, and the superscript $sorted$ indicating that the trajectories are sorted\\
            \hline
            $rps$ & The dictionary used to record the list of reachable points for each position \\
            \hline
            $PREF$ & The prefix categories generated by taking the longest proper prefix of each target pattern in $TP$ \\
            \hline
            $\mathcal{M}_i$ & The dictionary to record trajectories in $\Omega$ with their corresponding prefix categories \\
            \hline
            $\mathcal{M}_{selected}$ & The dictionary used to record those trajectories are selected with their corresponding prefix categories\\
            \hline
            $ancestor$ & The ancestor of the prefix $u_\tau$ refers to the categories in PREF that have $u_\tau$ as their suffix \\
            \hline
    \end{tabularx}
\end{table}

\subsection{More Experimental Results}\label{sec: More Experimental Results}
More experimental results will be placed here.
\begin{figure}[htbp]
\captionsetup{font=small}
\centering
\includegraphics[width=0.5\textwidth]{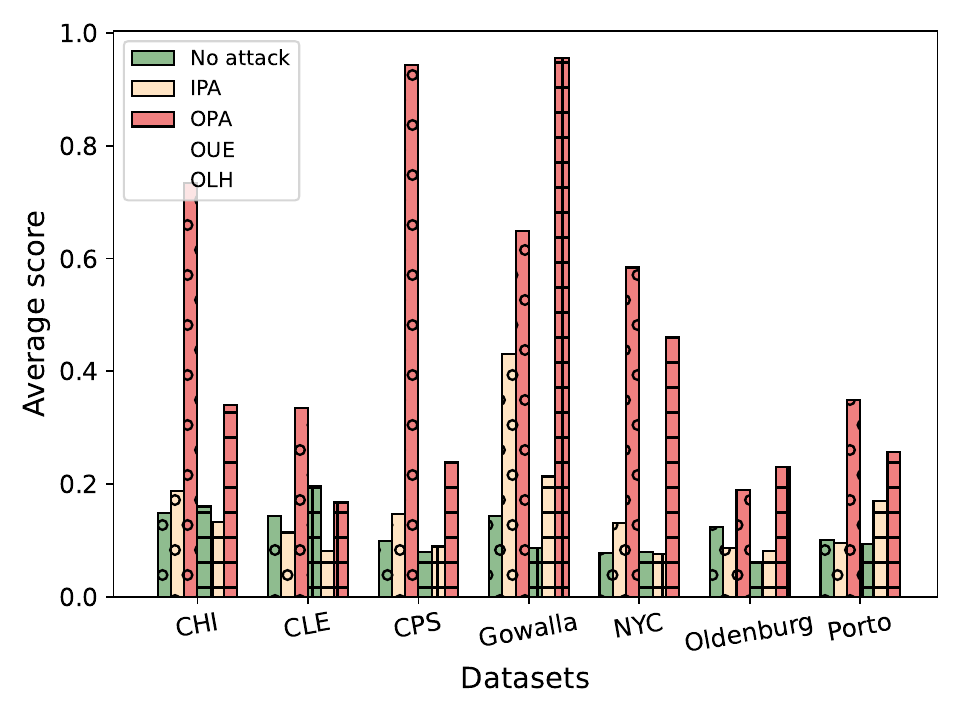}
\vspace{-12pt}
\caption{Average scores of PrivTC (OUE vs. OLH)}
\label{Fig:PrivTC_OUE_OLH_AvgScore}
\end{figure}

\begin{figure}[htbp]
\captionsetup{font=small}
\centering
\includegraphics[width=0.5\textwidth]{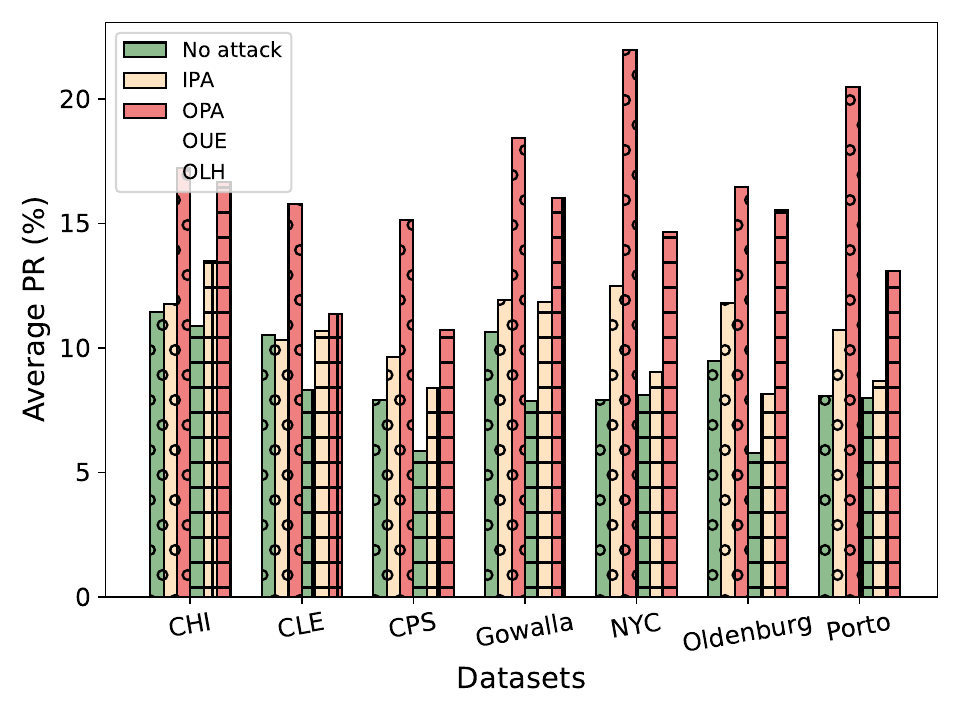}
\vspace{-12pt}
\caption{Average PRs of PrivTC (OUE vs. OLH)}
\label{Fig:PrivTC_OUE_OLH_AvgPR}
\end{figure}

\begin{figure}[htbp]
\captionsetup{font=small}
\centering
\includegraphics[width=0.5\textwidth]{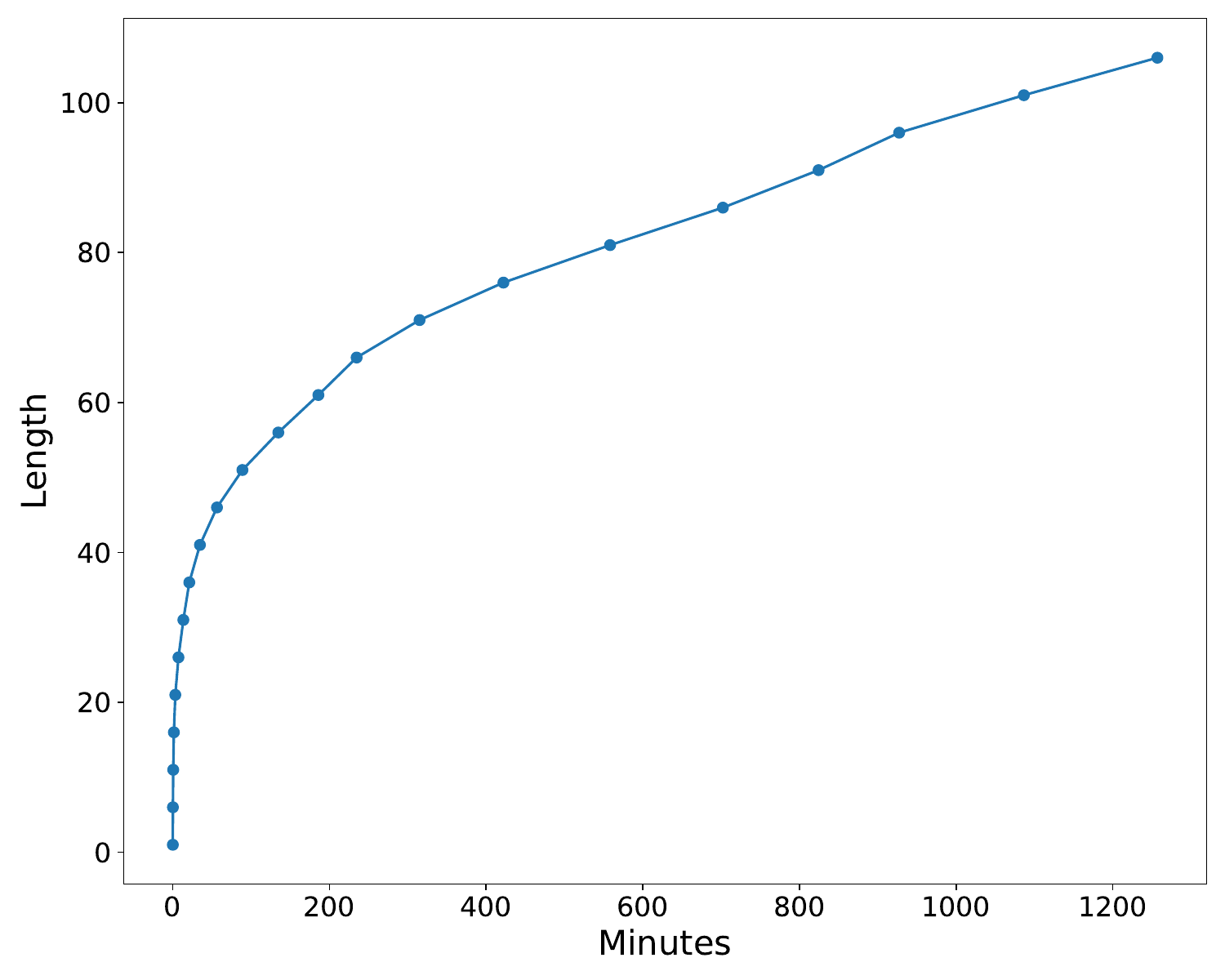}
\caption{Trajectory lengths that can be generated by prefix-suffix approach over 24 hours on Porto dataset}
\label{Fig:time-length}
\end{figure}

\begin{figure}[htbp]
\captionsetup{font=small}
\centering
\includegraphics[width=9cm]{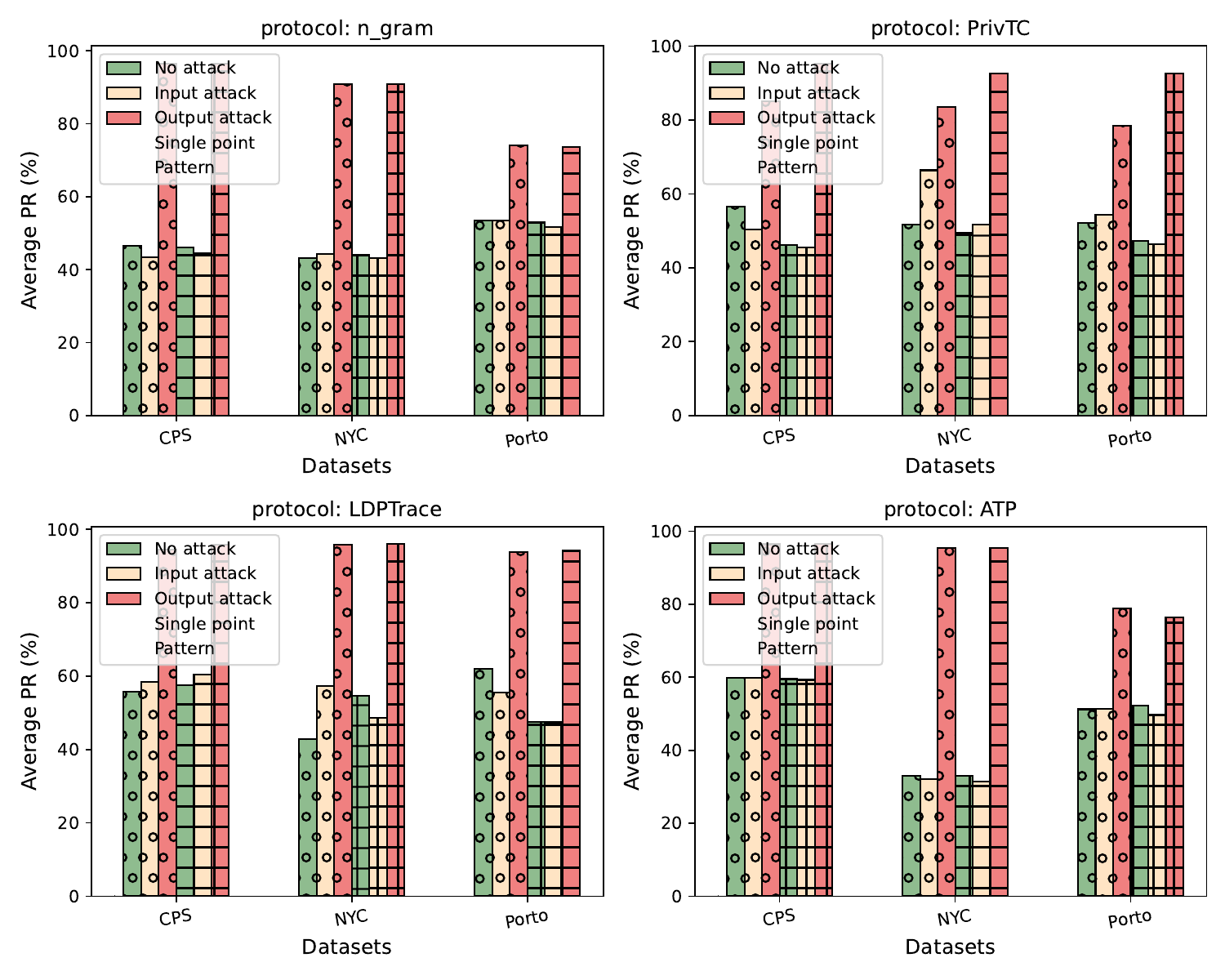}
\vspace{-13pt}
\caption{Average percentile rank (PR) of single-point-based attack and pattern-based attack across with $k_{\text{max}} = 1$.}
\label{Fig:Single_point_vs_Pattern_PR_kmax1}
\end{figure}

\begin{figure}[htbp]
\captionsetup{font=small}
\centering
\includegraphics[width=9cm]{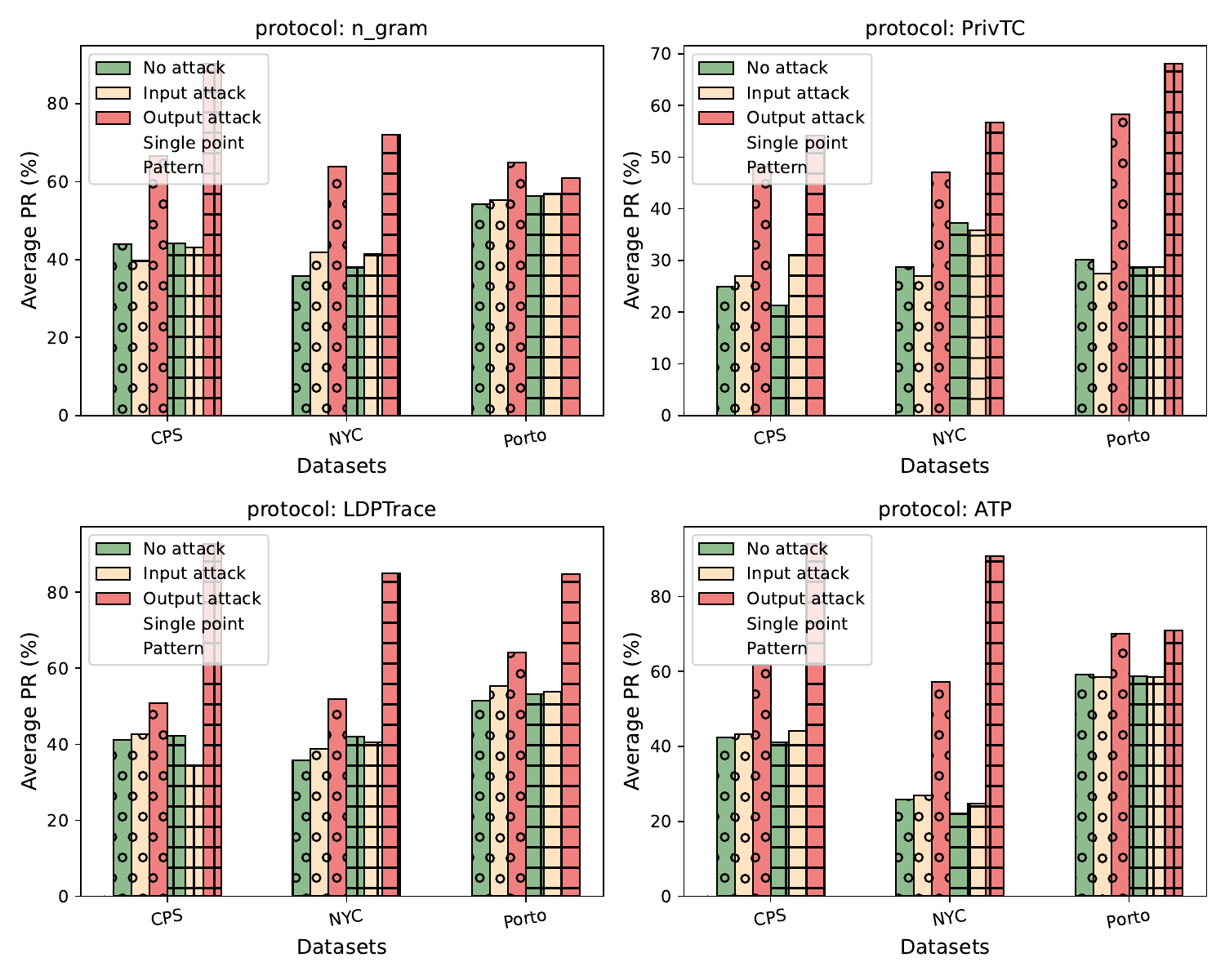}
\vspace{-13pt}
\caption{Average percentile rank (PR) of single-point-based attack and pattern-based attack across with $k_{\text{max}} = 2$.}
\label{Fig:Single_point_vs_Pattern_PR_kmax2}
\end{figure}

\begin{figure}[htbp]
\captionsetup{font=small}
\centering
\includegraphics[width=9cm]{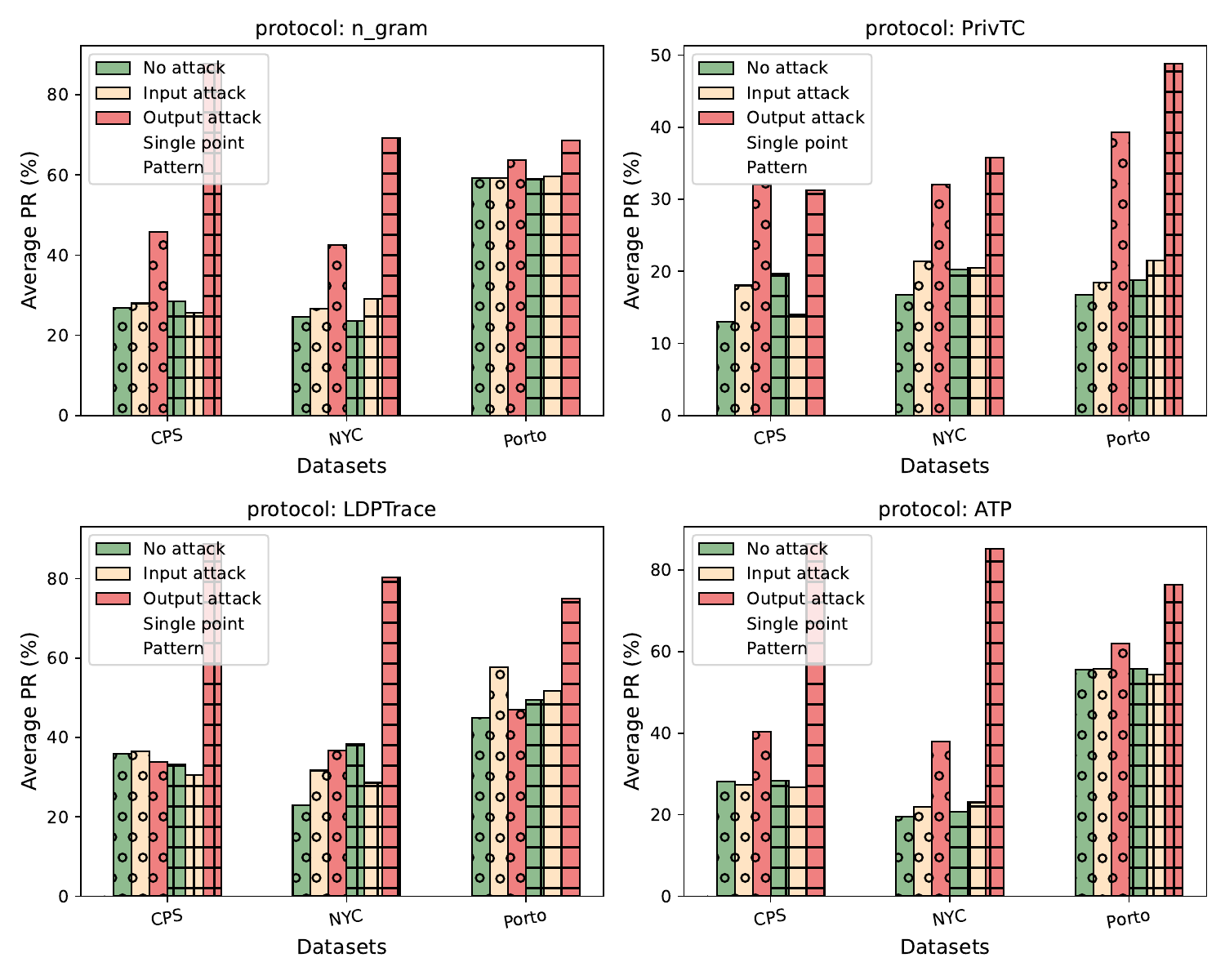}
\vspace{-13pt}
\caption{Average percentile rank (PR) of single-point-based attack and pattern-based attack across with $k_{\text{max}} = 3$.}
\label{Fig:Single_point_vs_Pattern_PR_kmax3}
\end{figure}

\begin{figure}[htbp]
\captionsetup{font=small}
\centering
\includegraphics[width=9cm]{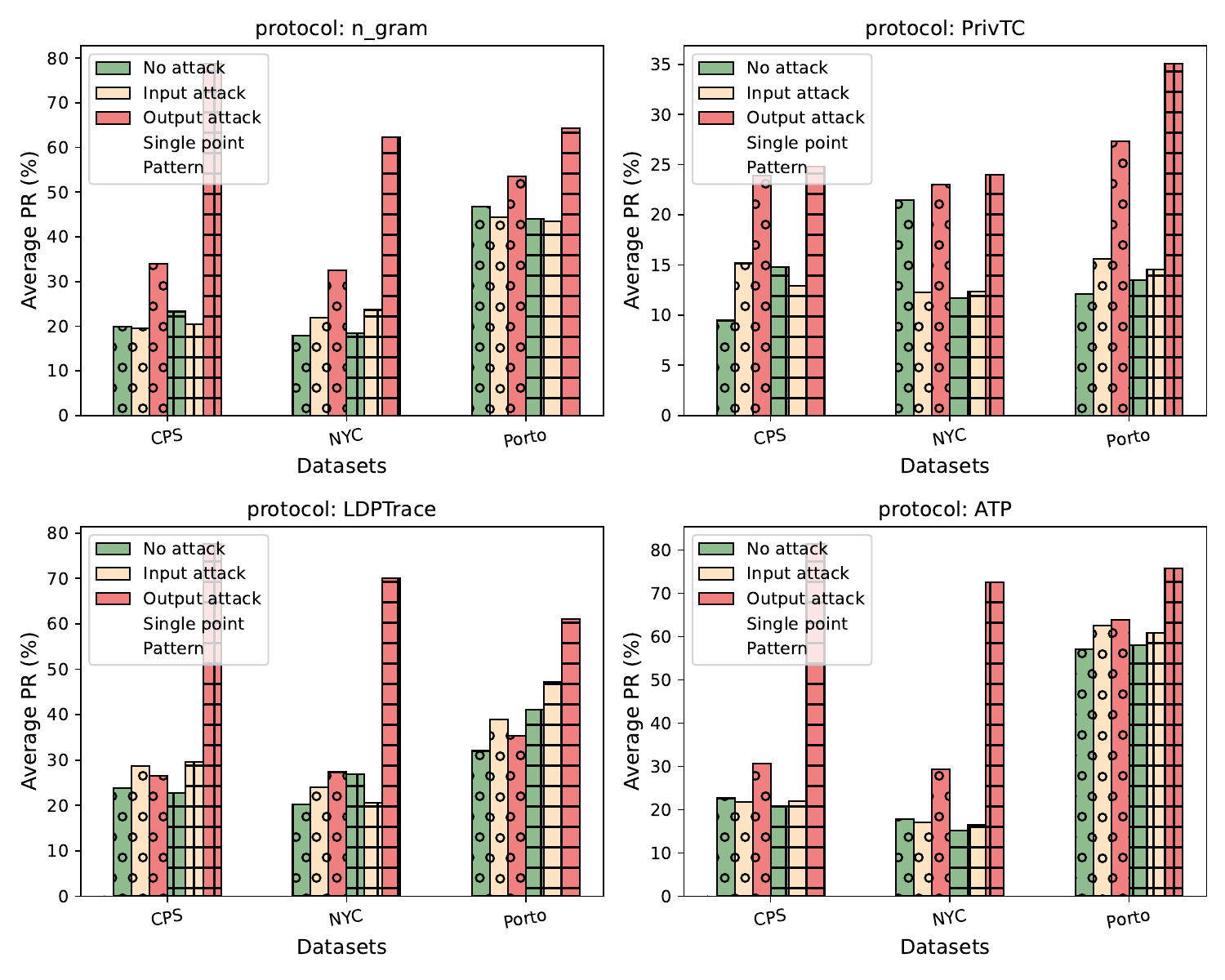}
\vspace{-13pt}
\caption{Average percentile rank (PR) of single-point-based attack and pattern-based attack across with $k_{\text{max}} = 4$.}
\label{Fig:Single_point_vs_Pattern_PR_kmax4}
\end{figure}

\begin{figure}[htbp]
\captionsetup{font=small}
\centering
\includegraphics[width=9cm]{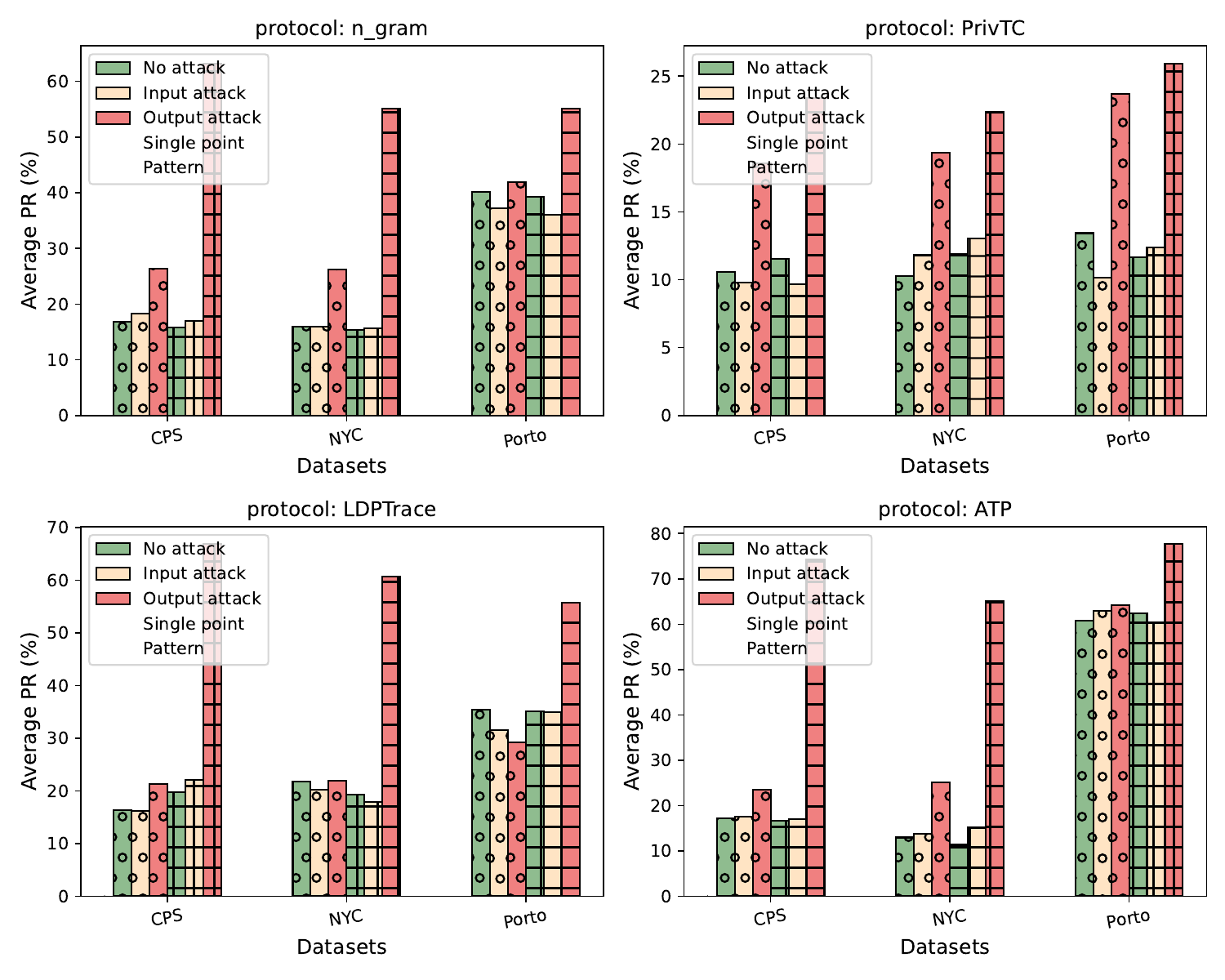}
\vspace{-13pt}
\caption{Average percentile rank (PR) of single-point-based attack and pattern-based attack across with $k_{\text{max}} = 5$.}
\label{Fig:Single_point_vs_Pattern_PR_kmax5}
\end{figure}

\begin{figure}[htbp]
\captionsetup{font=small}
\centering
\includegraphics[width=9cm]{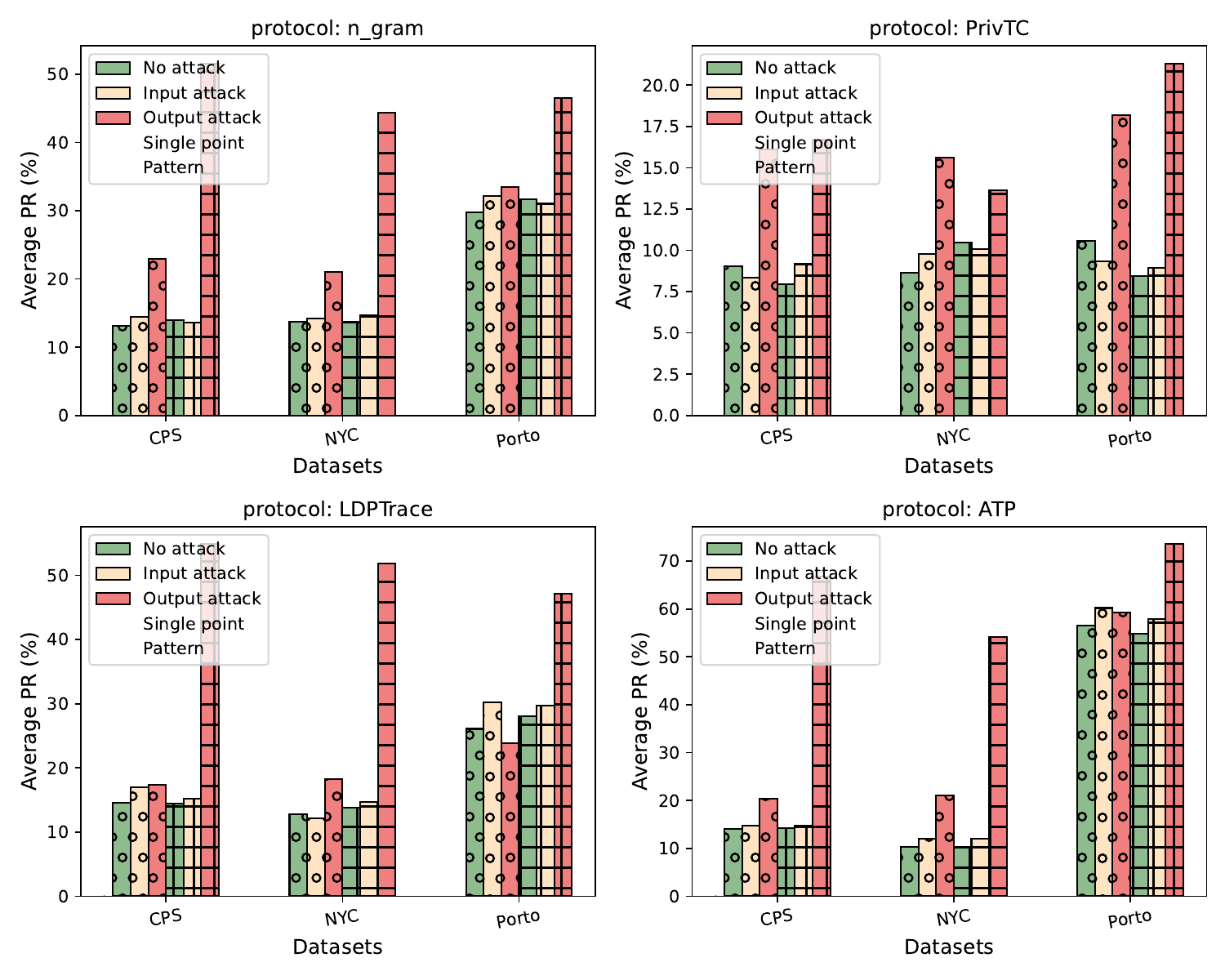}
\vspace{-13pt}
\caption{Average percentile rank (PR) of single-point-based attack and pattern-based attack across with $k_{\text{max}} = 6$.}
\label{Fig:Single_point_vs_Pattern_PR_kmax6}
\end{figure}

\begin{figure}[htbp]
\captionsetup{font=small}
\centering
\includegraphics[width=8.5cm]{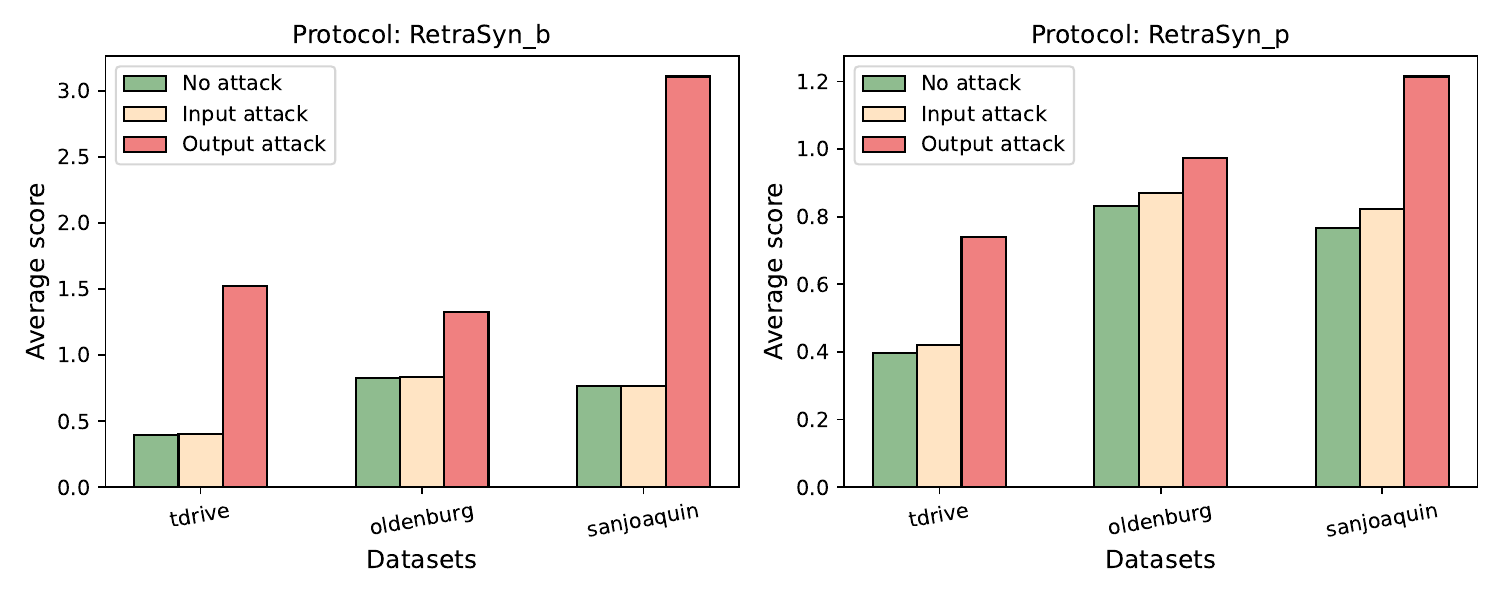}
\vspace{-13pt}
\caption{Average scores of RetraSyn with $\varepsilon = 1$.}
\label{Fig:RetraSyn_AvgScore}
\end{figure}

\begin{figure}[htbp]
\captionsetup{font=small}
\centering
\includegraphics[width=8.5cm]{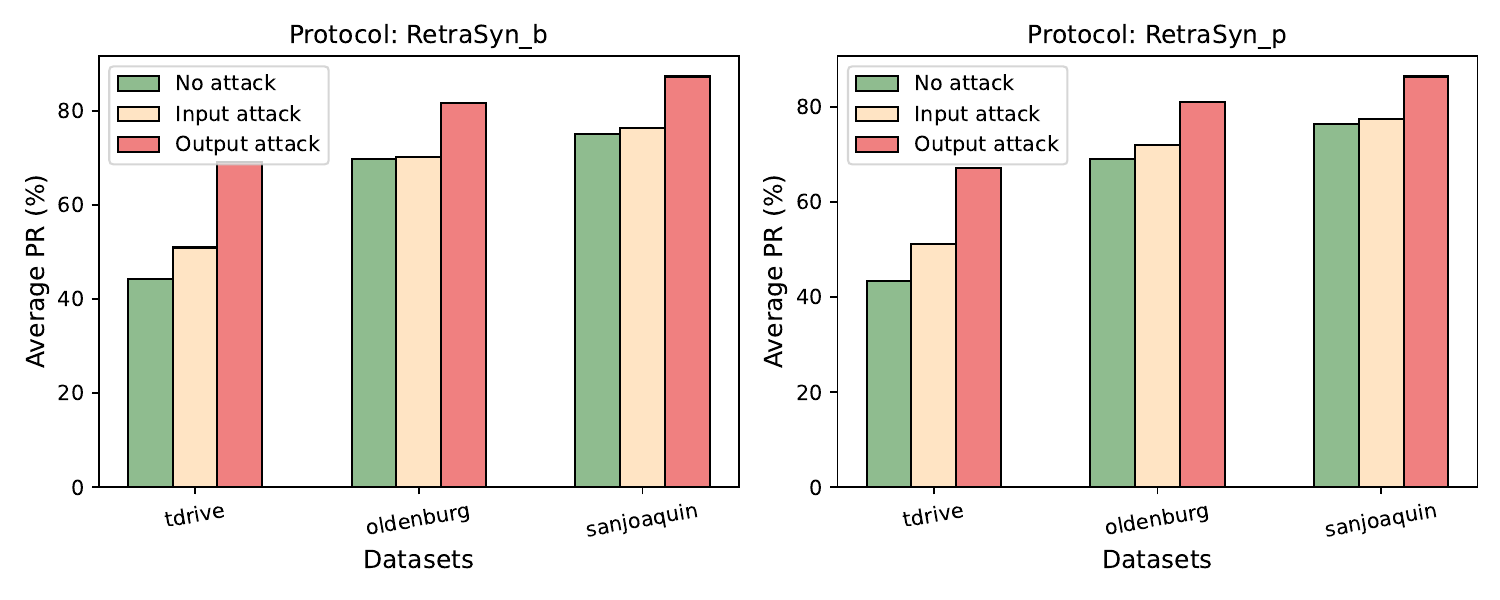}
\vspace{-13pt}
\caption{Average PRs of RetraSyn with $\varepsilon = 1$.}
\label{Fig:RetraSyn_AvgPR}
\end{figure}

\begin{figure}[htbp]
\captionsetup{font=small}
\centering
\begin{minipage}[t]{0.45\textwidth}
\centering
\includegraphics[width=8.5cm]{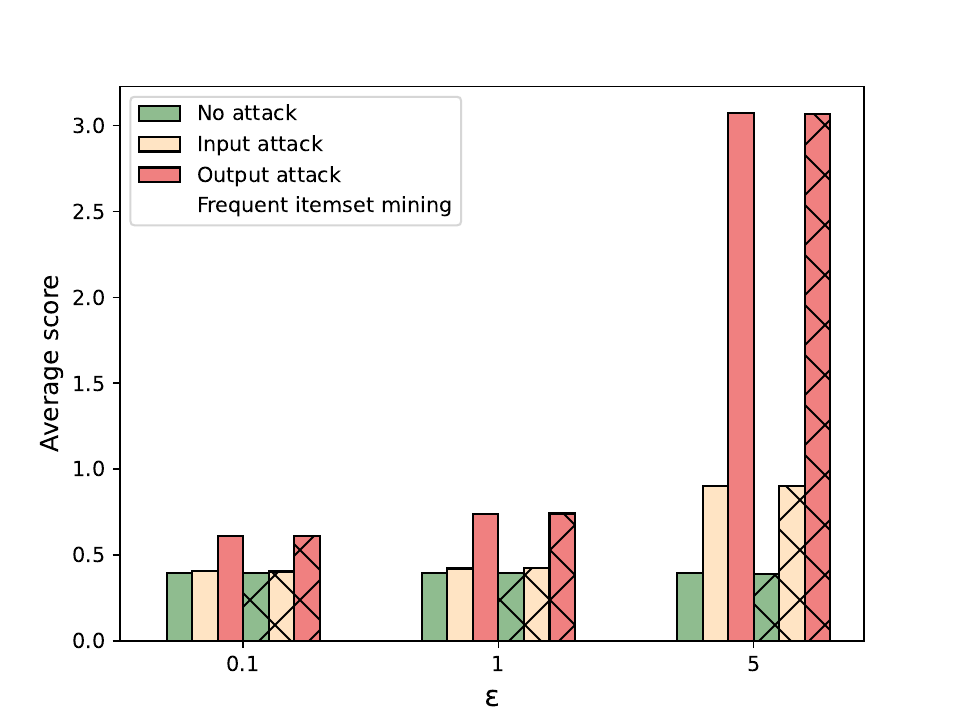}
\subcaption{\scriptsize $\text{RetraSyn}^\text{p}$ on T-Drive}
\end{minipage}
\begin{minipage}[t]{0.45\textwidth}
\centering
\includegraphics[width=8.5cm]{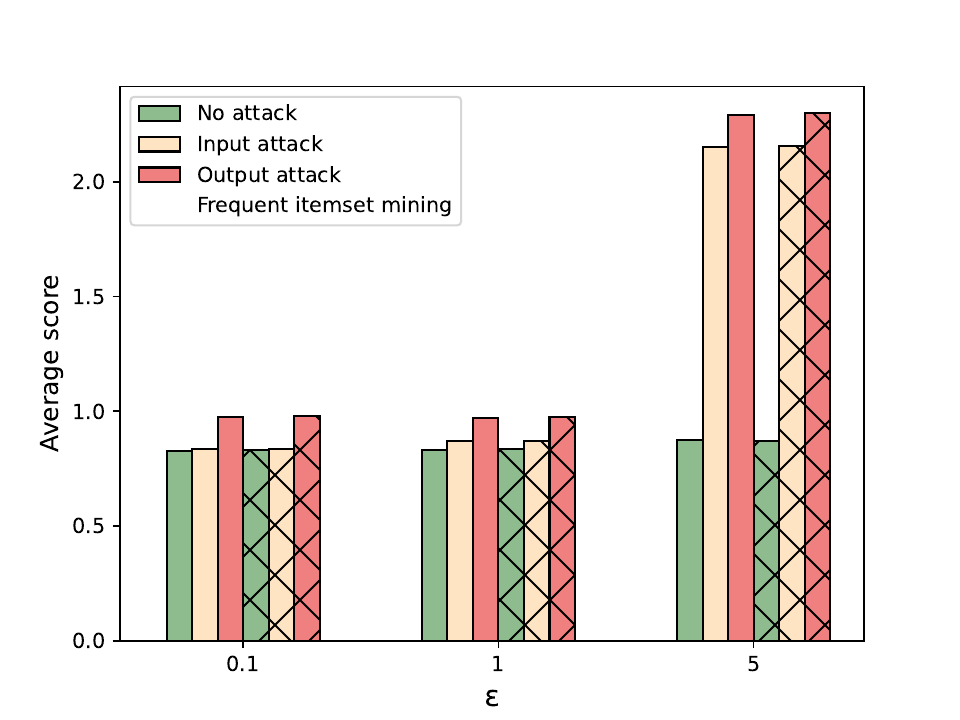}
\subcaption{\scriptsize $\text{RetraSyn}^\text{p}$ on Oldenburg}
\end{minipage}
\begin{minipage}[t]{0.45\textwidth}
\centering
\includegraphics[width=8.5cm]{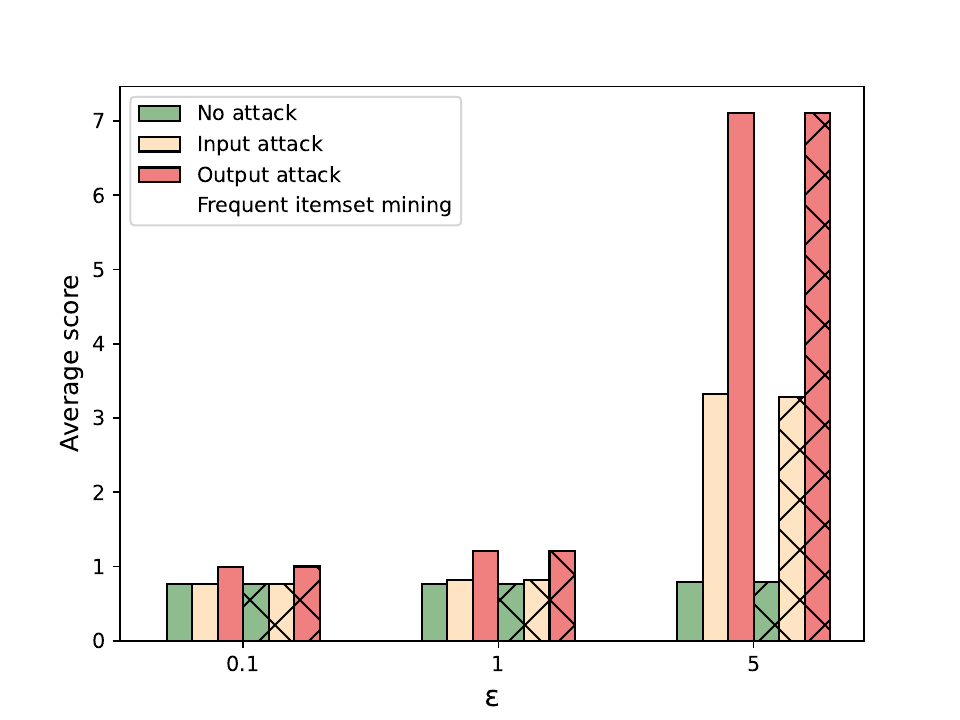}
\subcaption{\scriptsize $\text{RetraSyn}^\text{p}$ on SanJoaquin}
\end{minipage}
\vspace{-13pt}
\caption{Average scores of $\text{RetraSyn}^\text{p}$ with defenses.}
\label{RetraSyn_p_with_defense}
\end{figure}

\begin{figure*}[htbp]
\centering
\begin{minipage}[t]{0.48\textwidth}
\centering
\includegraphics[width=7cm, height=5cm]{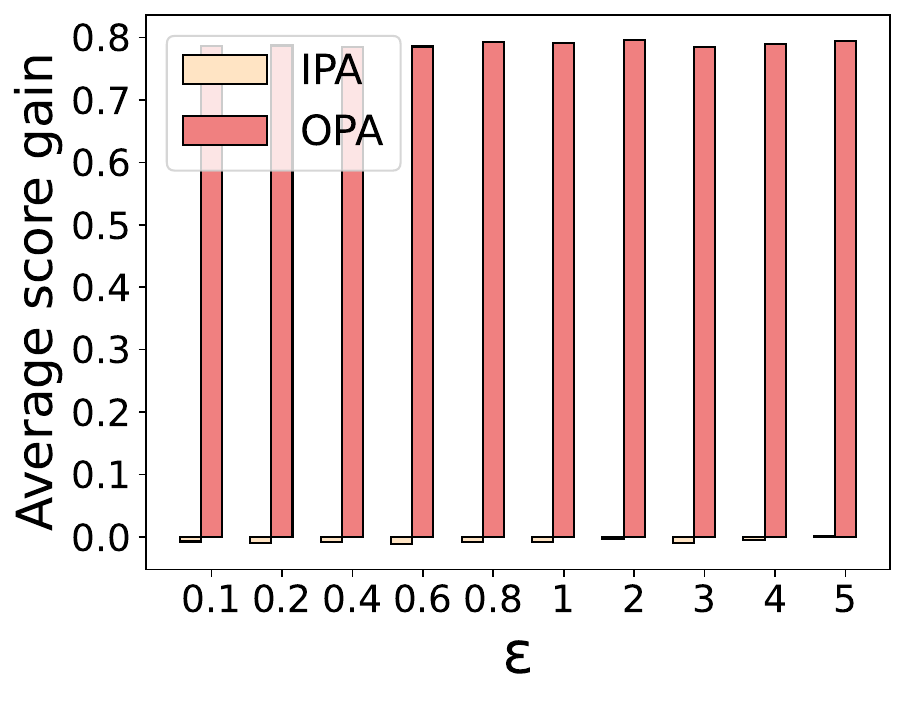}
\includegraphics[width=7cm, height=5cm]{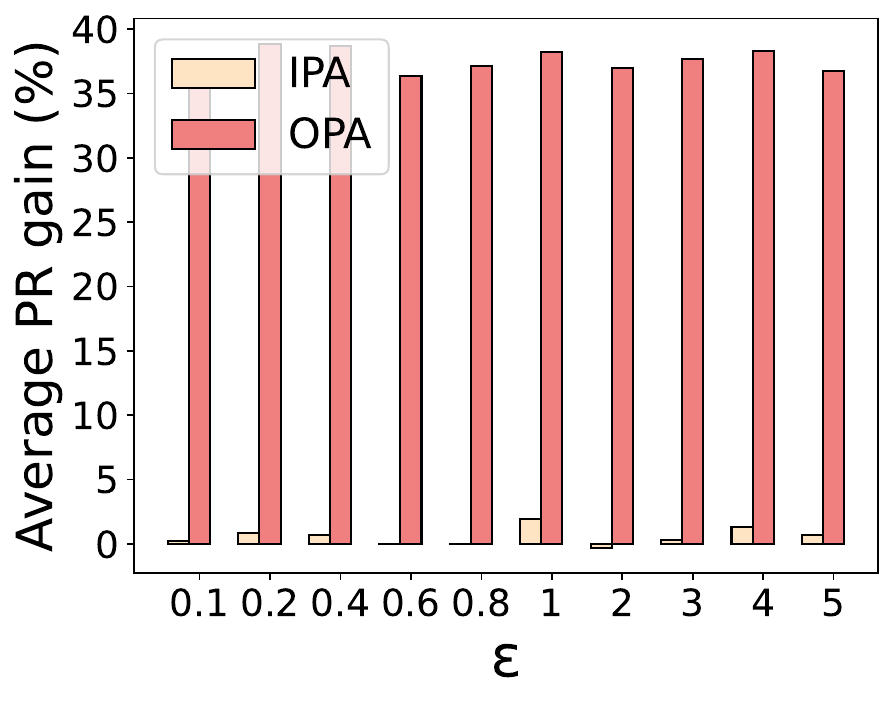}
\subcaption{$n$-gram on CPS}
\end{minipage}
\begin{minipage}[t]{0.48\textwidth}
\centering
\includegraphics[width=7cm, height=5cm]{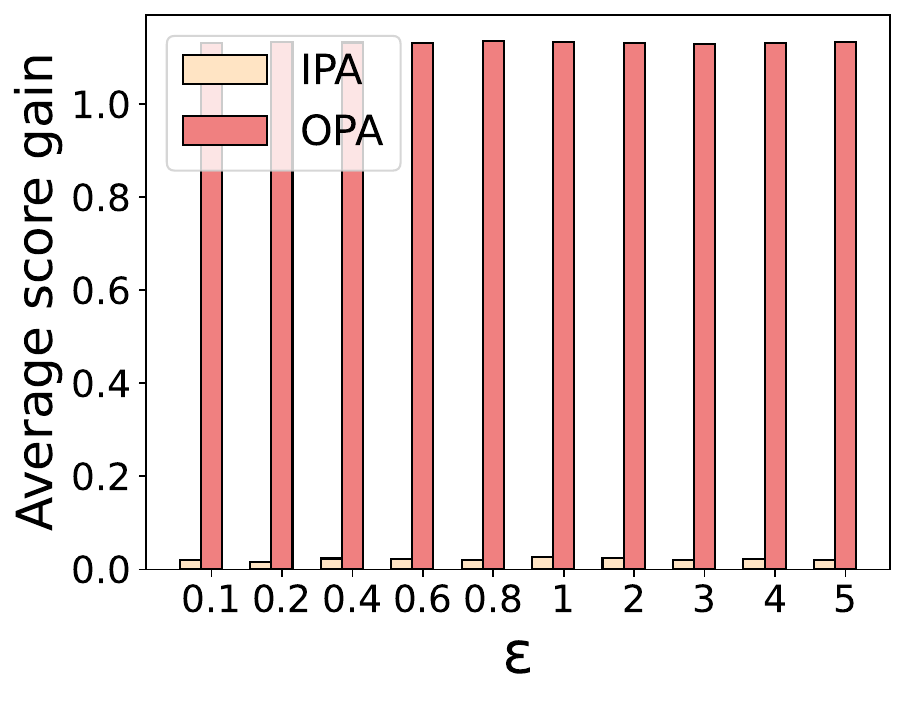}
\includegraphics[width=7cm, height=5cm]{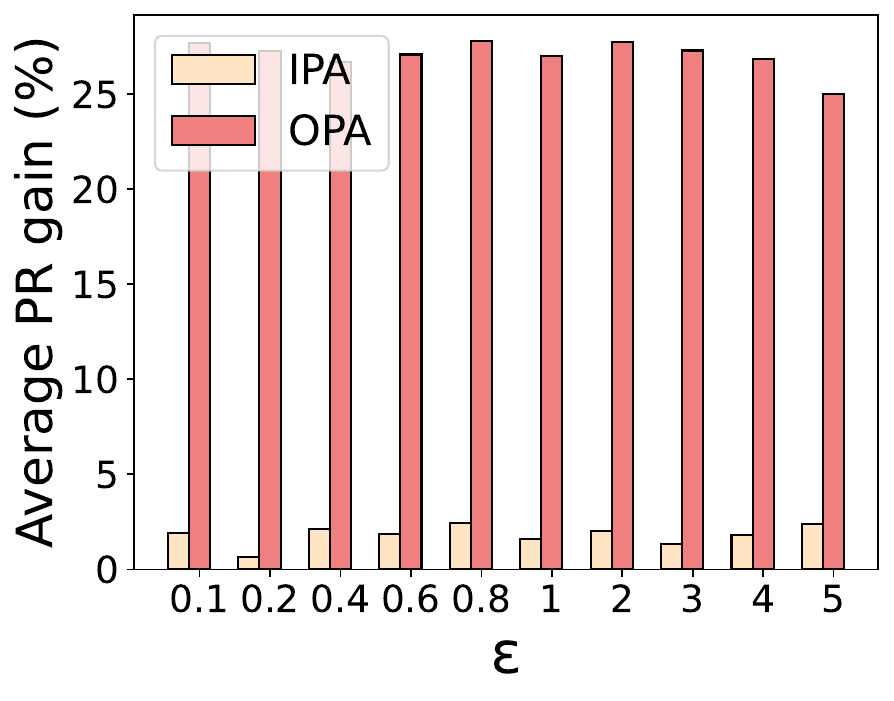}
\subcaption{$n$-gram on NYC}
\end{minipage}
\caption{Average score gain and PR gain of $n$-gram on CPS and NYC datasets with varying $\varepsilon$'s.}
\label{n_gram_All_eps}
\end{figure*}

\begin{figure*}[htbp]
\centering
\begin{minipage}[t]{0.48\textwidth}
\centering
\includegraphics[width=7cm, height=5cm]{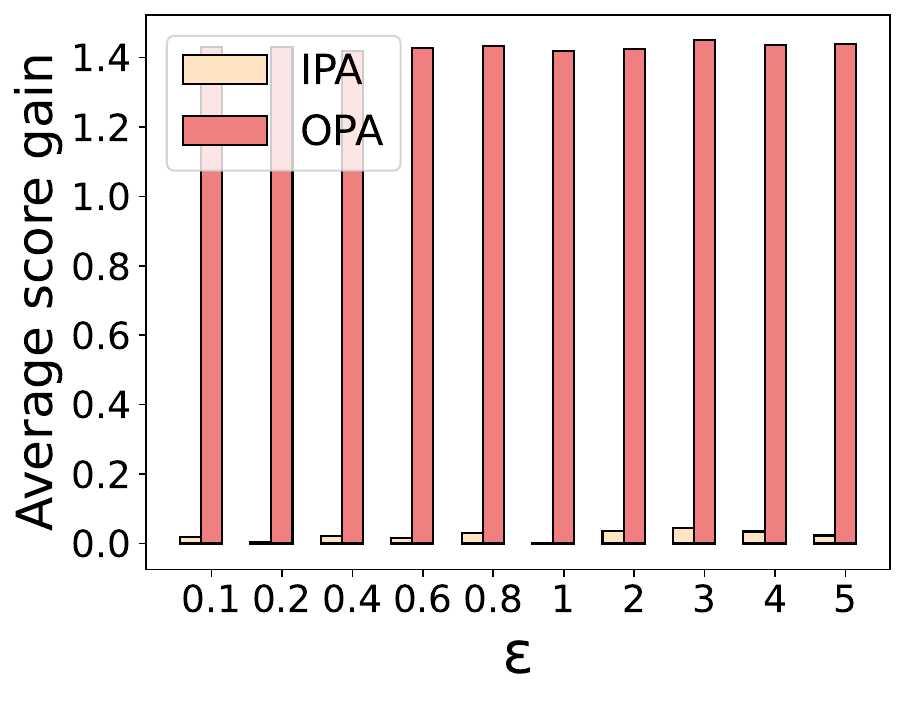}
\includegraphics[width=7cm, height=5cm]{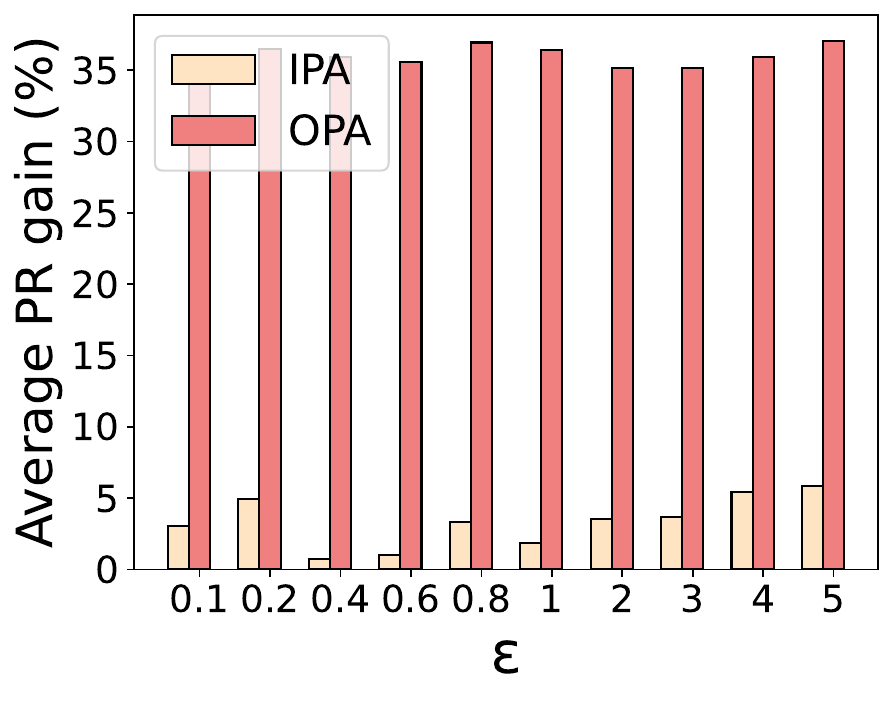}
\subcaption{ATP on CHI}
\end{minipage}
\begin{minipage}[t]{0.48\textwidth}
\centering
\includegraphics[width=7cm, height=5cm]{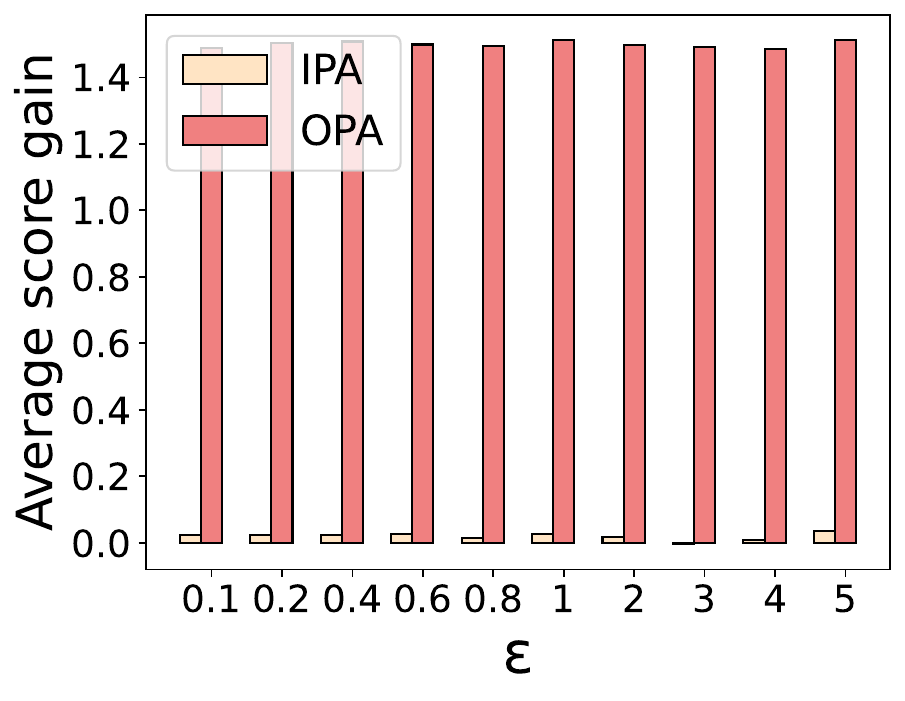}
\includegraphics[width=7cm, height=5cm]{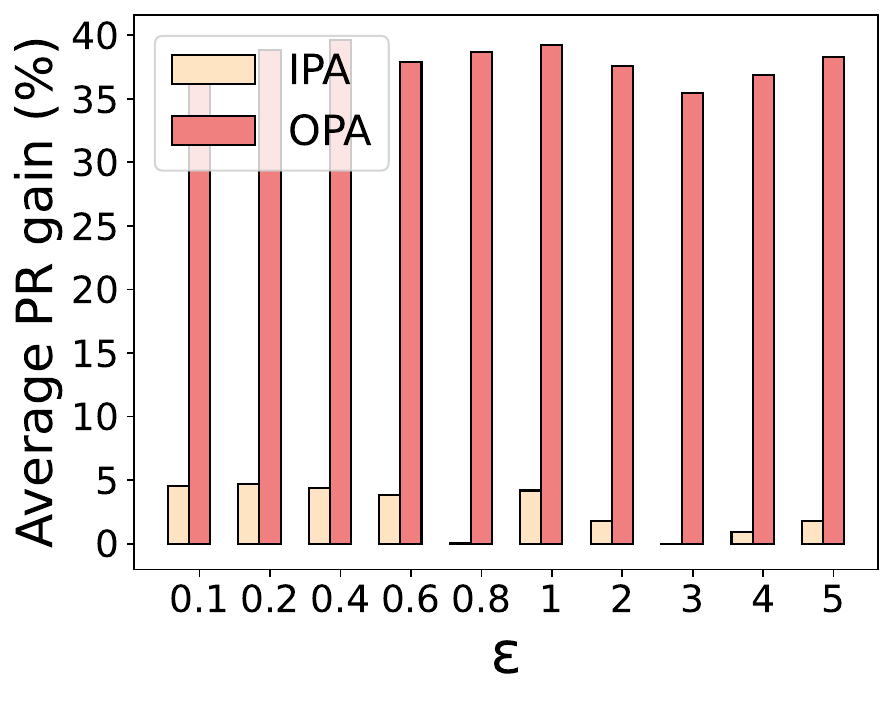}
\subcaption{ATP on CLE}
\end{minipage}
\begin{minipage}[t]{0.48\textwidth}
\centering
\includegraphics[width=7cm, height=5cm]{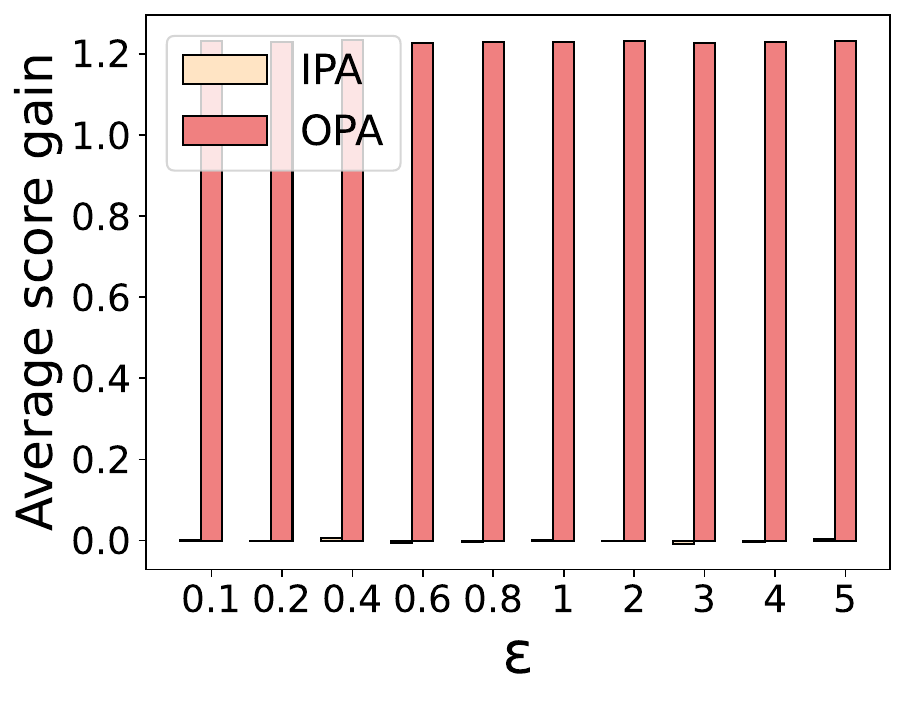}
\includegraphics[width=7cm, height=5cm]{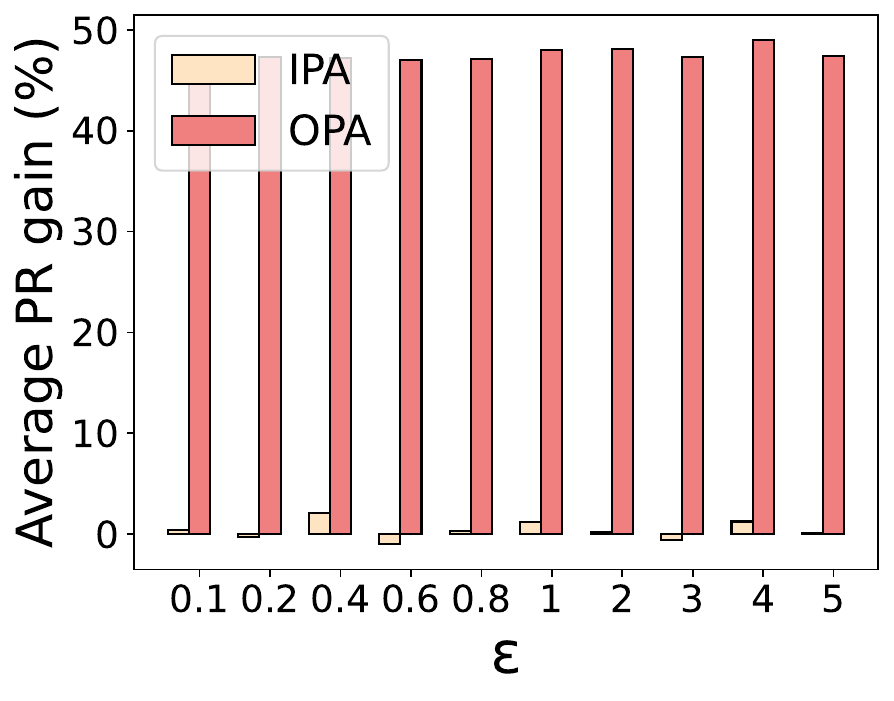}
\subcaption{ATP on CPS}
\end{minipage}
\begin{minipage}[t]{0.48\textwidth}
\centering
\includegraphics[width=7cm, height=5cm]{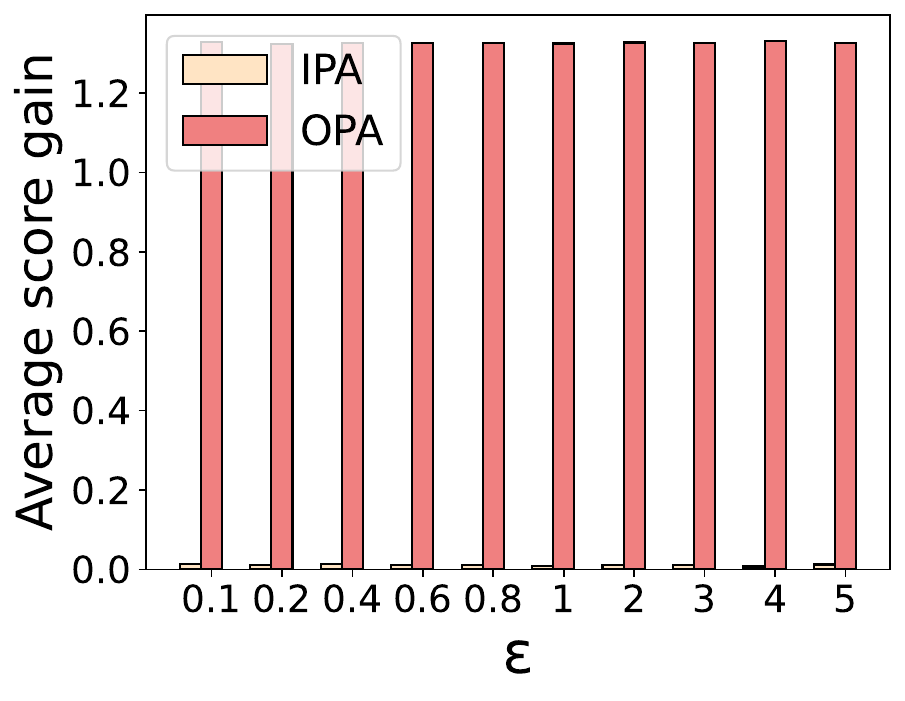}
\includegraphics[width=7cm, height=5cm]{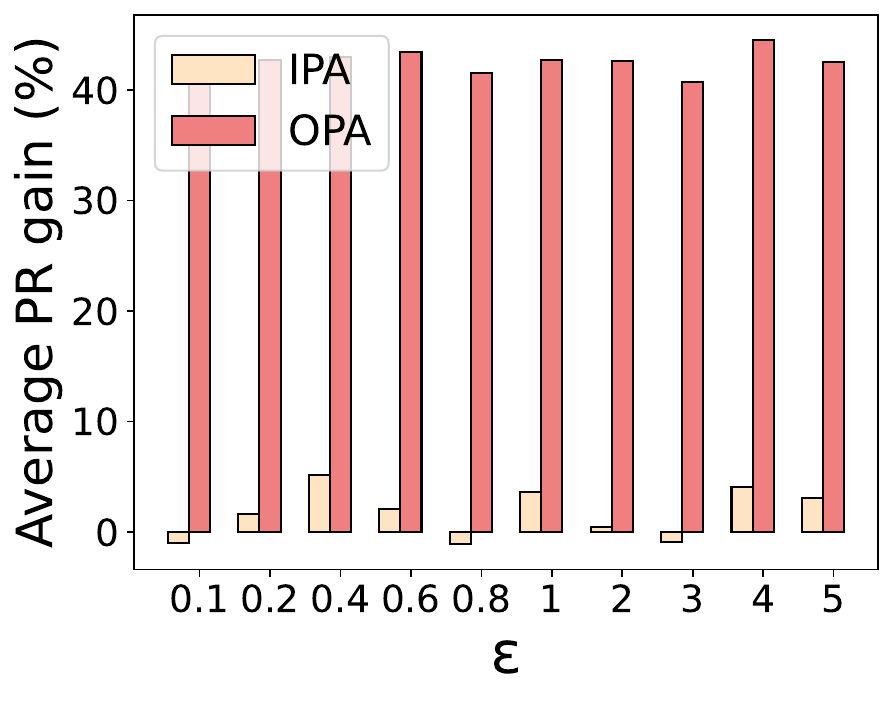}
\subcaption{ATP on NYC}
\end{minipage}
\caption{Average score gain and PR gain of ATP on CHI, CLE, CPS, and NYC datasets with varying $\varepsilon$'s.}
\label{ATP_All_eps}
\end{figure*}

\begin{figure*}[htbp]
\centering
\begin{minipage}[t]{0.32\textwidth}
\centering
\includegraphics[width=5.6cm, height=4cm]{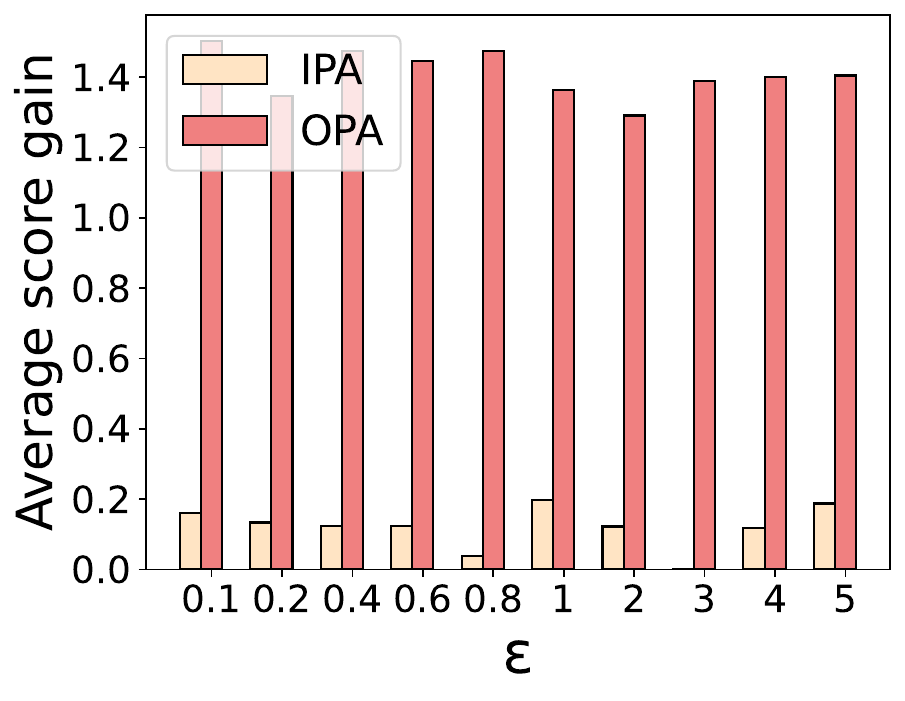}
\includegraphics[width=5.6cm, height=4cm]{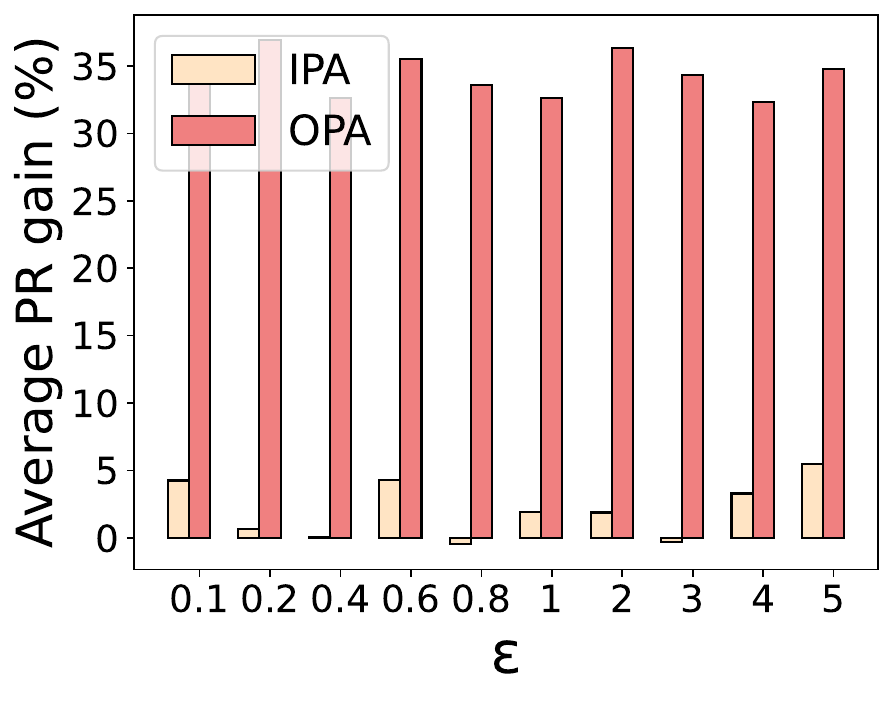}
\subcaption{LDPTrace on CPS}
\label{LDPTrace_All_eps_CPS}
\end{minipage}
\begin{minipage}[t]{0.32\textwidth}
\centering
\includegraphics[width=5.6cm, height=4cm]{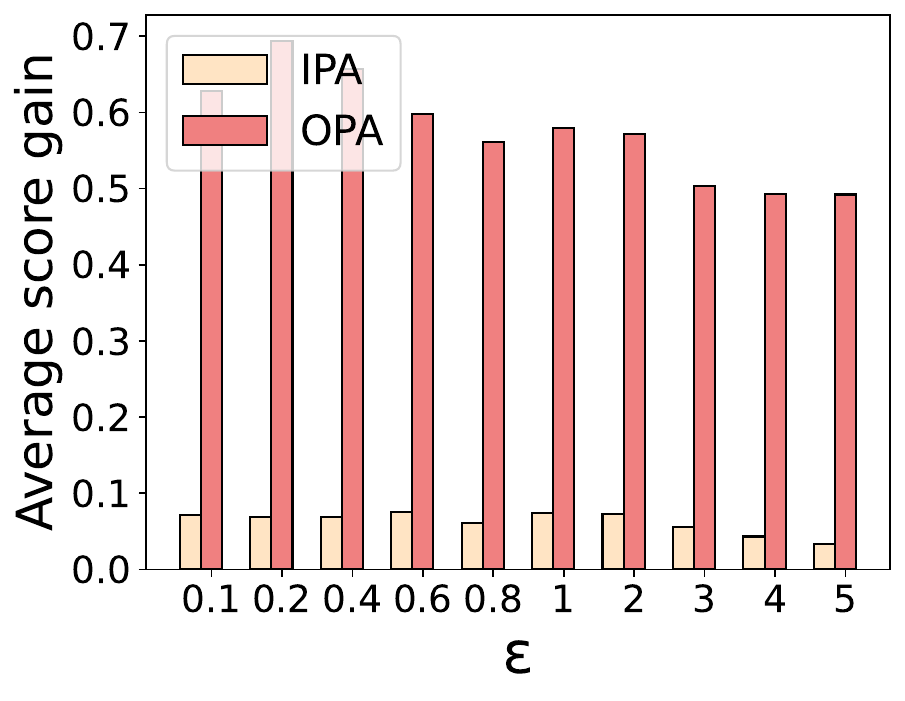}
\includegraphics[width=5.6cm, height=4cm]{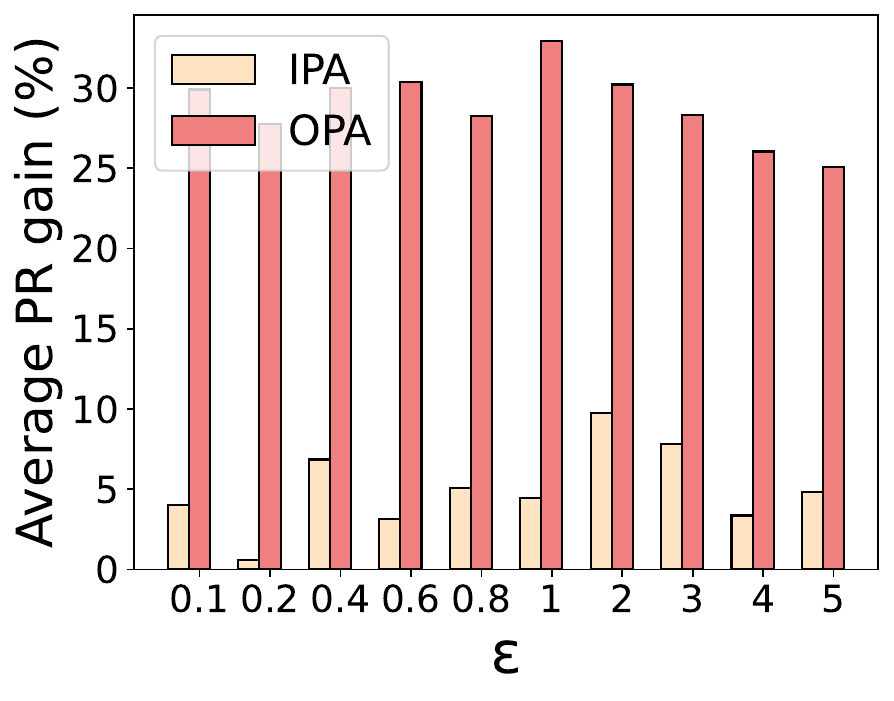}
\subcaption{LDPTrace on Oldenburg}
\label{LDPTrace_All_eps_Oldenburg}
\end{minipage}
\begin{minipage}[t]{0.32\textwidth}
\centering
\includegraphics[width=5.6cm, height=4cm]{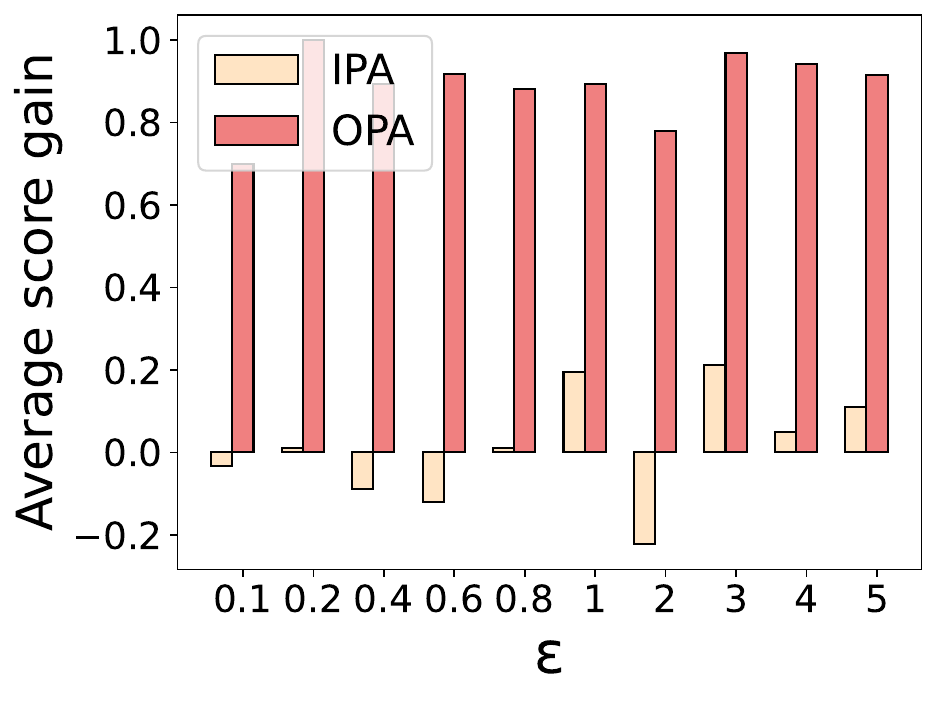}
\includegraphics[width=5.6cm, height=4cm]{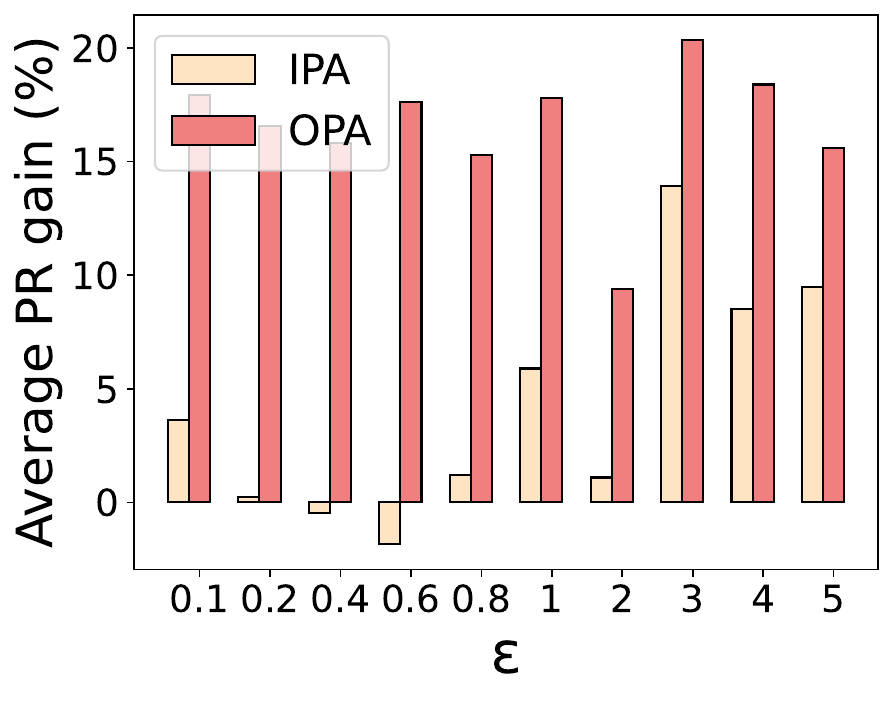}
\subcaption{LDPTrace on Porto}
\label{LDPTrace_All_eps_Porto}
\end{minipage}
\caption{Average score gain and PR gain of LDPTrace on CPS, Oldenburg, and Porto datasets with varying $\varepsilon$'s.}
\label{LDPTrace_All_eps}
\end{figure*}

\begin{figure*}[htbp]
\centering
\begin{minipage}[t]{0.48\textwidth}
\centering
\includegraphics[width=7cm, height=5cm]{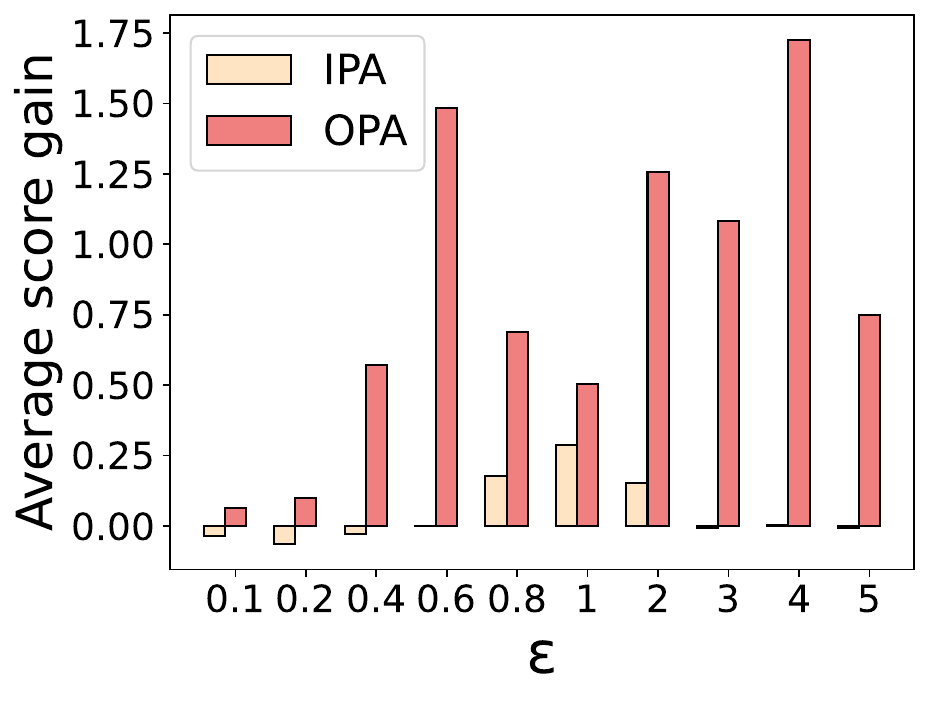}
\includegraphics[width=7cm, height=5cm]{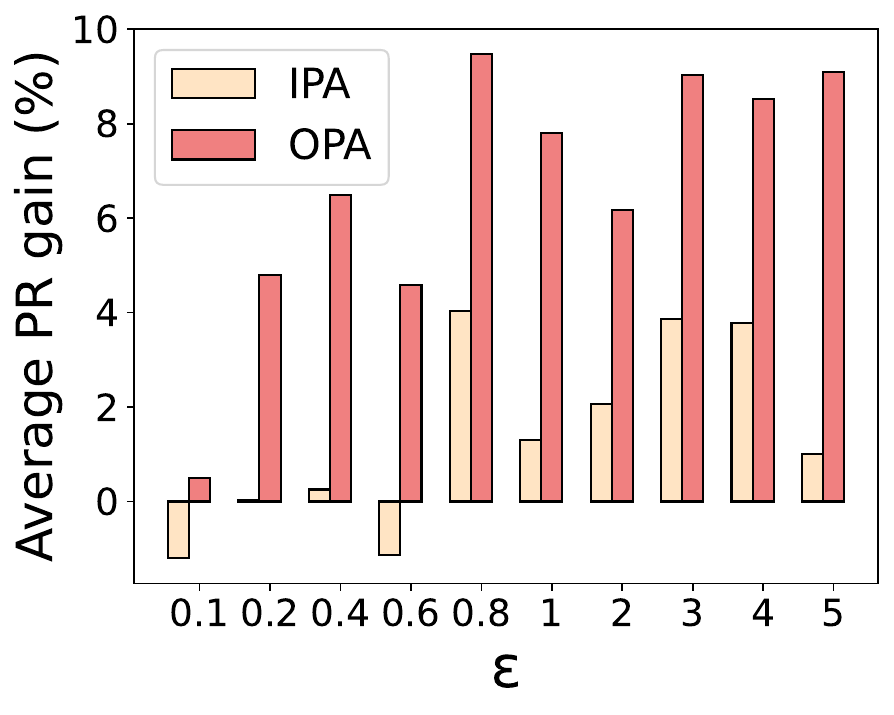}
\subcaption{PrivTC on Gowalla}
\end{minipage}
\begin{minipage}[t]{0.48\textwidth}
\centering
\includegraphics[width=7cm, height=5cm]{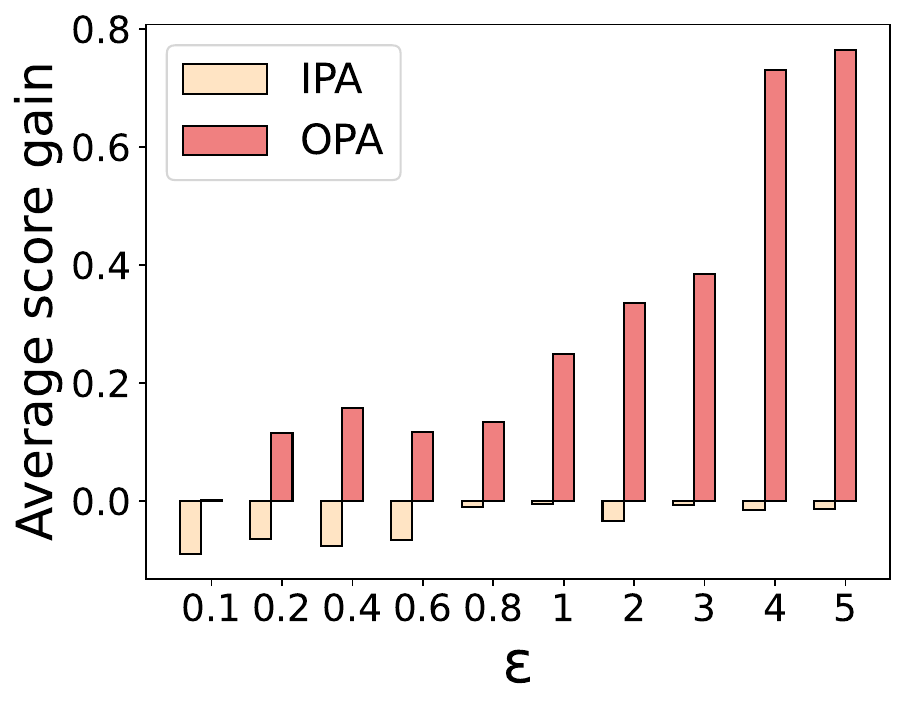}
\includegraphics[width=7cm, height=5cm]{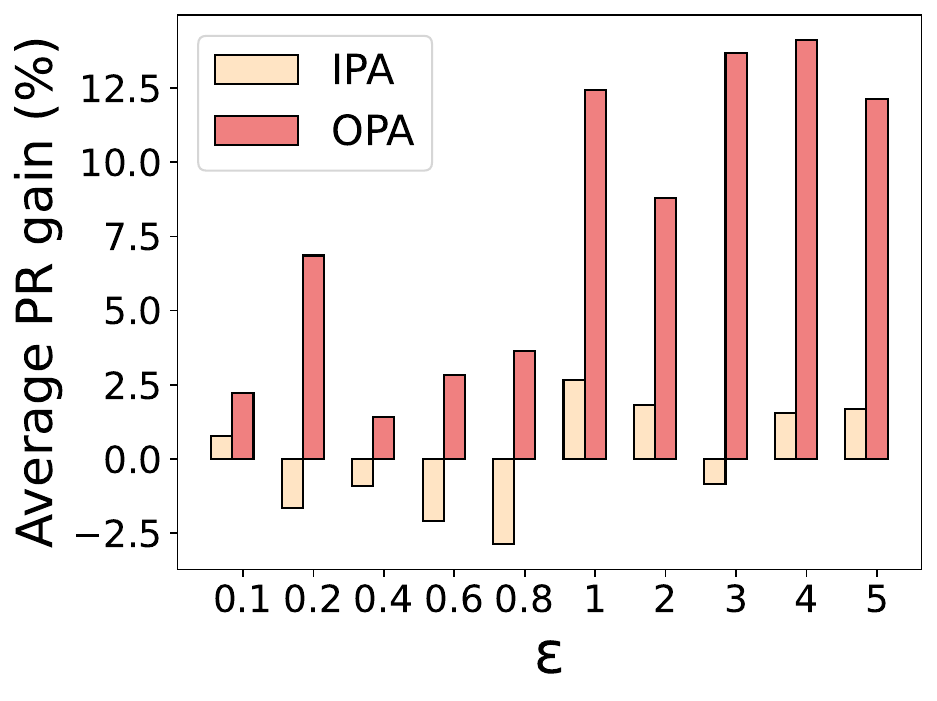}
\subcaption{PrivTC on Porto}
\end{minipage}
\caption{Average score gain and PR gain of PrivTC on Gowalla and Porto datasets with varying $\varepsilon$'s.}
\label{PrivTC_All_eps}
\end{figure*}

\begin{figure}[htbp]
\centering
\begin{minipage}[t]{0.5\textwidth}
\centering
\includegraphics[trim=11 23 40 40, clip, width=4.4cm]{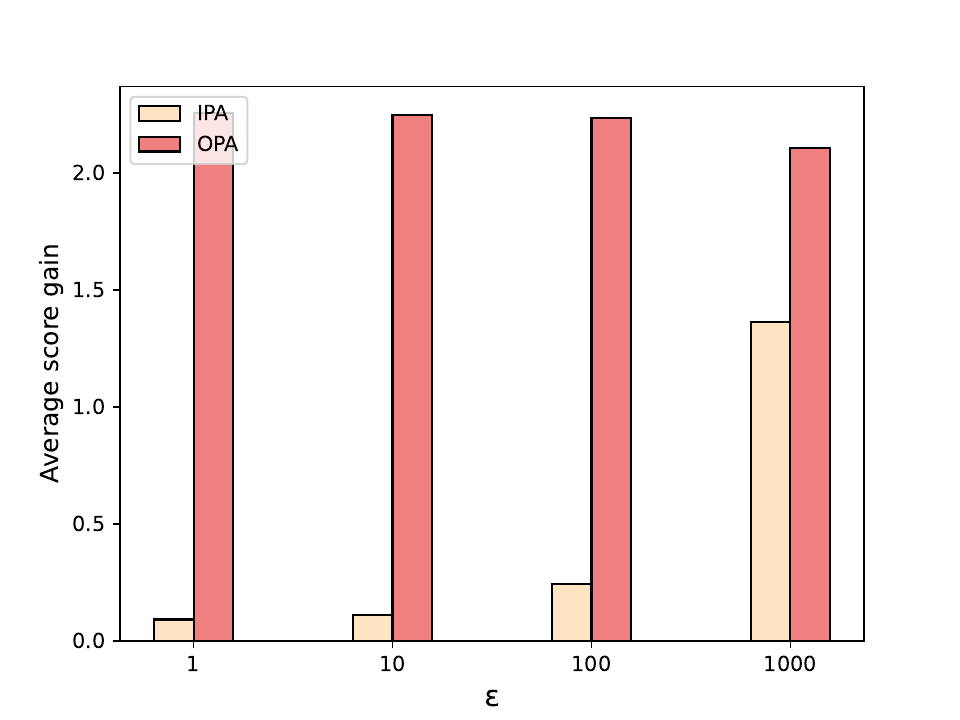}
\includegraphics[trim=11 23 40 40, clip, width=4.4cm]{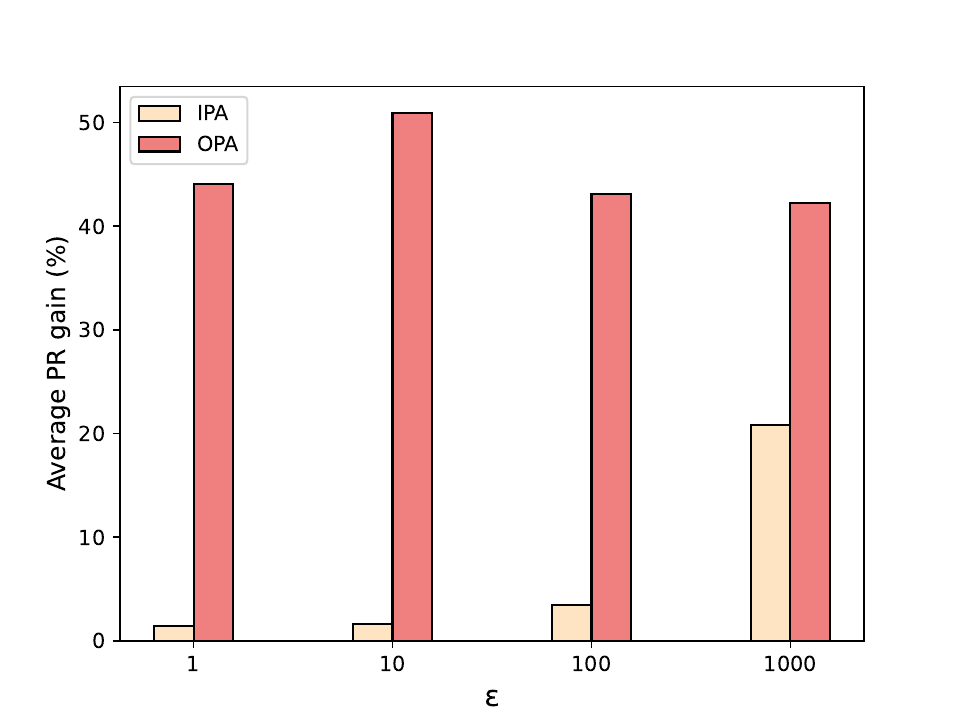}
\subcaption{$n$-gram}
\label{n_gram_more_eps}
\end{minipage}
\begin{minipage}[t]{0.5\textwidth}
\centering
\includegraphics[trim=11 23 40 40, clip, width=4.4cm]{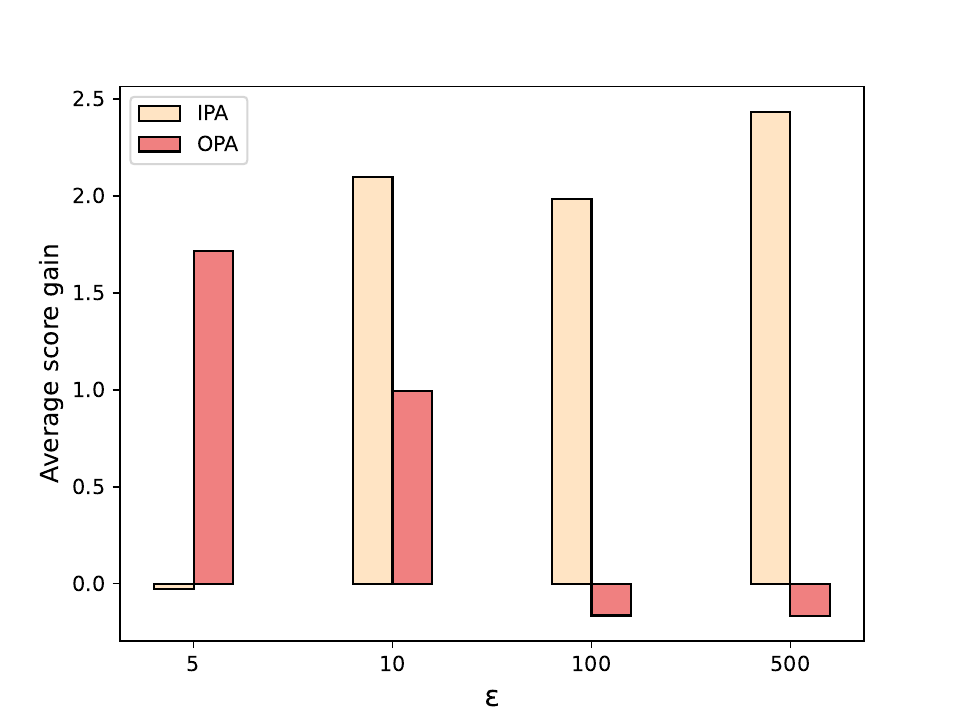}
\includegraphics[trim=11 23 40 40, clip, width=4.4cm]{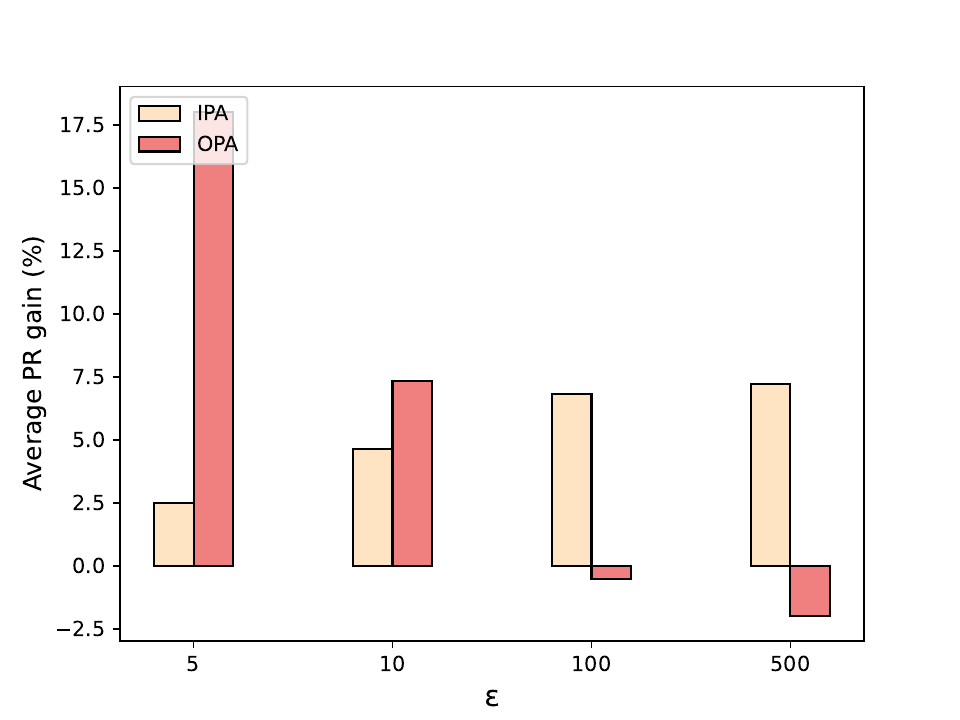}
\subcaption{PrivTC}
\label{PrivTC_more_eps}
\end{minipage}
\begin{minipage}[t]{0.5\textwidth}
\centering
\includegraphics[trim=11 23 40 40, clip, width=4.4cm]{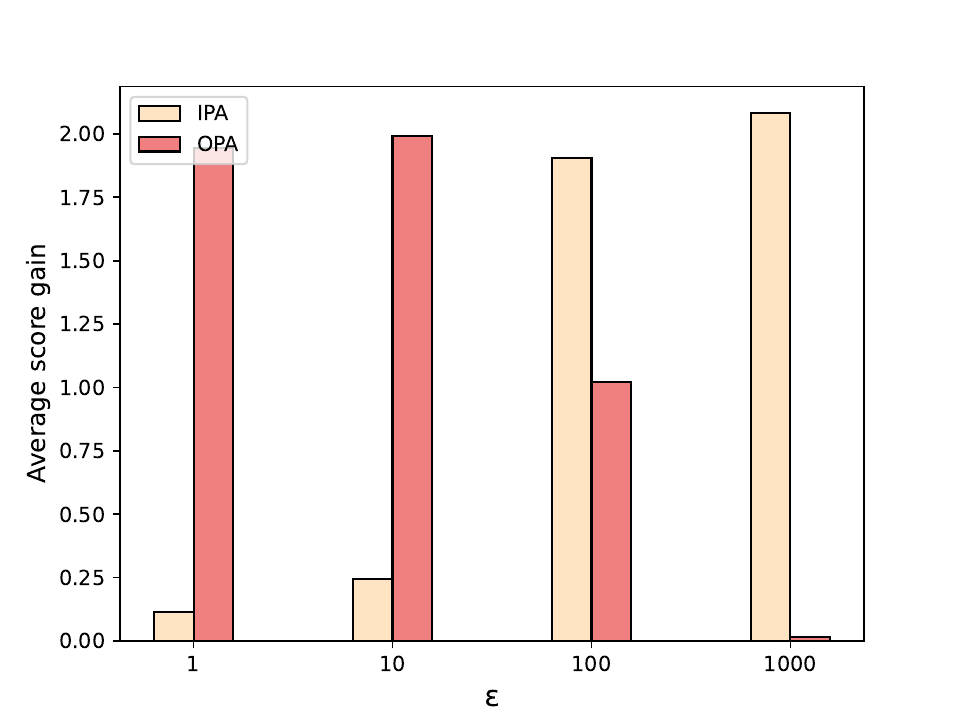}
\includegraphics[trim=11 23 40 40, clip, width=4.4cm]{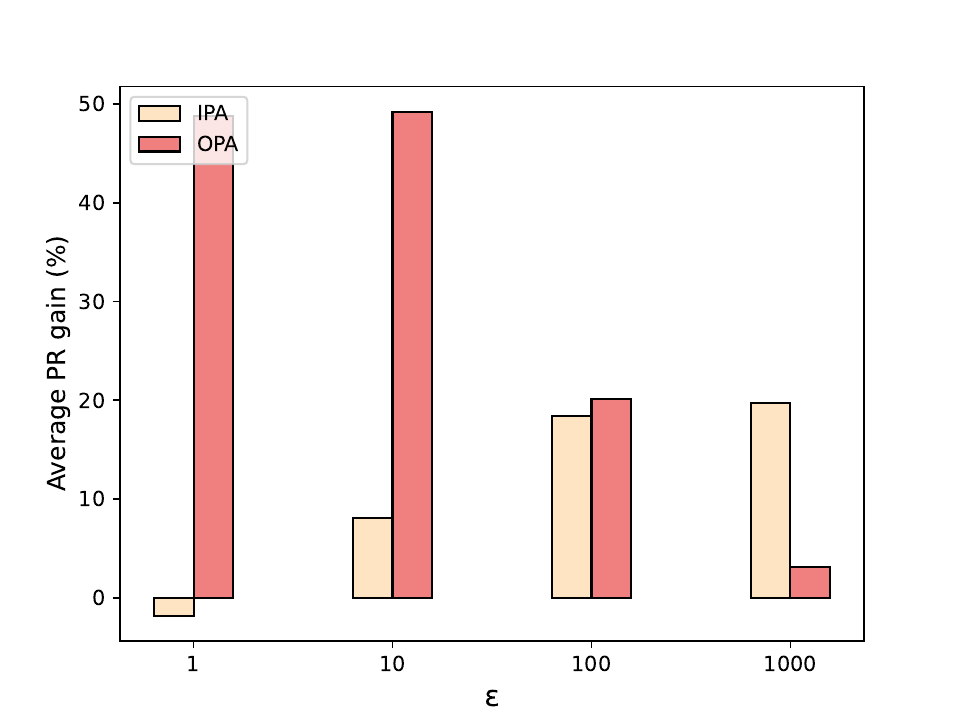}
\subcaption{LDPTrace}
\label{LDPTrace_more_eps}
\end{minipage}
\begin{minipage}[t]{0.5\textwidth}
\centering
\includegraphics[trim=11 23 40 40, clip, width=4.4cm]{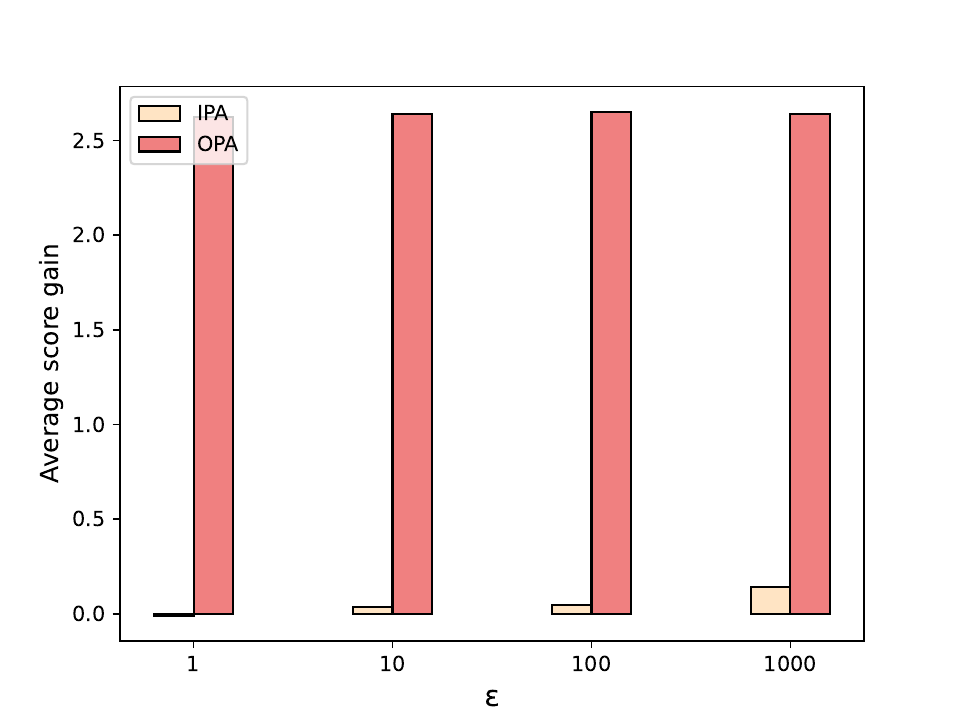}
\includegraphics[trim=11 23 40 40, clip, width=4.4cm]{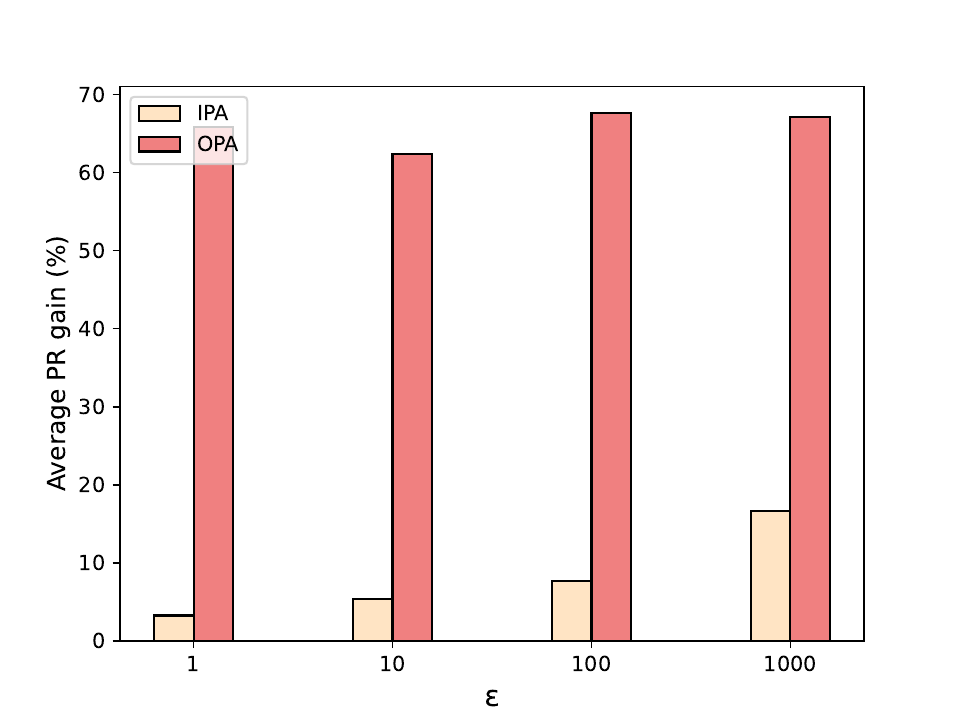}
\subcaption{ATP}
\label{ATP_more_eps}
\end{minipage}
\caption{Score gain and PR gain of $n$-gram, PrivTC, LDPTrace, and ATP averaged across all seven datasets.}
\label{More_eps}
\end{figure}

\subsection{An Example of \textsc{TraP}.}\label{sec: An Example of TraP}

To illustrate \textsc{TraP}, consider an example with 4 points - $a$, $b$, $c$, and $d$ within the point domain $\mathcal{P}$. The reachability is as follows: point $a$ is reachable to $b$; point $b$ is reachable to $a$, $c$, and $d$; point $c$ is reachable to $b$ and $d$; point $d$ is reachable to $b$ and $c$. The target set $\textit{TP}$ has scores: $\{(a, b): 1, (a, b, c): 2, (b, d): 1\}$. The total number of fake users is $m=12$, distributed as $m_{L=1}=3, m_{L=2}=5, m_{L=3}=4$. Additionally, the maximum allowable trajectory repetitions $\textit{max}_{\text{rep}}$ is limited to $2$.

Before proceeding with the trajectory instantiation, we first determine $\textit{PREF}=\{(~), (a), (a, b), (b)\}$. For $i=1$, we generate all length-$1$ trajectories: $(a)$, $(b)$, $(c)$, and $(d)$. Using the categorizing rules (longest match principle), we construct the table below, sorting trajectories in the same prefix class by their scores in descending order:

\begin{table}[H]
    \setlength{\belowcaptionskip}{3px}
    \centering
    \begin{tabular}{l|l}
		\textbf{Prefix} & \textbf{Trajectories whose suffix matches this prefix} \\
		\hline
        $(a)$ & $(a)_{\text{score=0}}$ \\
        $(a, b)$ & \\
        $(b)$ & $(b)_\text{score=0}$ \\
        $(~)$ & $\cancel{(c)_\text{score=0}}$, $\cancel{(d)_\text{score=0}}$ \\
    \end{tabular}
\end{table}

For $i=1$, three trajectories need to be included in the output set. Since all scores are $0$, we randomly select three trajectories from $\{(a), (b), (c), (d)\}$. To decide which trajectories to retain or discard for the next steps, we consider the following criteria:

\BlankLine

\noindent Prefix Category $(a)$ -
\begin{itemize}
\item With one trajectory $(a)$ in this category, which ends with point $a$, we check the reachable point set of $a$ to see how many points can be appended, which is 2 (points $a$ and $b$).
\item Since $\textit{max}_{\text{rep}}=2$, each trajectory can appear twice in the output set. Therefore, at least 4 trajectories can be generated from the length-$1$ trajectories in the $(a)$ category.
\item 
This category and its related ancestor categories do not yet have any trajectories. Hence, all length-$1$ trajectories in the $(a)$ category should be retained.
\end{itemize}

\noindent Prefix Category $(b)$ -
\begin{itemize}
\item Similar to the $(a)$ category, all length-$1$ trajectories in the $(b)$ category should be retained for the same reasons.
\end{itemize}

\noindent Prefix Category $(~)$ -
\begin{itemize}
\item The $(~)$ category is a suffix of any non-empty prefix. When all scores are equal, forming a target pattern by appending trajectory segments to length-$1$ trajectories in the $(a)$ and $(b)$ categories is faster than in the $(~)$ category.
\item Given that prefixes $(a)$ and $(b)$ can potentially generate at least $(2+4)*2=12$ longer trajectories, exceeding $m_{\text{max}}=5$, all trajectories in the $(~)$ category can be discarded.
\end{itemize}
With the task for $i=1$ completed, we proceed to $i=2$. In this step, we append all reachable points from the last point of each length-$1$ trajectory retained from the previous step.

The table below lists the prefixes and the corresponding trajectories whose suffix matches each prefix:

\begin{table}[H]
\setlength{\belowcaptionskip}{3px}
\centering
\begin{tabular}{l|l}
\textbf{Prefix} & \textbf{Trajectories whose suffix matches this prefix} \\
\hline
$(a)$ & $(a, a)_{\text{score=0}}$, $(b, a)_{\text{score=0}}$ \\
$(a, b)$ & $(a, b)_{\text{score=1}}$ \\
$(b)$ & $\cancel{(b, b)_\text{score=0}}$ \\
$(~)$ & $\cancel{(b, d)_{\text{score=1}}}$, $\cancel{(b, c)_\text{score=0}}$ \\
\end{tabular}
\end{table}

Recall that $m_{L=2} = 5$, indicating that there should be 5 length-$2$ trajectories in the output set. The optimal approach is to prioritize trajectories with higher scores. Therefore, $(a, b)$ and $(b, d)$ should be included twice each, as $\textit{max}_{\text{rep}}=2$. The remaining trajectory can be randomly selected from the zero-score trajectories.

Although all trajectories in prefix $(a)$ have scores of $0$, they must be retained since $(a)$ is not a suffix of any prefix. Adding trajectory segments could result in a higher-scoring trajectory than those from other prefixes. The trajectory $(a, b)$ must be preserved. Following the reachability and maximal repetition rules, $(a, b)$ can produce $4*2=8$ trajectories, surpassing $m_{\text{max}}=m_{L=3}=4$. Consequently, the trajectory $(b, b)$ should be removed as it has a lower score than $(a, b)$ and its category $(b)$ is a suffix of the category $(a, b)$. Note that although $(b, d)$ appears in the output set of length-$2$ trajectories, it is excluded in the subsequent instantiation round because its category $(~)$ is a suffix of the category $(a, b)$, and $(b, d)$'s score is not higher than $(a, b)$, which can produce eight length-$3$ trajectories.

The procedure is reiterated to create trajectories of length $i=3$, but it is excluded due to space limitations.

Note that Algorithm~\ref{alg:prefix-suffix} generates a poisonous trajectory set. For protocols that send transition probabilities of points, pairs, or triplets to the server instead of full trajectories, we make some adaptations to conduct attacks, as detailed in Section~\ref{sec:attack types}.

\subsection{Attacking Real-Time Trajectory Synthesis with Local
 Differential Privacy}\label{sec:RetraSyn}

In the preceding sections, we attacked and analyzed four different non-real-time LDP trajectory protocols. Here, we present our attack on the real-time LDP trajectory protocol, RetraSyn~\cite{hu2024retrasyn}.

The experimental settings are consistent with Section~\ref{sec:exp_setup}, except for the datasets, dataset size, and the $\textit{max}_{\text{rep}}$ parameter. RetraSyn is evaluated on three datasets: T-Drive, Oldenburg, and San Joaquin. These datasets contain trajectories with hundreds of timestamps, not all starting from timestamp 0. To ensure adequate point coverage across timestamps, we use 100,000 trajectories per dataset. Due to the increased dataset size, the $\textit{max}_{\text{rep}}$ parameter is set to 10 to allow more repetition per trajectory.

RetraSyn has two variants: $\text{RetraSyn}^\text{b}$ and $\text{RetraSyn}^\text{p}$, which differ in privacy budget allocation strategies. $\text{RetraSyn}^\text{b}$ uses a budget-division approach, where the total privacy budget $\varepsilon$ is split across individual timestamps, with all active users reporting at each timestamp using a fraction of the budget. In contrast, $\text{RetraSyn}^\text{p}$ employs a population-division approach, where the user population is divided, and only a subset of users reports at each timestamp, using the full budget $\varepsilon$. The trade-off between these variants lies in the frequency of reporting and the noise level: $\text{RetraSyn}^\text{b}$ provides more frequent updates with less noise per update, while $\text{RetraSyn}^\text{p}$ offers stronger privacy guarantees but with fewer reporting users.

Due to the time-intensive nature of the experiments, we perform one run per configuration. The average scores and PRs are shown in Figure~\ref{Fig:RetraSyn_AvgScore} and Figure~\ref{Fig:RetraSyn_AvgPR}, respectively. Results across all datasets confirm that OPA > IPA > no attack, demonstrating the effectiveness of our attack method.

Additionally, we test defense mechanisms. Similar to PrivTC and LDPTrace, RetraSyn incorporates a server-side step for aggregating and synthesizing trajectory datasets. RetraSyn also includes normalization as part of its process, so we only add frequent itemset mining as a defense strategy. Results are presented in Figures~\ref{RetraSyn_b_with_defense} and \ref{RetraSyn_p_with_defense}.

\subsection{Comparing the Attack Performance for OUE and OLH Versions of PrivTC}\label{sec:OUE_OLH_PrivTC}
For the OLH version of PrivTC, since only one item is randomly selected for transmission to the server, we randomly choose one item from the points, pairs, or triplets of poisoned trajectories for each user in the output attack. We perform five experimental runs to calculate the average result.

Intuitively, attacking the OLH version of PrivTC should be less effective than attacking the OUE version, and our results confirm this. \cref{Fig:PrivTC_OUE_OLH_AvgScore,Fig:PrivTC_OUE_OLH_AvgPR} present the results for $\varepsilon = 1$. On most datasets, the average scores and PRs of both IPA and OPA remain higher than those without any attack. Although the attack on OLH is slightly less effective than on OUE, it is still difficult to defend against attacks targeting the OLH version.

\subsection{Execution Time: Prefix-suffix vs. Brute-force Approach}\label{sec:execution_time_comparison}

To assess the efficiency of our prefix-suffix approach compared to the brute-force method, we conducted experiments using the parameters described in Section~\ref{sec:exp_parameters}, except for selecting the top 5000 highest-scoring trajectories for each length. We measured the maximum trajectory length that each method could process within a limited time frame. The CPS dataset, with 262 POIs, was chosen to mitigate the brute-force method's disadvantages. The experiments were run on hardware with eight Intel Xeon CPUs (2.20 GHz) and 53 GB of memory.

Using the brute-force method to generate poisoned trajectories of length 5 resulted in memory limitations, restricting it to generating only up to length 4, which took approximately 7 minutes and 30 seconds. We used this time limit to evaluate the prefix-suffix approach, which was able to generate trajectories up to length 183, demonstrating a substantial improvement in efficiency.

We also tested the prefix-suffix method on the unclustered Porto dataset, containing 25,382 POIs, and let it run for 24 hours. As shown in Figure~\ref{Fig:time-length}, it was able to generate trajectories up to length 110 within that time frame.

\subsection{Why Not Use Single Point Frequency as the Attack Target?}\label{sec:Baseline_single_point} 
When attacking LDP trajectory protocols, the most intuitive approach is to increase the frequency of a single point. However, we chose to target specific patterns as our attack objective. Patterns can include trajectory segments of any length, including single points (patterns of length 1). This choice stems from the fact that the targeted protocols do not solely collect single-point frequencies. Successfully attacking single points alone does not fully compromise the protocol. In contrast, targeting patterns of varying lengths provides greater flexibility, allowing us to manipulate frequencies from single points to entire trajectories, thus enabling a more comprehensive attack strategy.

We conducted experiments to compare single-point-based attacks with pattern-based attacks. The single-point-based attack focuses on increasing the frequency of target patterns of length 1, while the pattern-based attack targets patterns ranging from length 1 to $k_{\text{max}}$. The metric used to evaluate both attacks is the average PR, as defined in Eq.~\ref{AvgPR}, which calculates percentile ranks for target patterns of length 1 to $k_{\text{max}}$.

For $k_{\text{max}}$ values from 1 to 6, we performed five experiments and averaged the results, with a privacy budget $\varepsilon$ of 1. The results are shown in \cref{Fig:Single_point_vs_Pattern_PR_kmax1,Fig:Single_point_vs_Pattern_PR_kmax2,Fig:Single_point_vs_Pattern_PR_kmax3,Fig:Single_point_vs_Pattern_PR_kmax4,Fig:Single_point_vs_Pattern_PR_kmax5,Fig:Single_point_vs_Pattern_PR_kmax6}. When $k_{\text{max}} = 1$, both attack methods perform similarly. As $k_{\text{max}}$ increases, the average PR of the pattern-based attack slightly decreases due to the difficulty of preserving longer patterns, while the single-point-based attack's average PR drops significantly since it cannot generate target patterns longer than 1. Overall, as $k_{\text{max}}$ increases, the pattern-based attack outperforms the single-point-based attack more clearly.

\subsection{Supplementary Algorithms}\label{sec: Supplementary Algorithms}

The function $\text{CountPattern}(\textit{tp},\tau)$ returning the number of subarrays in trajectory $\tau$ that match the target pattern $\textit{tp}$ is shown in Algorithm~\ref{alg:CountPattern}. A brute-force algorithm for the optimization problem in Equation~(\ref{maximization_problem}) is shown in Algorithm~\ref{alg:brute-force}.

\begin{algorithm}[t]
    \caption{CountPattern (in Brute-force Approach)}
    \label{alg:CountPattern}
    \SetKwInOut{Input}{Input}
    \SetKwInOut{Output}{Output}
    \Input{Target pattern $\textit{tp}$, trajectory $\tau$}
    \Output{A number $\textit{num}$ indicating how many subarrays of $\tau$ matches $\textit{tp}$}  
    
    Initialize $\textit{num}$ to $0$;
    
    \For{$0 \le i \le |\tau|-|\textit{tp}|$}{
        Initialize $\textit{match}$ to $1$;
        
        \For{$1 \le j \le |\textit{tp}|$}{
        
            \If{$\tau[i+j] \ne \textit{tp}[j]$}{
            
                $\textit{match}=0$;
                
                Break;
            } 
        }

        $\textit{num} := \textit{num}+\textit{match}$;
    }
    \KwRet{$\textit{num}$};
\end{algorithm}

\begin{algorithm}[t]
    \caption{Brute-force Approach}
    \label{alg:brute-force}
    \SetKwInOut{Input}{Input}
    \SetKwInOut{Output}{Output}
    \Input{Point domain $\mathcal{P}$, target pattern set $\textit{TP}$, length distribution $m_{L=i} \hspace{0.5em} \forall L_{\min} \le i \le L_{\max}$, maximum permissible trajectory repetitions $\textit{max}_{\text{rep}}$}
    \Output{A trajectory set $\mathcal{T}^*$ containing $m$ trajectories}  
    
    Initialize $\mathcal{T}^*$ to $\emptyset$;
    
    \For{$L_{\min} \le i \le L_{\max}$}{
    
        Instantiate all trajectories that satisfy the reachability constraint with length $i$;
        
        \tcp{denote as \(\mathcal{T}_{L=i}\)}
        
        Calculate the score of each trajectory using Equation~(\ref{TrajScore});
        
        Sort $\mathcal{T}_{L=i}$ in descending order according to scores;
        
        \tcp{denote the sorted array as \(\mathcal{T}^{\text{sorted}}_{L=i}\)}
    
        Initialize $\mathcal{T}^*_{L=i}$ to $\emptyset$;

        \For{$1 \le j \le \lfloor m_{L=i} / \textit{max}_{\textnormal{rep}} \rfloor + 1$}{
            \For{$1 \le \textnormal{times} \le \textit{max}_{\textnormal{rep}}$}{
                \eIf{$|\mathcal{T}^*_{L=i}| < m_{L=i}$}{
                    Put $\mathcal{T}^{\text{sorted}}_{L=i}[j]$ into $\mathcal{T}^*_{L=i}$;
                }{
                    Break;
                }
            }
        }
        
        Add $\mathcal{T}^*_{L=i}$ to $\mathcal{T}^*$;
    }
    \KwRet{$\mathcal{T}^*$};
\end{algorithm}

\subsection{Missing Pseudocode}\label{sec: Missing Pseudocode}
Algorithms~\ref{alg:pick-high} and~\ref{alg:Delete} are the pseudocode for \verb"Pick-High" and \verb"Delete" in Section~\ref{sec: TraP: A Prefix-Suffix Approach}. 

\begin{algorithm}[h]
    \caption{Determine the Maximal Trajectory Score Set}
    \label{alg:pick-high}
    
    \SetKwFunction{Pick}{Pick-High}
    \SetKwProg{Fn}{Function}{:}{}
    \Fn{\Pick{$\Omega_{i+1}$, $m_{L=i}$, $\textit{max}_{\textnormal{rep}}$, $SC$}}{
        $\Omega_{i+1}^{sorted}$ = sort $\Omega_{i+1}$ by criteria=score with $SC$\\
        Initialize $\mathcal{T}^*_{L=i}$ to $\emptyset$\\

        \For{$j = 1$ \rm{to} $\lfloor m_{L=i} / \textit{max}_{\textnormal{rep}} \rfloor + 1$}{
            \For{$\textnormal{times} = 1$ \rm{to} $\textit{max}_{\textnormal{rep}}$}{
                \eIf{$|\mathcal{T}^*_{L=i}| < m_{L=i}$}{
                    Put $\Omega_{i+1}^{sorted}[j]$ into $\mathcal{T}^*_{L=i}$
                }{
                    Break
                }
            }
        }
    Return $\mathcal{T}^*_{L=i}$
    }
\end{algorithm}

\begin{algorithm}[h]
    \caption{Delete Hopeless Trajectories}
    \label{alg:Delete}
    \small
    
    \SetKwFunction{Delete}{Delete}
    \SetKwProg{Fn}{Function}{:}{}
    \Fn{\Delete{$\Omega_{i+1}$, $m_{\text{max}}$, $\textit{max}_\text{rep}$, $\textit{rps}$, $\textit{SC}$, $\textit{PREF}$, $\mathcal{M}_i$}}{

        Initialize $\mathcal{M}_{i}^{selected}$ = $\{u_{\tau}:[\enspace] | u_{\tau}\in PREF\}$\\
        \For{$u_{\tau} \in PREF$}{
            \tcc{The ancestor of the prefix $u_\tau$ refers to the categories in PREF that have $u_\tau$ as their suffix}
            ancestor = look for ancestor of $u_\tau$\\
            \If{$|\rm{ancestor}| \neq 0$}
            {
                \For{$\tau \in \mathcal{M}_{i}[u_{\tau}]$}{
                    \If{$SC[\tau] > \max_{\tau_{j} \in \rm{ancestor}}\{SC[\mathcal{M}_{i}[\tau_{j}][0]]\}$}
                    {
                        \tcc{$\because \mathcal{M}_{i}$ is sorted $\therefore$ the first trajectory in each prefix category has largest score}
                        Append $\tau$ to $\mathcal{M}_{i}^{selected}[u_\tau]$
                    }\Else{
                        $accumulate = 0$\\
                        \For{$\tau_{j} \in \rm{ancestor}$}{
                            \For{$\tau_{k} \in \mathcal{M}_{i}^{selected}[\tau_{k}]$}{
                                \If{$SC[\tau_{k}] \geq SC[\tau]$}{
                                    $accumulate \mathrel{+}= rps[\tau_{k}[-1]]\times \textit{max}_\text{rep}$
                                }
                            }
                        }
                        \If{$accumulate \le m_{\text{max}}$}{
                            Append $\tau$ to $\mathcal{M}_{i}^{selected}[u_\tau]$
                        }\Else{
                            Remove $\tau$ from $\Omega_{i+1}$
                        }
                    }
                }
            }
            \Else{
                $accumulate = 0$\\
                \For{$\tau \in \mathcal{M}_{i}[u_{\tau}]$}
                {
                    \If{$accumulate \le m_{\text{max}}$}
                    {
                        $accumulate \mathrel{+}= rps[\tau[-1]]\times \textit{max}_\text{rep}$
                    }\Else{
                        Remove $\tau$ from $\Omega_{i+1}$ 
                    }
                }
            }
        }
    }
\end{algorithm}

\end{document}